\documentclass[useAMS,usenatbib]{mn2e}
\usepackage{graphicx,color,oldgerm}
\usepackage{amssymb,amsmath}
\usepackage[breaklinks=true]{hyperref}

\let\sun=\odot
\newcommand{\SgrA}{Sgr~A$^\ast$}
\newcommand {\MBH}{\ensuremath{M_{\mathrm{BH}}}}
\newcommand {\Msun}{\ensuremath{{\rm M}_{\odot}}}
\newcommand {\Mstar}{\ensuremath{M_{\ast}}}
\newcommand {\Rsun}{\ensuremath{{\rm R}_{\odot}}}
\newcommand {\Mpthree}{\ensuremath{\Msun\,\mathrm{pc}^{-3}}}
\newcommand {\RS}{\ensuremath{R_{\mathrm{S}}}}
\newcommand {\kms}{\ensuremath{\mathrm{km\,s}^{-1}}}
\newcommand {\peryr}{\ensuremath{\mathrm{yr}^{-1}}}
\newcommand{\HOP}{{\texttt{HOP}}}
\newcommand{\Removed}[1]{} %
\newcommand{\rem}[1]{} 
\newcommand{\Comment}[1]{{\color{red}[COMMENT:] #1}}
\newcommand{\Added}[1]{{ #1}}

\title[Comprehensive set of main sequence stellar collisions\\] {%
A comprehensive set of simulations of high-velocity collisions
between main sequence stars.}

\author[M. Freitag and W. Benz]{
Marc Freitag$^{1,2}$\thanks{E-mail: m-freitag@northwestern.edu. Present address: Department of Physics and Astronomy, Northwestern University, 2145 Sheridan Road, Evanston, IL 60208-0834, USA }, 
and Willy Benz$^{3}$\thanks{E-mail: willy.benz@phim.unibe.ch}\\
$^{1}$Observatoire de Gen\`eve, Chemin des Maillettes 51, CH-1290 Sauverny, Switzerland\\
$^{2}$Astronomisches Rechen-Institut, M\"onchhofstra{\ss}e 12-14, D-69120 Heidelberg, Germany\\
$^{3}$Universit\"at Bern, Sidlerstrasse 5, CH-3012 Bern, Switzerland}

\begin{document}

\date{Accepted. Received; in original form}

\pagerange{\pageref{firstpage}--\pageref{lastpage}} \pubyear{2004}

\maketitle

\label{firstpage}

\begin{abstract}
  We report on a very large set of simulations of collisions between
  two main sequence (MS) stars. These computations were done with the
  ``Smoothed Particle Hydrodynamics'' method. Realistic stellar
  structure models for evolved MS stars were used. In order to sample
  an extended domain of initial parameters space (masses of the stars,
  relative velocity and impact parameter), more than 14\,000
  simulations were carried out. We considered stellar masses ranging
  between 0.1 and $75\,\Msun$ and relative velocities up to a few
  thousands {\kms}. To limit the computational burden, a resolution of
  1000--32\,000 particles per star was used. The primary goal of this
  study was to build a complete database from which the result of any
  collision can be interpolated. This allows us to incorporate the
  effects of stellar collisions with an unprecedented level of realism
  into dynamical simulations of galactic nuclei and other dense
  stellar clusters. We make the data describing the initial condition
  and outcome (mass and energy loss, angle of deflection) of all our
  simulations available on the Internet. We find that the outcome of
  collisions depends sensitively on the stellar structure and that, in
  most cases, using polytropic models is inappropriate.  Published
  fitting formulae for the collision outcomes, established from a
  limited set of collisions, prove of limited use because they do not
  allow robust extrapolation to other stellar structures or relative
  velocities.
\end{abstract}

\begin{keywords}
hydrodynamics -- methods: numerical -- stars: interior -- galaxies: 
nuclei, star clusters -- Galaxy: centre
\end{keywords}

\section{Introduction} 
\label{sec:intro}


\subsection{Stellar collisions in galactic nuclei}
\label{sec:StellCollInNat}

\begin{figure*}
\begin{center}
  \resizebox{0.9\hsize}{!}{%
  \includegraphics{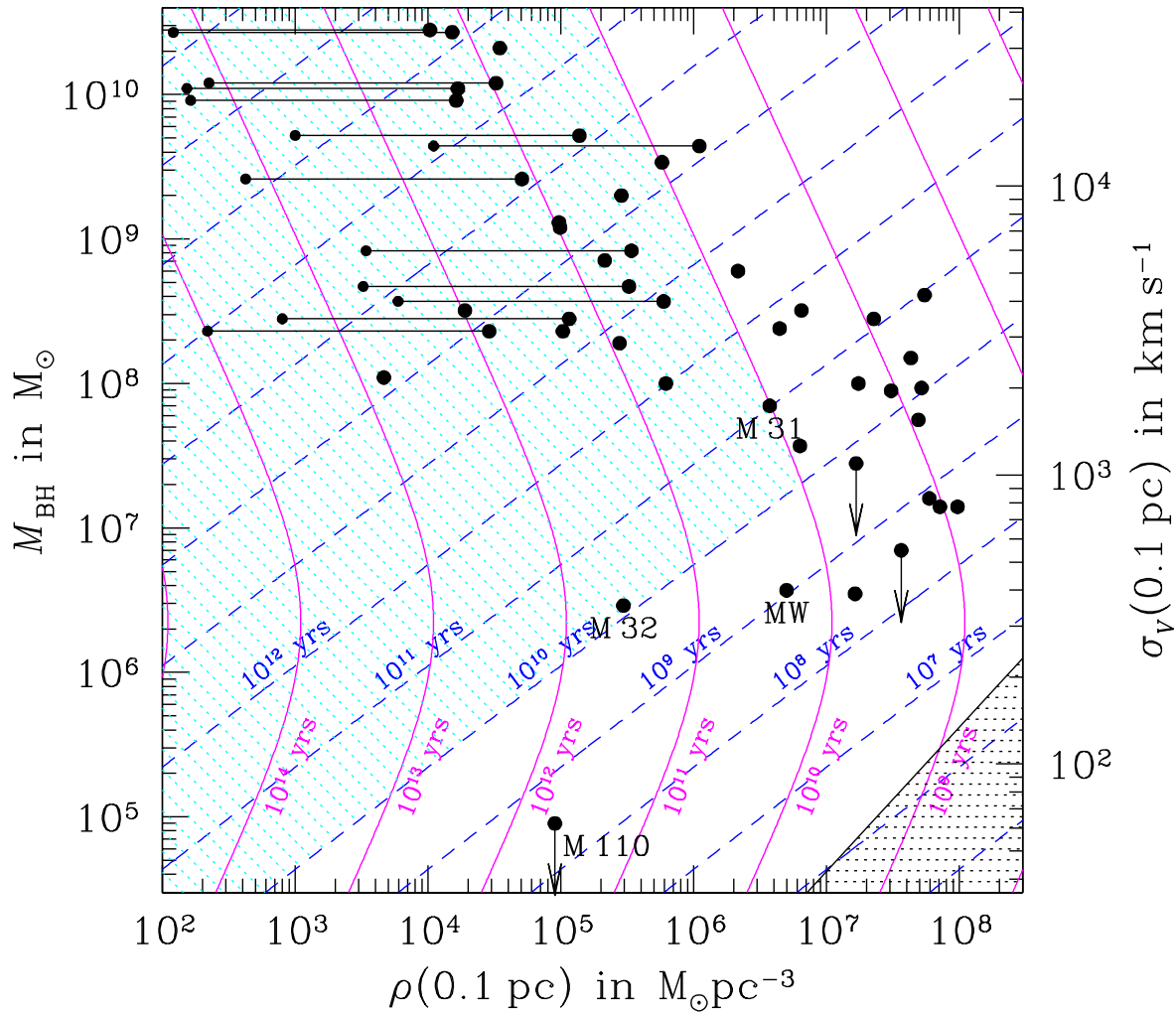} }
\end{center}
  \caption{ Relaxation and collision times at 0.1\,pc from a massive
    black hole in the centre of a galactic nucleus (inspired from a
    similar diagram by \citealt{Arab97}). We plot curves of
    iso-$T_{\mathrm{relax}}$ (dashes, blue in colour version) and
    iso-$T_{\mathrm{coll}}$ (solid lines, magenta in colour version)
    in a plane parameterised by the stellar density at 0.1\,pc and the
    mass of the central black hole. The right ordinate scale indicates
    the stellar velocity dispersion (Keplerian value,
    $\sigma_{V}=\sqrt{GM_{\rm BH}/r}$).  All stars are assumed to be
    of solar type. The left shaded sector (cyan in colour version)
    corresponds to conditions for which both $T_{\mathrm{relax}}$ and
    $T_{\mathrm{coll}}$ are larger than Hubble time, so that such
    nuclei are not expected to show signs of secular evolution. In the
    shaded lower right corner, the black hole does not dominate the
    kinematics and the effect of the cluster's self gravitation should
    be included in the computation of the characteristic time-scales.
    \newline Large black dots show the estimated conditions for
    observed galactic nuclei. In most cases, the estimation of the
    stellar density at 0.1\,pc requires important extrapolation from
    the data, as such a small radius is resolved only for a few
    galaxies of the local group (the Milky Way, M\,31 and M\,32). For
    this extrapolation, we use a power-law cusp of the form
    $\rho(r)=\rho_0(r_0/r)^\gamma$. The values of $\rho_0$, $r_0$ and
    $\gamma$ are taken from \citet{GebhardtEtAl03} or \citet{Faber97}.
    The densities for M\,31 and M\,32 are from \citet{Lauer98} and the
    Milky Way's value from \protect\citet{GTKKTG96}. The values of
    $M_{\mathrm{BH}}$ are estimates by \protect\citet{Mag98},
    \protect\citet{vdM99c} or better constrained values gathered by
    Kormendy (\citeyear{Kormendy04}, see
    \url{http://chandra.as.utexas.edu/~kormendy/bhsearch.html} for
    these data and a list of original references). In some cases,
    $\rho_0$ and $r_0$ have already been extrapolated from larger
    radii! Cases with an horizontal line connected to a second smaller
    dot are nuclei for which the slope $\gamma$ is observational
    compatible with $0$, according to \protect\citet{Faber97}. The
    second point indicates the density value at 0.1\,pc if constant
    $\rho$ is assumed up to $r_0$.  }
  \label{fig:TrelTcollNG}
\end{figure*}

In the recent years, the study of stellar collisions has received
renewed interest from researchers studying the dynamics of dense
stellar systems, either open/globular clusters or galactic central
regions (see the contributions in \citealt{Shara02}). Our own
motivation is to perform simulations of the long-term evolution of
dense stellar systems, particularly galactic nuclei, with a new Monte
Carlo stellar dynamics code which incorporate collisions as
``micro-physics'' \citep{FB01a,FB02b}.

Before going into a brief description of the astrophysical
motivations of these works, some clarification about the notion of
``stellar collision'' is called for. In this paper we shall use this
term to refer to a process during which two stars, previously unbound
to each other, get so close that not only gravitational forces but
also hydrodynamical ones come into play to determine the outcome of
the interaction. So, strictly speaking, collisions are not only
contact encounters but also comprise tidal interactions. However, for
reasons to be exposed in Sec.~\ref{subsec:tid_bin}, we restrict here
to events leading to physical contact at first periastron passage.

To assess the importance of collisions in a given astrophysical
context, the key quantity to monitor is the collision time, which we
define here as the average time needed for ``test-star'' 1 to
experience a collision with any ``field-star'' 2,
\begin{eqnarray}
  T_\mathrm{coll} &=& \left( n_2 S v_\mathrm{rel} \right)^{-1} 
  \label{eq:Tcoll}\\
  \mbox{with\ } S &=& \pi d_\mathrm{coll}^2\left( 1 +
    \frac{2G(M_1+M_2)}{d_\mathrm{coll}v_\mathrm{rel}^2} \right), 
  \label{eq:Scoll}
\end{eqnarray}
where $n_2$ is the number density of the field stars, $v_\mathrm{rel}$
the relative velocity and $d_\mathrm{coll}$ the pericentre distance
leading to a collision ($d_\mathrm{coll} = R_1 + R_2$ for contact at
first passage, neglecting tidal deformation). $S$ is the collisional
cross-section. At low relative velocity, it is greatly enhanced over
the geometric value by gravitational attraction. This effect, dubbed
``focusing'' is expressed by the second term in the brackets of
Eq.~\ref{eq:Scoll}. In most astronomical contexts, the velocity
dispersion is much smaller than the stellar escape velocity
$V_{\star}=\sqrt{2GM_{\star}/R_{\star}}$ ($\simeq 620$\,{\kms} for
sun-like stars) and gravitational focusing dominates. In these cases,
integrating over a Maxwellian distribution for relative velocities
yields \cite[Eq.~8--125]{BT87}:
\begin{eqnarray}
   T_\mathrm{coll} &\simeq& 7\times10^{14}\,\mathrm{yrs}\,\times \\
   && \nonumber
   \left(\frac{R_{\star}}{{\Rsun}}\right)^{-1}
   \left(\frac{M_{\star}}{{\Msun}}\right)^{-1}
   \left(\frac{n}{\mathrm{pc}^{-3}}\right)^{-1} 
   \left(\frac{\sigma_v}{\mathrm{km}\,\mathrm{s}^{-1}}\right).
\end{eqnarray}

In systems with $T_\mathrm{coll}$ smaller than typical stellar ages,
collisions have expectedly imprinted not only the stellar population
but also the global dynamical structure. Very high densities are
necessary for such situations to take place but even when collisions
occur at frequencies too low to be of dynamical relevance, they still
can be of great astrophysical interest {\it per se} because they are
suspected to lead to the formation of unusual individual stellar
objects, such as blue stragglers or stripped giants \citep[and
references therein]{Davies96,Shara99}. Collisions are unimportant in
the bulk of a galaxy; the probability for the sun to suffer a
collision during its 10~Gyr main sequence life, amounts to no more
than $10^{-7}$! Only in stellar clusters and galactic nuclei, is there
a non-vanishing probability for at least some stars to experience
collisions. For reviews about the role of collisions in various
environments, we refer to the various papers in \citet{Shara02}.


Among known stellar environments, galactic nuclei are those in which
the most extreme values of the stellar density and velocity dispersion
are attained. The best known case is our own Galaxy. Inside a sphere
of radius 0.4\,pc at the Galactic centre, the stellar density exceeds
$4\times10^6\,{\Msun}\mathrm{pc}^{-3}$ and a velocity dispersion of
order 500\,km\,s$^{-1}$ has been reported at a distance of 0.01\,pc of
the $2-3\times 10^6\,{\Msun}$ central black hole
\citep{GTKKTG96,GPEGO00}. Most other galactic nuclei are not resolved
yet so we can only produce very uncertain estimates of
$T_{\mathrm{coll}}$ for these systems.  Some of them are indicated in
Fig.~\ref{fig:TrelTcollNG}. As a bias toward our own interests, we
treat only the situation of a massive black hole dominating the
kinematics of the surrounding stars.

From this diagram, we see that there are very few galaxies for which
we can be certain that collisions played a important dynamical role.
Using Nuker model fits to represent the density profiles and the
empirical relation between the mass of the central object and the
velocity dispersion in the spheroidal component \citep{TremaineEtAl02}
as a proxy for the BH's mass, \citet{Yu03} estimated the collision
times for a series of observed galactic nuclei. She found only a few
cases with $T_{\mathrm{coll}}$ possibly shorter than the Hubble time
and that, in present-day nuclei, collisions do not produce observable
colour gradients in the stellar populations. It may be that the
importance of these processes has been somewhat overestimated in the
past \citep{vdBergh65,Sanders70a}.

The centre of the Galaxy is a particularly complex and fascinating
environment \citep{GenzelEtAl03,GhezEtAl03,SchoedelEtAl03}. The ``SO''
stars orbiting the $3-4\times 10^6\,\Msun$ BH {\SgrA} at distances
smaller than 0.04\,pc seem to be on the MS with masses of at least
10\,{\Msun} \citep{GhezEtAl03b}. Recent stellar formation at this
place seems impossible and scenarios to bring them from a few pc away
in less than their short lifetime require considerable fine tuning
\citep[][and references therein]{KFM04}. Consequently, it is tempting
to hypothesise they were created in a sequence of mergers of older,
lighter MS stars \citep{GenzelEtAl03}.  Using simple Fokker-Planck
modelling (not including a central BH), \citet{Lee94,Lee96} concluded
that mergers can not account for the formation of the massive stars
found near the centre. On the other hand, whether collisions are
responsible for the observed relative depletion of red giants at the
Galactic centre is still a debated issue
\citep{Gerhard94,DBBS98,Alexander99,BD99}. Clearly, more detailed
stellar dynamical models, that take into account the presence of the
central BH and include a realistic treatment of collisions and stellar
evolution are called for to establish the role of collisions in the MW
central cluster.

\defcitealias{SS66}{SS66}

There are however strong theoretical motivations to believe that
stellar encounters may have taken place in large numbers in the past
evolution in many galactic centres with sufficiently high stellar
densities. The main reason is the presence of massive compact dark
objects in the centre of many, if not most, galaxies. These mass
concentrations are most probably supermassive black holes (SMBHs) with
masses $10^5-5\times 10^9\,{\Msun}$ (\citealt{KR95,Barth04};
\citealt*{BHS04}; \citealt{FF04,Kormendy04,Pinkney03}). From a series of
works published in the 70's \cite[among
others]{Peebles72,SL76,BW76,BW77,DO77a,DO77b,Young77a}, it is known
that a SMBH-surrounding stellar system whose long-term evolution is
driven by 2-body gravitational encounters will relax to a density
profile, close to a power-law $\rho(r)\propto r^{-\gamma}$, which
yields a constant flux of stars toward the centre where they are
destroyed either by tidal disruptions or energetic collisions. If all
stars have the same mass, the exponent is $\gamma=1.75$. In the
innermost regions of such a cusp, a high collision rate is expected.
But the collisions themselves could act as a feed-back mechanism on
the evolution of the stellar system and the growth of the black hole
so that the actual formation of relaxational cusp is questionable.
From analytical considerations, \citet{Frank78} concludes that
collisions in the cusp are never of importance, when compared to tidal
disruptions, but this statement is seriously challenged by other
studies and, in particular, more recent numerical simulations
(\citealt{YSW77,Young77b,DS83,DDC87a,DDC87b}; \citealt*{MCD91};
\citealt{Rauch99}).  Unfortunately, the discussion of the contribution
of various dynamical processes to the evolution of galactic nuclei has
been blurred by uncertainties about the precise outcome of these
individual processes.  For instance the amount of gas that is accreted
by the SMBH following a tidal disruption is still debated \cite[and
references therein]{ALP00}. As for stellar collisions, most previous
works relied on quite unrealistic prescriptions, like complete
destruction \citep{YSW77,Young77b,McMLC81,DS83} or on a
semi-analytical recipe proposed by \citet[hereafter SS66]{SS66} that
completely neglects the real hydrodynamical nature of the process
\citep{Sanders70b,DDC87a,DDC87b,MCD91}. The work of \citet{Rauch99} is
a noticeable exception; he used the results of a set of hydrodynamical
simulations of stellar collisions by Davies to derive fitting formulae
for the quantitative outcome of these events.  The present work
originated in our wish to get rid of these annoying uncertainties
about the role of collisions in dynamical simulations of galactic
nuclei \citep{FB01d,FB02b}.

Many of the papers we have just cited were not only concerned with the
past evolution of galactic nuclei but also (or mainly) with scenarios
to feed SMBH and provide quasars' luminosities. Gas-dynamical
processes are now favoured candidates for the fuelling of active
galactic nuclei (AGN) and the dense cluster hypothesis seems somewhat
out-of-fashion \citep[and references therein]{SBF90,Combes01}. On the
other hand, AGN models have been proposed in which large luminosities
in electro-magnetic radiation and/or relativistic particles are
emitted by the hot gas clouds created by very energetic stellar
collisions.  First propositions along that line
\citep{Woltjer64,Sanders70a} postulated that the stars' velocities
were due to the cluster's self gravity.  More recent models
\citep{Keenan78,DKT93,CPW96,TCFCP00} invoke a SMBH to provide velocity
dispersions ranging from a few $10^3$\,km\,s$^{-1}$ to a few
$10^4$\,km\,s$^{-1}$. These non-standard AGN models may be successful
in reproducing observed luminosity-variability relations that are
otherwise difficult to explain, but they should be re-examined in the
light of a more refined treatment of stellar collisions and stellar
dynamics. A third possibility for collisions to contribute directly to
the luminosity of AGN is to boost the rate of supernovae through
creation of massive stars by mergers \citep{Colgate67,SW78}.

Finally, even though they are not the dominating luminosity source in
AGNs, stellar collisions may be responsible for the formation of
massive black holes in dense galactic nuclei, either by run-away
merging (\citealt{Sanders70b,QS90,PZMcM02};
\citealt*{RFG03,GFR04,FGR04b,FGR04c,FGR05}) or by creating a massive gas cloud that subsequently
evolves to a black hole \citep{SS66,BR78,LFSY90}.

\subsection{Previous simulations of collisions between main sequence stars.}
\label{subsec:prev_works}

{
\begin{table*}
  \caption{Hydrodynamical simulations of high-velocity collisions between MS stars in
    the literature.}
    \setlength\tabcolsep{3pt}
    \begin{center}
  \begin{tabular}{lclllll}
    \hline\hline
    Reference&Abbrev.&Stellar models&$q=M_1/M_2$ & $V_{\mathrm{rel}}^{\infty}/V_{\ast}$$^{\mathrm{(a)}}$
    &$M$--$R$ relation&Method\\ \hline
    \protect\citet{SC72} && polytropes $n=3$ & 1 & 0, 1.6, 3.2
    & & Head-on, 2D finite diff. \\
    \protect\citet{BH87}  &BH87& polytropes $n=1.5$ & 1 
    & 0--2.33 & & SPH 1000 part. \\
    \protect\citet{BH92}  &BH92& polytropes $n=1.5$ & 0.2
    & 0--1.5$^{\mathrm{(b)}}$ & $R_{\ast}\propto M_{\ast}^{0.85}$ & SPH 7000
    part. \\
    \protect\citet{LRS93}$^{\mathrm{(c)}}$ &LRS93& polytropes $n=3$$^{\mathrm{(d)}}$ &
    0.1--1 & 0.2--3.8 & $R_{\ast}\propto M_{\ast}^{0.8}$ & SPH 8000
    part. \\
    Davies \protect\citep{Rauch99}$^{\mathrm{(e)}}$ &R99&  polytropes $n=3$ & 0.25, 0.5, 1 & 1--25 & 
    $R_{\ast}\propto M_{\ast}$ & SPH $\sim 40\,000$ part.\\
    This work && realistic & 0.0013--1 & 0.1--30 & realistic & SPH
    4000--36\,000 part. \\
    \hline
    \multicolumn{6}{l}{
      $^{\mathrm{(a)}}$See symbols definition in Sec.~\protect\ref{subsec:def}.
      }\\
    \multicolumn{6}{l}{
      $^{\mathrm{(b)}}$Up to 5 for head-on collisions.
      }\\
    \multicolumn{6}{l}{
      $^{\mathrm{(c)}}$Fitting formulae are given.
      }\\
    \multicolumn{6}{l}{
      $^{\mathrm{(d)}}$Eddington models.
      }\\
    \multicolumn{6}{l}{
      $^{\mathrm{(e)}}$Results only given as fitting formulae.
      }\\
  \end{tabular}
\end{center}
  \label{tbl:coll_simul}
\end{table*}
}

Table~\ref{tbl:coll_simul} lists the previous computations of
high-velocity collisions between MS stars. We only mention
``realistic'', multi-dimensional hydrodynamical simulations. This
excludes early calculations that were based either on semi-analytical
methods \citepalias{SS66} or on one-dimensional numerical schemes
\citep{Mathis67,DeYoung68}. Such approaches were clearly
over-simplifications in which the real 3D hydrodynamical nature of the
problem was not properly accounted for. The importance of these
``pre-hydrodynamics'' works should not be underestimated, however.
For instance, even though it was always deemed too simplistic to yield
better than order-of-magnitude estimates, the \citetalias{SS66} method
had been adapted and used in a few key simulation works. We postpone a
presentation of this ``historical'' method to
Sec.~\ref{subsec:comp_lit} where we compare our results to predictions
of this approach.

With the historical exception of \citet{SC72}, all cited works were
realised using SPH (Smoothed Particle Hydrodynamics). When featured
with a tree to compute gravitation, SPH is a grid-less method which
can cope with any asymmetrical three dimensional geometry.  It ignores
void spaces completely, imposes no physical limits beyond which matter
cannot be tracked, does not come into trouble with large dynamic range
as long as variable smoothing lengths are implemented.  SPH is better
suited to highly dynamical problems than to near-equilibrium
configurations \citep{SM93}. For all these reasons, SPH is
particularly well suited to the simulation of stellar collisions. 

From Table~\ref{tbl:coll_simul}, it is clear that the study of the
outcome of high-velocity collisions has not attracted much interest in
the last few years, in contrast to parabolic encounters \citep[][among
others]{LRS96,SL97,SFLRW00,SADB02}. As a consequence, the resolutions
used seem very modest, by present-day standards; for instance,
\citet{SADB02} present a parabolic collision simulated with $10^6$
particles. Obviously, the simulations presented in this work do not
correspond to a break-through in terms of resolution. This reflects
the fact that most computations were realised a few years ago, when
computing power was more limited and, most importantly, that we had to
cover a huge parameter-space, requiring more than 10\,000 simulations
(see Sec.~\ref{subsec:buildingtable}). This sheer quantity, combined
with the use of realistic stellar models instead of polytropes,
represent the main improvements over previous works.

\defcitealias{BH87}{BH87}
\defcitealias{BH92}{BH92}
\defcitealias{LRS93}{LRS93}
\defcitealias{Rauch99}{R99}

For simplicity, in the remaining of this article, we refer to the work
of \citet{BH87} as \citetalias{BH87}, to \citet{BH92} as BH92, to
\citet*{LRS93} as \citetalias{LRS93} and to \citet{Rauch99} as
\citetalias{Rauch99}. For a more comprehensive list of references on
simulations of all kind of stellar collisions, see the web site
maintained by MF in the framework of the ``MODEST''
collaboration\footnote{``MODEST'' stands for Modelling DEnse STellar
  systems, see
  \url{http://www.manybody.org/modest/}.
  For the collision ``working group'', go to
  \url{http://www.manybody.org/modest/WG/wg4.html}.}.

\subsection{Collisions with non-main-sequence stars}

\begin{figure*}
  \resizebox{\hsize}{!}{%
   \includegraphics{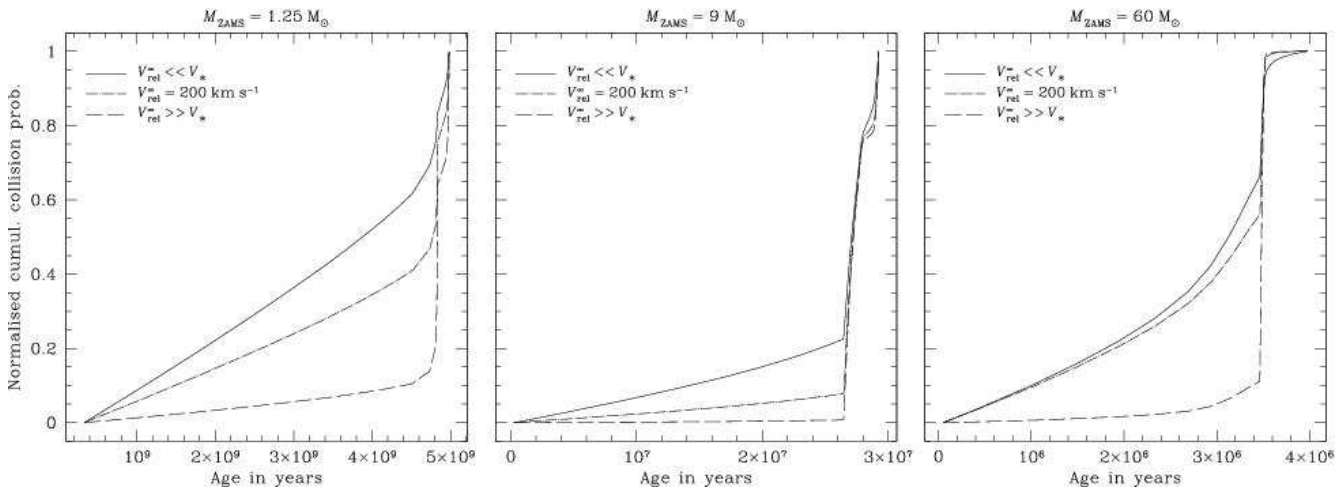} }
  \caption{Cumulative collision probability (normalised to 1)
  integrated over the lifetime of three stellar models. The second
  star is assumed to be sun-like. In each case, three velocity regimes
  are considered. At low relative velocities
  ($V_{\mathrm{rel}}^\infty\ll V_{\ast}$), the collisional
  cross-section scales like $R_{\ast}+\Rsun$ (solid lines); at very
  high velocities ($V_{\mathrm{rel}}^\infty\gg V_{\ast}$), we recover
  the geometrical cross section, $\propto (R_{\ast}+\Rsun)^2$ (dashed
  lines). We also plot the case for a relative velocity of
  200\,{\kms}, an intermediate value typical for galactic nuclei. The
  end of the MS phase corresponds to the point where the slope of the curves 
  increases suddenly. Stellar evolution models are from the compilation of
  \citet{LS01}, available on-line at
  \url{http://vizier.cfa.harvard.edu/viz-bin/VizieR?-source=VI/102}.}
  \label{fig:pcoll_vs_age}
\end{figure*}

In this work, we only treat collisions between two main-sequence (MS)
stars. The motivations for this choice was to keep the number and
variety of collisions to consider at a manageable level and that the
present version of our Monte Carlo code only includes simplified
stellar evolution which skips over the giant phase and turns MS stars
directly into remnants. However, in a real stellar system, MS-MS
encounters may not dominate the global collision rate. Indeed, a given
star of mass $> 1\,{\Msun}$ may have a smaller probability for
colliding with another star during its MS life than during its red
giant (RG) phase despite the latter being about 10 times shorter
\cite[for instance]{BFBC93}. This is made very clear by integrating
the collisional cross-section over the lifetime of the star, as we did
in Fig.~\ref{fig:pcoll_vs_age}. In many cases, the collision
probability during the red giant phase exceeds its MS counterpart for
high relative velocities. RG-RG collisions are less likely than RG-MS
events. Indeed, the ratio of probabilities can be estimated as follows
\[
\frac{P_{\mathrm{RG-MS}}}{P_{\mathrm{RG-RG}}} \sim
\frac{n_{\mathrm{MS}}}{n_{\mathrm{RG}}} \cdot
\frac{R_{\mathrm{RG}}^2}{(2R_{\mathrm{RG}})^2} \simeq 0.25
\frac{T_{\mathrm{MS}}}{T_{\mathrm{RG}}} \simeq 3-10.
\]
\Added{Although probably more common than MS-MS encounters, RG-MS
  collisions may not be more important. RG envelopes have very low
  densities so only little mass is lost in most cases and the RG
  recovers its appearance. At relative velocities found in galactic
  nuclei, the MS star cannot be captured unless it is aimed nearly
  directly at the RG centre \citep{BD99}. Furthermore, as giants will
  loose their envelope anyway through winds and a planetary nebula or
  SN phase, collision with giants will probably make little difference
  as far as the feeding of a central SMBH is concerned.
  
  Due to mass segregation in clusters and nuclei, collisions between
  compact remnants (CRs) and MS (or RG) stars are probably much more
  common than the small CR fractional number would suggest. For
  instance, the innermost 0.1\,pc of the {\SgrA} cluster is likely
  dominated by invisible stellar BHs \citep{MEG00} which may
  collisionally destroy MS and RG stars. CR-MS and CR-RG collisions
  may also be of great interest as a way to produce exotic objects,
  such as cataclysmic variables. Unfortunately, due to the high
  dynamical range involved, the hydrodynamical simulation of these
  events is challenging and comprehensive predictions for their
  outcome are still lacking.}

Now that the astrophysical stage is set, we can proceed with a
description of our simulation work. In Sec.~\ref{sec:approach}, we
describe the choice and setting of initial conditions and present the
numerical methods we use to compute and analyse stellar collisions. In
Sec~\ref{sec:results}, results are reported and we explain how to
exploit them in stellar dynamical simulations.  Finally, in
Sec~\ref{sec:concl}, some general conclusions and a discussion of
further work to be done are presented.

\section{Description of the approach}
\label{sec:approach}

\subsection{Definitions, basic formulae and units}
\label{subsec:def}

\begin{figure}
  \resizebox{\hsize}{!}{%
    \includegraphics{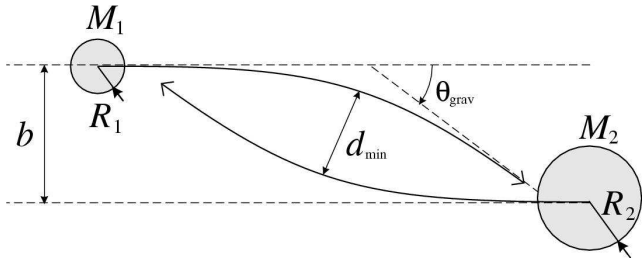}%
    }
  \caption{%
    Sketch of a gravitational 2-body hyperbolic encounter in the
    centre of mass reference frame. 
 }
  \label{fig:hyp_encounter}
\end{figure}
  
As some quantities will be referred to again and again, we find it
useful to define them once for all at the beginning of this article.
Collisions between two main sequence (MS) stars are considered. In the
centre of mass frame, the collision is completely determined by 4
quantities: the masses $M_1$ and $M_2$ (in our work, we made the
unconventional notation choice: $M_1\leq M_2$), the impact parameter
$b$ (see Fig.~\ref{fig:hyp_encounter}) and the relative velocity at
infinite separation, $V_\mathrm{rel}^{\infty}$. The stellar radii are
$R_1$ and $R_2$.  Instead of applying a simple but unrealistic
power-law mass-radius relation, the values for the radii are taken
from the stellar models described in Sec.~\ref{subsec:sm}.

We shall often refer to the situation of a 2 point-mass hyperbolic
encounter where all finite-size (hydrodynamical) effects are
neglected. In this case, we define the periastron distance,
\begin{equation}
  d_\mathrm{min} = (R_1+R_2)\,\frac{ 2\hat{b}^2\hat{v}^2 }{
    1+\sqrt{1+4\hat{b}^2\hat{v}^4} }
  \label{eq:dmin}
\end{equation} 
with $\hat{b}=b/(R_1+R_2)$ and
$\hat{v}=V_\mathrm{rel}^{\infty}/V_{\star}$ (see Eq.~\ref{eq:vstar}).
When gravitational focusing is important, $d_\mathrm{min}$ is a more
convenient parameter than $b$.\footnote{Furthermore, $b$ is not
  defined for parabolic encounter whereas $d_\mathrm{min}$ still is.}
Ignoring tidal effects such as deformations and trajectory
modification until strong hydrodynamical interactions begin, we assume
that only collisions with $d_\mathrm{min} < R_1+R_2$ lead to contact
between stars at the first periastron passage.

In case both stars survive the encounter and are left unbound to each
other, we define a collisional deflection angle
$\theta_\mathrm{coll}$. This is the angle between the direction of the
initial relative velocity (at infinite separation) with the direction
of the final relative velocity (at $\infty$). To assess the importance
of finite-size hydrodynamical effects, it is useful to compare
$\theta_\mathrm{coll}$ with the value for a Keplerian hyperbolic orbit
$\theta_\mathrm{grav}$,
\begin{equation}
  \tan \left(\frac{\theta_\mathrm{grav}}{2}\right) =
  \frac{G(M_1+M_2)}{b (V_\mathrm{rel}^{\infty})^2}.
\end{equation}

A natural velocity scale for collisions is the relative velocity at contact for
stars initially at rest at infinity,
\begin{equation}
  V_{\star}=\sqrt{\frac{2G(M_1+M_2)}{R_1+R_2}}.
\label{eq:vstar}
\end{equation}

The structure of MS stars with $M_{\ast}>0.5\,{\Msun}$ is very
concentrated (see Fig~\ref{fig:MRrel} and the appendix) and the
total radius is not a good indicator of the extension of the stellar
matter. It is thus often useful to normalise quantities with reference
to the half-mass radius $R_{\ast}^{\mathrm{(h)}}$, i.e.\ the radius of
a sphere that contains half the stellar mass. These radii can be read
from the 50\,\% curve in Fig.~\ref{fig:MRrel}\footnote{We could also
  use radii at 75, 90 or even 95\,\% of the mass. It's only the very
  dilute outer 5\,\% of the stellar mass that increases $R_{\ast}$ so
  much in high mass MS stars.}. We can then define a ``half-mass
velocity'' scale through
\begin{equation}
  V_{\star}^{\mathrm{(h)}} =
  \sqrt{\frac{2G(M_1+M_2)}{R_1^{\mathrm{(h)}}+R_2^{\mathrm{(h)}}}}.
\label{eq:vstar_half}
\end{equation}
This quantity gives a better idea of the relative velocity when
strong hydrodynamical effects begin to play an important role.
Note that we use total masses in this definition. We often
normalise initial parameters by these half-mass quantities so handy
definitions are
\begin{equation}
  \lambda =
  \frac{d_{\mathrm{min}}}{R_1^{\mathrm{(h)}}+R_2^{\mathrm{(h)}}}
  \mbox{~~and~~} \nu =
  \frac{V_\mathrm{rel}^{\infty}}{V_{\star}^{\mathrm{(h)}}}.
\end{equation}

Typical scales are set by ``solar units'', i.e.:
\begin{eqnarray*}
  {\Rsun} &:=&7\times10^{8}\,\mathrm{m}, \\
  {\Msun} &:=&2\times10^{30}\,\mathrm{kg}, \\
  {\rm V}_{\sun} &:=&\left(G{\Msun}/{\Rsun}\right)^{1/2}=436.5\,\kms, \\
  {\rm T}_{\sun}
  &:=&\left({\Rsun}^3/(G{\Msun})\right)^{1/2}=1604\,\mathrm{s}.
\end{eqnarray*} 
These values are also referred to as ``code units''.

\subsection{Stellar models}
\label{subsec:sm}

\begin{figure}
  \resizebox{\hsize}{!}{%
    \includegraphics{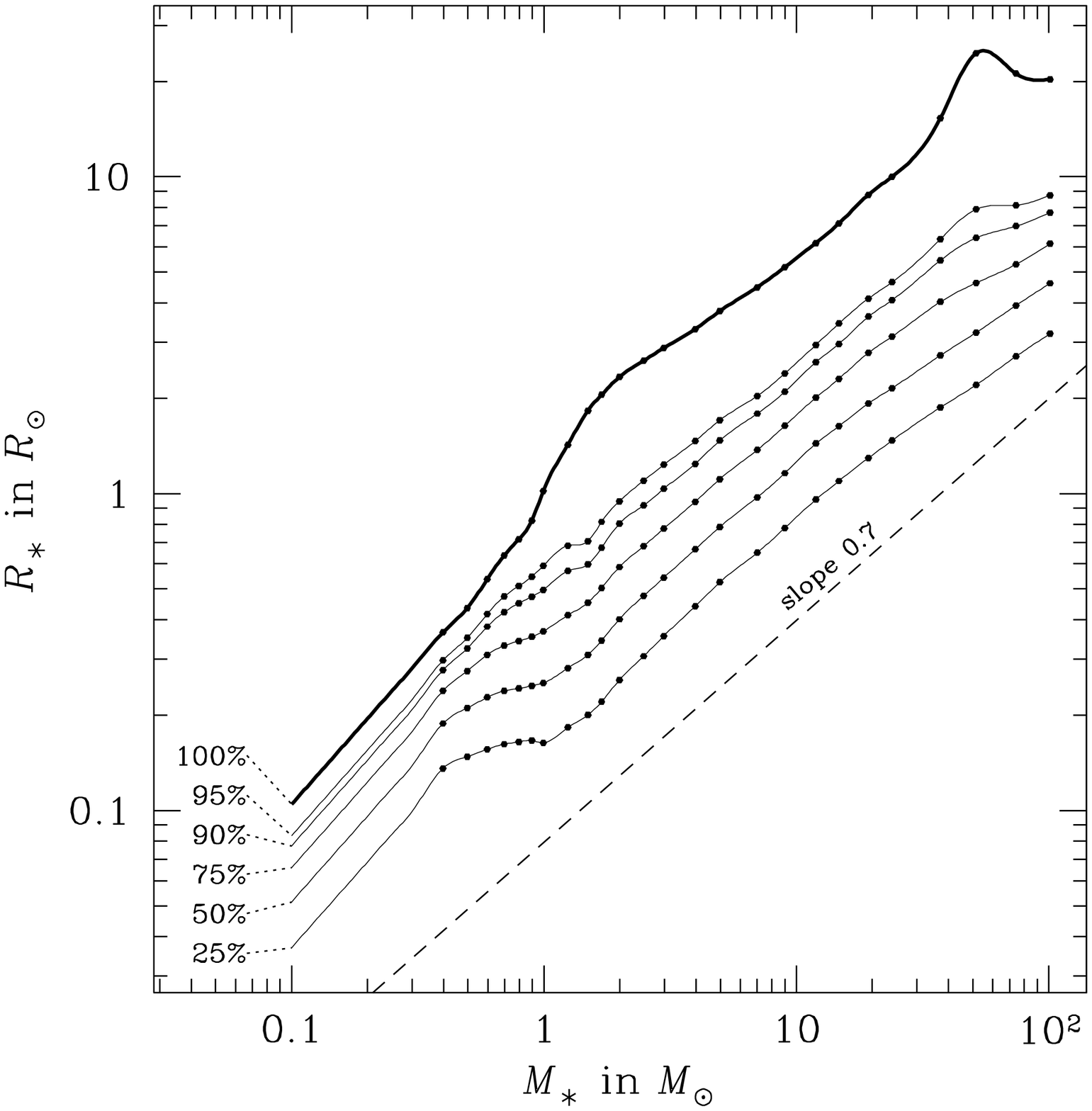}%
    }
  \caption{%
    Mass-Radius relation for the stellar models. The thick line shows
    the total radius as a function of the mass of the star. Thin lines
    are for inner Lagrangian radii containing 25 to 95$\,$\% of the
    mass. Data for $M_{\star}\geq 0.4\,{\Msun}$ (dots) is from
    realistic stellar structure models
    \protect\citep{SSMM92,MMSSC94,CDSBMMM99}. Below $0.4\,{\Msun}$, a
    polytropic ($n=1.5$) structure is applied with a power-law M-R
    relation extrapolated from higher masses. The run of various
    Lagrangian radii makes it obvious that stellar models for
    different stellar masses cannot be deduced from each other through
    any homologous scaling. The ``real'' mass, determined through
    relation~\protect\ref{eq:mod_choice}, is given in abscissa, not
    the ZAMS value.  }
  \label{fig:MRrel}
\end{figure}
  
In our simulations, we use realistic main sequence (MS) models to set
up the initial stellar structures. Models from the Geneva stellar
evolution group \citep{SSMM92,MMSSC94} have been applied for ZAMS
masses ranging from $0.8$ to $85\,{\Msun}$, and models by
\citet{CDSBMMM99} for masses down to $0.4\,{\Msun}$. For each
(initial) stellar mass, we had to select one particular model among
those spanning the main sequence evolutionary track. We chose the
instant $t_\mathrm{model}$ which divides the MS life in two parts with
approximately equal collision probabilities. Assuming strong
gravitational focusing and neglecting any mass loss, the collision
probability per unit time is $\mathrm{d}P_\mathrm{coll}/\mathrm{d}t
\propto R_{\star}(t)$, so that,
\begin{equation}
  \int_{0}^{t_\mathrm{model}} \!\! R_{\star}(t)\,\mathrm{d}t 
  \simeq
  \int_{t_\mathrm{model}}^{\min(t_\mathrm{MS},12\,\mathrm{Gyr})}
  R_{\star}(t)\,\mathrm{d}t 
  \label{eq:mod_choice}
\end{equation}
with $t=0$ on the ZAMS. For high-mass stars ($\geq 20\,{\Msun}$) mass
loss by stellar winds is already important on the MS
\citep{SSMM92,MMSSC94} so that the adopted models have real masses
lower than their nominal ({\it i.e.}  ZAMS) masses, for instance, the
largest star we consider, a ``$85\,{\Msun}$'' model, has an actual
mass of only $74.3\,{\Msun}$. The mass-radius relation is shown in
Fig.~\ref{fig:MRrel}. For $M\geq0.4\,{\Msun}$, it is given by the
stellar models just discussed. For smaller masses, we simply
extrapolated a power-law relation from the $0.4\,{\Msun}$ and
$0.5\,{\Msun}$ points. It appears that this gave radii in good
agreement with detailed structure models by \citet{CB97} who yield
$R\simeq 0.12\,{\Rsun}$ at $0.1\,{\Msun}$ and $R\simeq
0.22\,{\Rsun}$ at $0.2\,{\Msun}$.

We used models with solar composition ($Y=0.30$, $Z=0.02$). A
population~II metallicity ($Y=0.24$, $Z=0.001$) would introduce
significant alterations in the stellar structures. Most noticeably,
low-$Z$ stars are initially more compact, with radii smaller by
10--40$\,$\%, and have larger convective cores for
$M_{\star}>1\,{\Msun}$ \citep{KW94}. For high-mass stars, the most
important difference is probably the much weaker mass-loss at lower
metallicity \citep{Maeder92}. We made no attempt to assess the impact
of these effects on collision outcomes. We hope that they can be
partially scaled out by a proper dimensionless parameterisation of the
initial conditions and results of the collisions (see
Sec.~\ref{sec:results}).  While the structure of stars less massive
than $0.5\,\Msun$ is very close to that of an $n=1.5$ polytrope, more
massive evolved MS stars do not match any polytropic model. In
particular, stars with $\Mstar\ge 1\,\Msun$ are more concentrated than
$n=3$ polytropes (see appendix for density
profiles).
  
The lowest stellar masses considered are 0.1 and $0.2\,\Msun$. For
such objects we didn't use detailed stellar structure models like
those by \citet{CB97} because they rely on a very complex equation of
state (EOS) accounting for degeneracy and electrostatic effects.
\Added{Such an EOS was not available to us for use in the SPH code at
  the time we embarked on this project. Also, solving this kind of
  complicated EOS (for each particle at each time step) is done using
  an iterative scheme and represents a significant computational
  burden.} Instead, we note that the interior of stars with masses
lower than $\sim 0.4\,{\Msun}$ is nearly completely convective, so
their internal structure is very close to that of a $n=1.5$ polytrope
\citep{HK94,CB00}. Given the mass and radius, we can build an initial
polytropic star in hydrostatic equilibrium using the EOS for a fully
ionised ideal gas. \Added{For $0.2\,\Msun$, we have compared our
  simple polytropic model with ideal-gas EOS to a state-of-the-art
  stellar structure provided by Isabelle Baraffe and found that
  discrepancies in the density and temperature profiles are below
  10\,\% except for the outermost envelope, a thin layer which is not
  represented in the SPH structure. Inspecting the realistic
  $0.2\,\Msun$ model, we see that only of order 0.01\,\% of the
  stellar mass has temperatures below $10^5$\,K for which incomplete
  molecule dissociation and ionisation may be important. Neglecting
  molecules and partially ionised gas may lead to a slight
  overestimate of the mass loss because some of the available kinetic
  energy has to be used to break up molecules and ionise atoms. This
  is certainly a very small effect as the energy required to
  completely ionise one gram of stellar matter of solar composition is
  $1.5\times 10^{6}$\,J \citep{KW94} but the kinetic energy at
  $500\,\kms$ (a typical contact velocity for a parabolic collision)
  is of order 100 times larger.}

\subsection{The SPH code}
\label{subsec:SPH}

Smoothed Particle Hydrodynamics is a Lagrangian particle-based method
that has been widely used to tackle all kinds of astrophysical
problems, from planetesimal fragmentation to cosmological structure
formation. For a description of the method and of its achievements, we
refer to reviews by \citet{Benz90} and \citet{Monaghan92}. See also
\citet{SM93} for a critical examination of the pros and cons of the
method and \citet{Monaghan99} for a presentation of its most recent
developments.

We used a version of the SPH code that corresponds to the description
in \citet{Benz90}. The kernel function is the standard spline
introduced by
\citet{ML85}. This code implements a binary tree to compute gravitational 
forces and find neighbours \citep{Press86,BCPB90}. ``Bulk'' and
von~Neumann-Richtmyer artificial viscosity terms are included with
$\alpha=2.5$ and $\beta=1.0$.

For the stellar matter, we assume the EOS of a completely ionised
mono-atomic ideal gas with account of the radiation pressure:
\begin{eqnarray}
  \rho & = & \frac{\mu}{\mathfrak{R}}\frac{1}{T}\left(P-\frac{a}{3}T^4\right) \\
  u & = & \frac{3}{2}\frac{\mathfrak{R}}{\mu}T+\frac{aT^4}{\rho},
\end{eqnarray}
where $\rho$ is the mass density, $T$ the temperature, $P$ the total
pressure, $u$ the specific internal energy, $\mu$ the mean molecular
weight, $a=7.56 \times
10^{-16}\,\mathrm{J}\,\mathrm{m}^{-3}\mathrm{K}^{-4}$ and
$\mathfrak{R}=8314\,\mathrm{J}\,\mathrm{K}^{-1}\mathrm{kg}^{-1}$. The
molecular weight of each particle is attributed from the initial
stellar structure (see next subsection). It remains constant during
the complete SPH simulation. In hydrostatic main sequence stars, the
radiation pressure becomes important for masses larger than
$5-10\,{\Msun}$ \citep{KW94}.
 
Release of nuclear energy has been shown to have none or very small
hydrodynamical influence \citep{Mathis67,RYBMH89}. We thus do not
include nuclear reactions in the energy equation. We also neglect
radiative transport. As long as the gas is optically thick (and the
bulk of it certainly is during the whole collisional process), energy
transport by radiation is a diffusion process which time scale, for a
sun-like star is $T_{\mathrm{KH}} \simeq 1.6\times 10^7$\,years
\cite[Kelvin-Helmholtz time]{KW94}. This is enormously larger that the
hydrodynamical time-scales (a few tens of hours, at most). For a gas
cloud of radius $R$ and mass $M$, the diffusion time is:
\begin{equation}
  t_{\mathrm{diff}} \simeq \frac{l}{c}N \mbox{~~with~~}
  N=\left(\frac{R}{l}\right)^2
\end{equation}
where $l$ is the mean free path of photons. It is connected to the
opacity $\kappa$ by $l=(\kappa \rho)^{-1}$. Thus,
\begin{equation}
  t_{\mathrm{diff}} \approx \frac{\kappa}{c} \frac{M}{R} =
  120\,{\rm yrs}\times\left(\frac{\kappa}{\kappa_{\mathrm{es}}}\right)
  \left(\frac{M}{{\Msun}}\right)
  \left(\frac{R}{100\,{\Rsun}}\right)^{-1},
\end{equation}
where $\kappa_{\mathrm{es}}\simeq 0.04\,\mathrm{kg}^{-1}\mathrm{m}^2$
is the opacity due to electron scattering (a lower bound for $\kappa$
in ionised gas). It follows that radiation cooling is negligible even
long after the end of the collision simulation.

\subsection{Building of initial SPH stellar models}
\label{subsec:SPH_init_models}

Building an SPH star from a given stellar structure model is not
completely straightforward. First, the spatial positions of the
particles have to be chosen. Then, each particle must be given a mass
and smoothing length in such a way that the total mass is respected
and the model's density profile $\rho_{\star}(R)$ is well approximated
by the SPH interpolate.  A second thermodynamical variable (the
internal energy $u(R)$, in our case), as well as the chemical
composition, is also specified by the structure model. These
quantities determine the pressure through the EOS.  If the EOS is
similar to the one used in the stellar structure code, the SPH
structure should be in hydrostatic equilibrium.

If all particles are attributed the same mass, their number density
must closely follow $\rho_{\star}(R)$, which, unless a huge number of
particles are used, results in a severe under-sampling of the outer
regions where the ``action'' takes place during most collisions. On
the other hand, using a constant particle density throughout the star,
by placing them on a periodic grid for instance, will lead to a very
inaccurate core representation as a small set of very massive
particles. This could expectedly yield unstable initial models and
noisy collisional results in case these few heavy particles strongly
participate to the hydrodynamics of the encounter. We thus had to find
some compromise between these two extreme options. If we neglect
particles overlap, the relation $\rho_{\star}(R) \simeq
m_\mathrm{part}(R) n_\mathrm{part}(R)$ holds, with $m_\mathrm{part}$
being the {\emph local} mass of each SPH particle at distance $R$ from
the star's centre and $n_\mathrm{part}$ their number density at that
position. Thus, we decided to impose
\begin{equation}
  n_\mathrm{part} \propto \rho_{\star}^\alpha \mbox{\ \ and\ \ } 
  m_\mathrm{part} \propto \rho_{\star}^{(1-\alpha)}
\end{equation}
with $\alpha=0.5$ (the two above mentioned extremes correspond to
$\alpha=1$ and $\alpha=0$ respectively). To obtain a $R$-dependent
$m_\mathrm{part}$, we place particles on concentric spheres with
variable spacing. On each sphere particles are arranged on constant
``latitude'' circles. When this is done, the smoothing lengths $h_i$
are adjusted until each particle overlaps approximately with the same
number ($\simeq 40$) of neighbours' centres. Finally, particles'
masses are iteratively adjusted in order to bring the SPH interpolate
for the density (at the centre of particles) closer to the model's
$\rho_{\star}$. This is done by repeating the assignments
\[
m_i \leftarrow m_i \left(0.3+0.7\frac{\rho_{\star}(R_i)}{\rho_i}\right)\ \ 
i=1\ldots N_{\rm part}
\]
20 times. As this procedure doesn't conserve the total mass $\Mstar$
exactly, all $m_i$ are then slightly rescaled by a uniform factor to
obtain the required $\Mstar$. This method is fast and effective to
give a good match to $\rho_{\star}$ for the bulk of the stellar
interior, as testified by the profiles shown in the appendix. But, despite the use of lighter particles to
represent the gas in the stellar envelope, the outermost layers of the
star are poorly modelled. In particular the SPH realisation fails at
precisely reproducing the stellar radius. This had to be expected in
models with a limited number of particles.

\begin{figure}
  \resizebox{\hsize}{!}{%
    \includegraphics{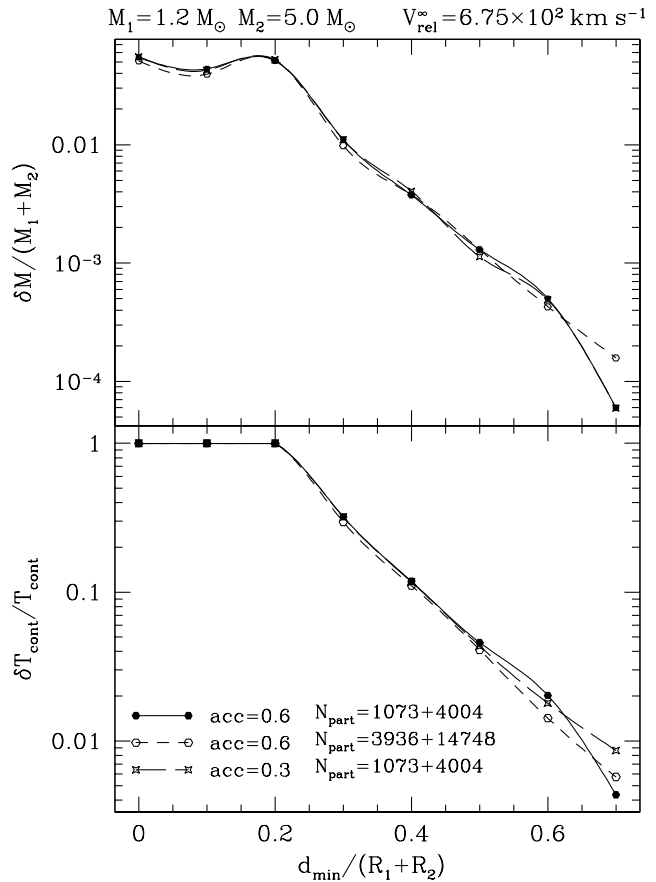}%
    }
  \caption{%
    Fractional losses of mass (top panel) and energy (bottom) as
    functions of the distance of closest approach (in Keplerian
    approximation) for fixed $(M_1,M_2,V_{\mathrm{rel}}^{\infty})$.
    Three sets of simulations are reported. In the first one (solid
    lines) we use a relatively low number of SPH particles and our
    standard (rather large) value for the binary tree accuracy
    parameter \texttt{acc} \protect\citep{BCPB90}. In the second
    series, we used about four times more particles. In the third, we
    used a lower \texttt{acc} value which gives a more precise
    computation of gravitational forces. The results are nearly
    coincident.  $T_{\mathrm{cont}}$ is the kinetic energy at contact,
    i.e. for a separation between centres of $R_1+R_2$. }
  \label{fig:dM_resol}
\end{figure}

Nonetheless, in grazing collisions, our use of low-mass SPH particles
to represent the outer parts of a star apparently leads to a reliable
determination of fractional mass losses as small as
$10^{-4}$--$10^{-3}$. This claim is grounded on diagrams like
 Fig.~\ref{fig:dM_resol} which shows the fractional mass
loss for two sets of simulations, the first one with the ``normal''
(low) resolution and the second one with a number of particles about
four times larger. The differences are obviously very small for all
cases but the most distant interactions.

An extended coverage of the four dimensional
$(M_1,M_2,V_\mathrm{rel},b)$ parameter space requires a huge number of
collisions to be computed. On the other hand, we don't need very
accurate results; a relative precision of about 10\,\% on mass and
energy loss should be sufficient for our purposes. More precise
results would not make much sense anyway as any application will
probably require some sort of inter- or extrapolation from our
simulation data, to apply it to stars with different masses or
metallicities, for instance. Thus we decided to tune numerical
parameters to values that allow relatively fast computations while
ensuring reasonable accuracy. This means that we generally used
1000--8000 SPH particles for each stellar model (a few collisions have
been computed with the most massive star having $\sim$16\,000 or
$\sim$32\,000 particles). In most simulations the total number of
particles ranges between 2000 and 10\,000. Thus, a collision is
computed in a few hours to a few days on a run-of-the-mill
workstation. We use a higher number of particles in high mass stars in
an attempt to resolve both the high density centre that contains most
of the mass and the low density envelope which is more likely to
interact with the other star. We also adapt the number of particles of
both stars in order to get spatial and mass resolution not too
dissimilar for stars of unequal sizes. As an example, for equal mass
stars we generally use 2000$+$2000 particles for low masses ($\le
1\,\Msun$) and 4000$+$4000 particles for high masses. These numbers
are certainly not impressive by present-day standards but already
corresponded to considerable computational burden given the large
number of collision to simulate and the typical speed of computers at
the time this study was initiated, more than five years ago. We now
discuss the test computations we made to ensure these particle numbers
were sufficient for our aims.

\subsection{Determination of the required resolution}

\begin{figure*}
  \resizebox{\hsize}{!}{%
    \includegraphics{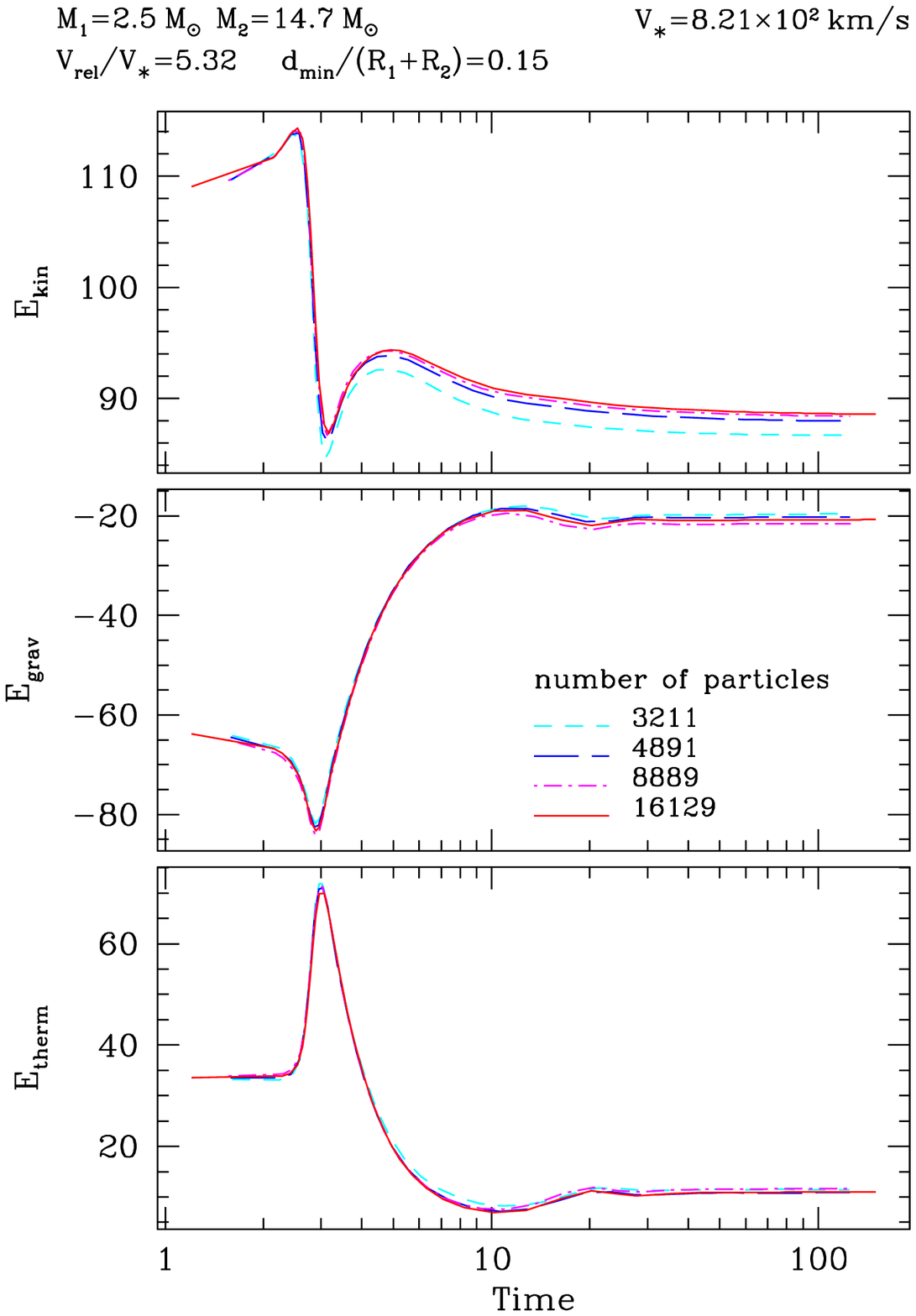}%
    \hspace{0.5cm}
    \includegraphics{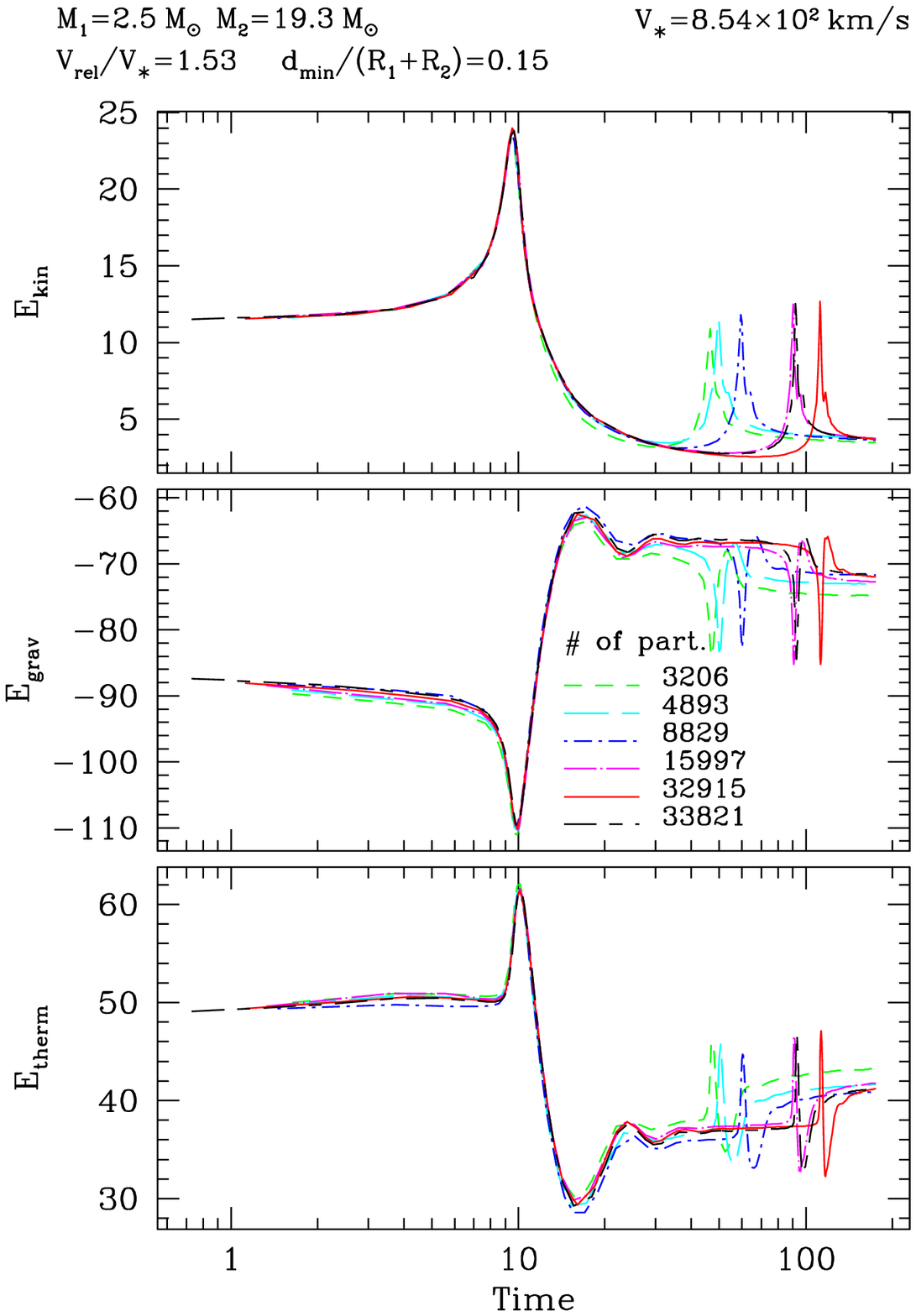}%
    }
  \caption{%
    Evolution of the system energies during two collisions. Each
    collision has been computed with different resolutions (3000 to
    34\,000 particles). Code units are used here. The quantities
    describing the collisions are specified on top of the diagrams.
    See text for further comments. The left column is a ``fly-by''
    encounter. The right column corresponds to a merger (see
    Fig.~\protect\ref{fig:evol_ResColl_1}).  }
  \label{fig:evol_ener_Npart}
\end{figure*}

To determine the minimal desirable number of particles to be used in
our simulations, we computed the same physical collision with various
$N_\mathrm{part}$. Fig.~\ref{fig:evol_ener_Npart} shows the evolution
of the energies during two typical collisions simulated with
increasing numbers of particles.  In all these cases but one, the
number of particles in the small star is $\sim$1000 while the large
star consist of $\sim$2000 to $\sim$32\,000 particles. The first
collision is a high-velocity quasi-hyperbolic ``fly-by''. In the
second example, the stars are left bound to each other after the first
periastron passage.  A second collision ensues that leads to a rapid
merging of the small star into the centre of the big one. In the
``fly-by'' case, the energy evolution curves for the various
$N_\mathrm{part}$ values are very similar to each other as soon as
more than 2000 particles are used to represent the large star.
Contrariwise, in the ``merger'' case, the time between the two
successive periastron passages exhibits a strong dependency on
$N_\mathrm{part}$. Not only does it not converge to some ``real''
value as the resolution is increased but the opposite occurs! This
intriguing behaviour casts important doubts on the ability of this
version of the SPH code to follow reliably the formation and evolution
of tidal binaries. However we note that a simulation with $32000+2000$
particles (the last one in the right panel of
Fig~\ref{fig:evol_ener_Npart}) gives nearly exactly the same energy
evolution as another one with $16000+1000$ particles (the fourth one).
This is a hint at the importance of a good resolution in
\emph{both} stars and not only the larger. Convergence in the results
is attained only if we increase both resolutions. \Added{It is not
  clear to us why this is so but it is obviously connected to the
  poorly resolved envelope structure which determines how much orbital
  energy is dissipated at first passage. This turns out to have very
  little implication for the amount of mass loss, the only quantity we
  want to determine in case of a merger. As shown on
  Fig.~\ref{fig:dM_resol}, it is only for grazing fly-bys that one get
  significant relative discrepancy between different resolutions, as
  long as energy and mass loss are concerned. This also is due to
  inaccuracies in the SPH realisation of the outer layers. But given
  the very small absolute values of these losses, this can only have
  very small impact when SPH results are used to implement collisional
  effects in simulations of high-velocity stellar systems.}

In any case, in our work, we are mostly interested in the final
outcome of collisions in terms of global quantities, like the mass and
energy loss. The dependency of these quantities on $N_\mathrm{part}$
turns out to be very weak. The fractional mass and energy losses and
the fractional deviation from the Keplerian deflection angle typically
vary by less than 20\,\% over the whole range of particle numbers used
in these tests (see appendix for diagrams). While
the lowest $N_\mathrm{part}$ produce results that are somewhat off, as
compared to higher resolution runs, 1000$+$2000 particles seem to be
already sufficient.

\subsection{Starting, ending and analysis of a collision simulation}

Any collision to be computed requires the specification of both
stellar models and of the relative velocity at infinity
$V_\mathrm{rel}^{\infty}$ and the impact parameter $b$. We neglect any 
finite-size effect until the separation between the stars' centres is
$3(R_1+R_2)$. Hence we analytically advance the stars to this situation 
on hyperbolic trajectories. At this point, we start the SPH simulation 
and set $t=0$.

The computation stops at $t_\mathrm{end}\ge 20\,T_\mathrm{dyn}$ with
$T_\mathrm{dyn} = \sqrt{(R_1+R_2)^3/(G(M_1+M_2))}$. If, at this time,
two surviving stars are present with separation less than
$3(R_1+R_2)_\mathrm{ini}$ or if the amount of gas with uncertain fate
(see next paragraph) exceeds 1$\,$\% of the total mass, the simulation
is integrated further (by setting $t_\mathrm{end} \gets 2
t_\mathrm{end}$) until it passes these tests or some maximal
integration time is reached.  In practise, no collision required
integration past $t=2500 \sqrt{{\Rsun}^3/(G{\Msun})}$.
Unfortunately, these simple termination prescriptions are probably
inadequate when a bound binary forms after first periastron passage.
They can indeed force integration for many orbital periods although we
expect the SPH scheme (at least when used with our set of numerical
parameters) to lose reliability in that regime whose outcome is very
likely to be a merger (see Sec.~\ref{subsec:results_overall}). A wiser
approach would have been to identify such encounters, stop computation
after first passage at periastron and, if needed, to rely on other
theoretical considerations to assess the outcome.

\Added{One monitors the energy (non-)conservation with $\delta_E =
  \left|E_{\rm end}-E_{\rm ini}\right|/E_{\rm norm}$ with $E_{\rm
    norm}=E_{\rm kin}^\infty+E_{\rm bind}$, where $E_{\rm ini}$,
  $E_{\rm end}$ are the initial and final total energies, $E_{\rm
    kin}^\infty$ the initial kinetic energy at infinity (orbital
  energy) and $E_{\rm bind}$ the sum of the binding energies of the
  stars (positive definite). Using the total energy, $E_{\rm
    ini}=E_{\rm kin}^\infty-E_{\rm bind}$ for normalisation isn't
  appropriate because it may be very close to zero, leading to
  misleadingly large values of $\delta_E$. $E_{\rm norm}$ gives a
  natural energy scale for the problem. Using $E_{\rm bind}$ for
  normalisation doesn't change much the $\delta_E$ statistics. The
  worst non-conservation amongst all simulations is $\delta_E \simeq
  0.06$; all but 15 simulations have $\delta_E < 0.01$; 66\,\% of all
  runs have $\delta_E < 0.001$.}

The SPH code yields quantities describing every particle at the end of
the computation, i.e. their positions, velocities, internal
energies\ldots This raw ``microscopic'' data has to be analysed to
provide a useful description of the outcome of the collision in terms
of ``macroscopic'' quantities, i.e. properties of outgoing star(s) (if
any). Namely, we want to know how many stars survive (0, 1 or 2) and
what their masses, positions and velocities at the end of the
computation are. This, and any other aspect of the structure of the
star(s) (see Sec.~\ref{subsec:use_results}) and of the ejected gas,
can be easily determined if we manage to build a list stating to which
star each particle belongs or whether it is unbound to any surviving
star. This data is provided by an analysis algorithm which proceeds in
two steps:
\begin{enumerate}
\item A first guess attribution is realised by a code which tries to
  identify, through density and proximity criteria, zero, one or two
  concentrated lumps of particles. To this end, we use the freely
  available {\HOP} algorithm by \citet{EH98}.  These groups are
  regarded as first approximations of stars to be refined in the
  second stage of the method.
\item We then iteratively cycle through all particles to compute the
  energies of each one relative to both stars (``A'' and ``B''). For
  iteration $k$, the energy of particle $i$ relative to star A as
  identified at the previous iteration (``A$_{k-1}$'') reads
\begin{eqnarray}
  \left.E^\mathrm{A}_i\right|_k &=& u_i +
  \frac{1}{2}m_i
  \left(\vec{v}_i-\vec{V}_{\mathrm{A}_{k-1}}\right)^2 \nonumber \\
&& - Gm_i\!\sum_{j\in \mathrm{A}_{k-1}}\!
 \frac{m_j}{\left\|\vec{x}_i-\vec{x}_j\right\|}.
\end{eqnarray}
In this formula, $u_i$, $m_i$, $\vec{v}_i$ and $\vec{x}_i$ are the
internal (thermal) energy, mass, velocity and position of particle
$i$. $\vec{V}_{\mathrm{A}_{k-1}}$ is the velocity of ``star''
A$_{k-1}$, i.e.
\[
  \vec{V}_{\mathrm{A}_{k-1}} =
  \sum_{\mathrm{A}_{k-1}}\! m_j\vec{v}_j \left/
    \sum_{\mathrm{A}_{k-1}}\!m_j \right..
\]
If either $\left.E^\mathrm{A}_i\right|_k$ or
$\left.E^\mathrm{B}_i\right|_k$ is negative, the particle is ascribed
to the ``star'' relative to which its energy is the most negative.
Particles with positive energies relative to both stars but with negative total
energy in the collision centre-of-mass reference frame (CMRF) are
tagged as ``doubtful''.  At the end of the iteration we thus get new
sets of particles making up ``stars'' A$_k$ and B$_k$. We go on
iterating until no modification in the composition of these sets
occurs anymore.
\end{enumerate}

This procedure deserves a few comments.
\begin{itemize}
\item The energy criterion may fail to predict the correct
  attributions. For instance, a particle with high velocity
  \emph{toward} a given star may happen to have positive energy
  relative to this star even if it will impact it and thus merge into
  it.  Furthermore, even without resorting to hydrodynamical
  processes, we learn from studies of the gravitational 3-body problem
  that the eventual fate of a particle submitted to the gravitational
  forces of two massive bodies can not generally be predicted just
  through energy consideration.  However, if we carry on the SPH
  integration to a physical time large enough for the stars to have
  moved away from each other to a large separation and/or for the
  large amplitude hydrodynamical processes to have ceased, we expect
  the final SPH configuration to be essentially free of such
  problematical particles and the energy criterion to be reliable.
\item ``Doubtful'' particles generally lie between the two stars so
  that they gain negative gravitational contributions to their total
  energy from both potential wells even though they are not bound to
  any one star. In such cases, their number should decrease as the
  distance between the stars increases. Another situation that can
  leave a relatively important doubtful mass fraction (i.e. $>1$\,\%
  of the total mass) occur in high velocity head-on collisions that
  result in an expending gas cloud. Its central part, lying in the
  potential well of the surrounding gas has negative total energy but
  nevertheless expands to infinity. Although these cases seem to have
  genuine physically interpretations, there are situations where a high
  $M_\mathrm{doubt}$ is indicative of some error in the analysis. One
  such case consist of a close tidal binary being erroneously
  identified as a single particle group in the first step. The
  iterative steps then progressively interpret one of the stars as a
  group of doubtful particles while retaining only the other lump as a
  ``real'' star. 
\item This last example illustrates how critical the first attribution
  stage is. Its failure to detect independent stars cannot be
  recovered by the iterative process! There is probably room for
  improvement in this part of our analysis procedure and the use of
  {\HOP}, an algorithm aimed at finding structures in large
  cosmological simulations, is arguably an inefficient overkill. A
  simple-minded approach that first divides particles in the same two
  groups that built up the pre-encounter stars proves to allow
  convergence to real stars in cases that confuse {\HOP}. To
  account for mergers, when the distance between the centres of the
  two groups is much smaller than some typical size, we can ascribe
  all particles to a single group to be then ``eroded'' down to the
  bound star (if it exists) by the iterative energy test. 
\end{itemize}

\Added{All simulations were first analysed using {\HOP} to produce the
  initial particle attributions. The results of the iterative
  procedure were then visually inspected by plotting $\log\rho$ versus
  spatial coordinate $x$ for all SPH particles and using different
  colours to code the attributions. Errors are immediately spotted in
  such a diagram allowing one to integrate the simulation for a longer
  time if the separation between ``stars'' (density peaks) is deemed
  too small or switching to the just-mentioned simple-minded scheme for
  initial attributions in the few cases {\HOP} clearly made a wrong
  guess.}

\begin{figure*}
    \begin{tabular}{cc}
      \multicolumn{1}{l}{\large (a)} & \multicolumn{1}{l}{\large (b)} \\
      \resizebox{0.48\hsize}{!}{\includegraphics{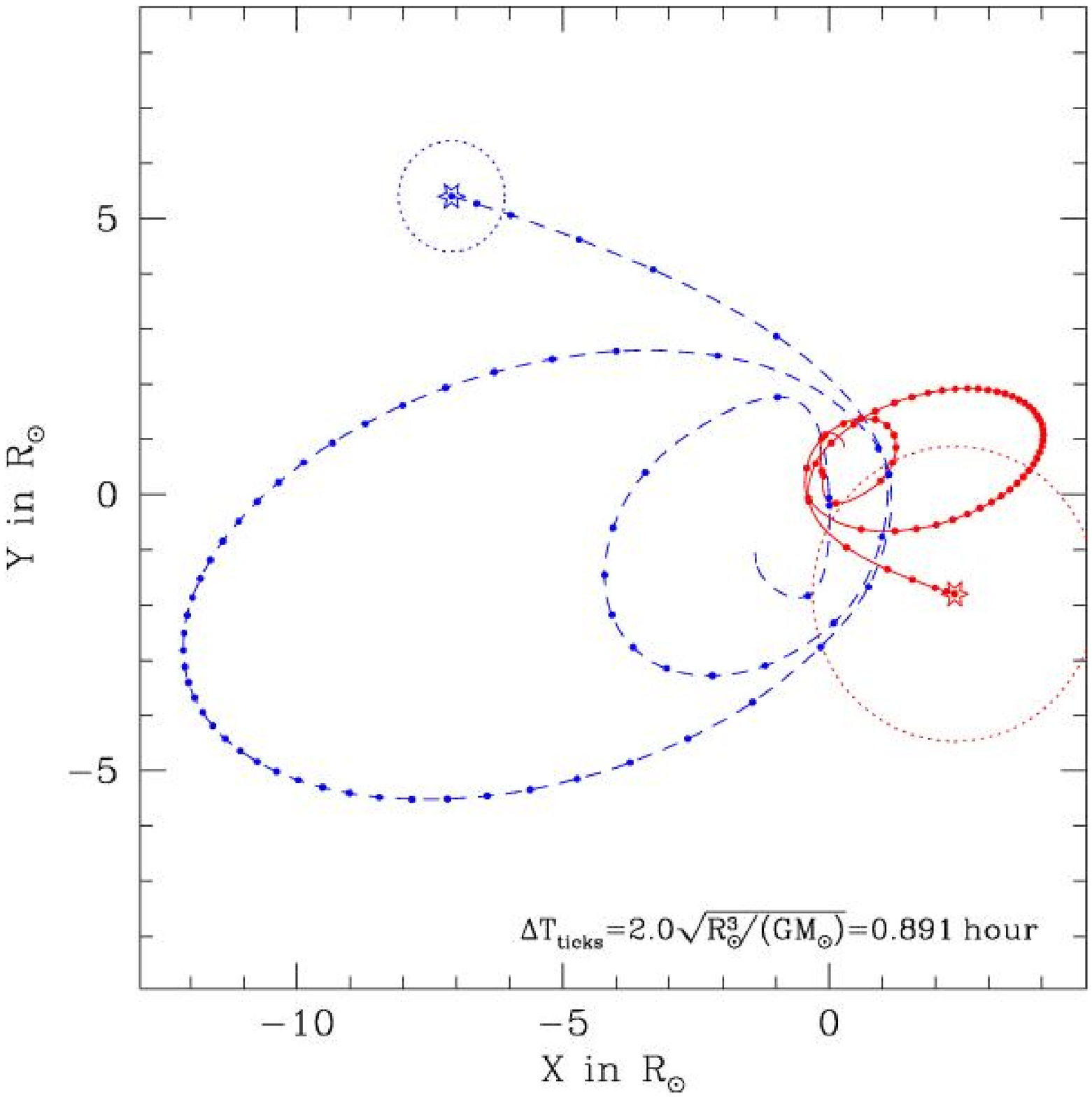}} &
      \resizebox{0.48\hsize}{!}{\includegraphics{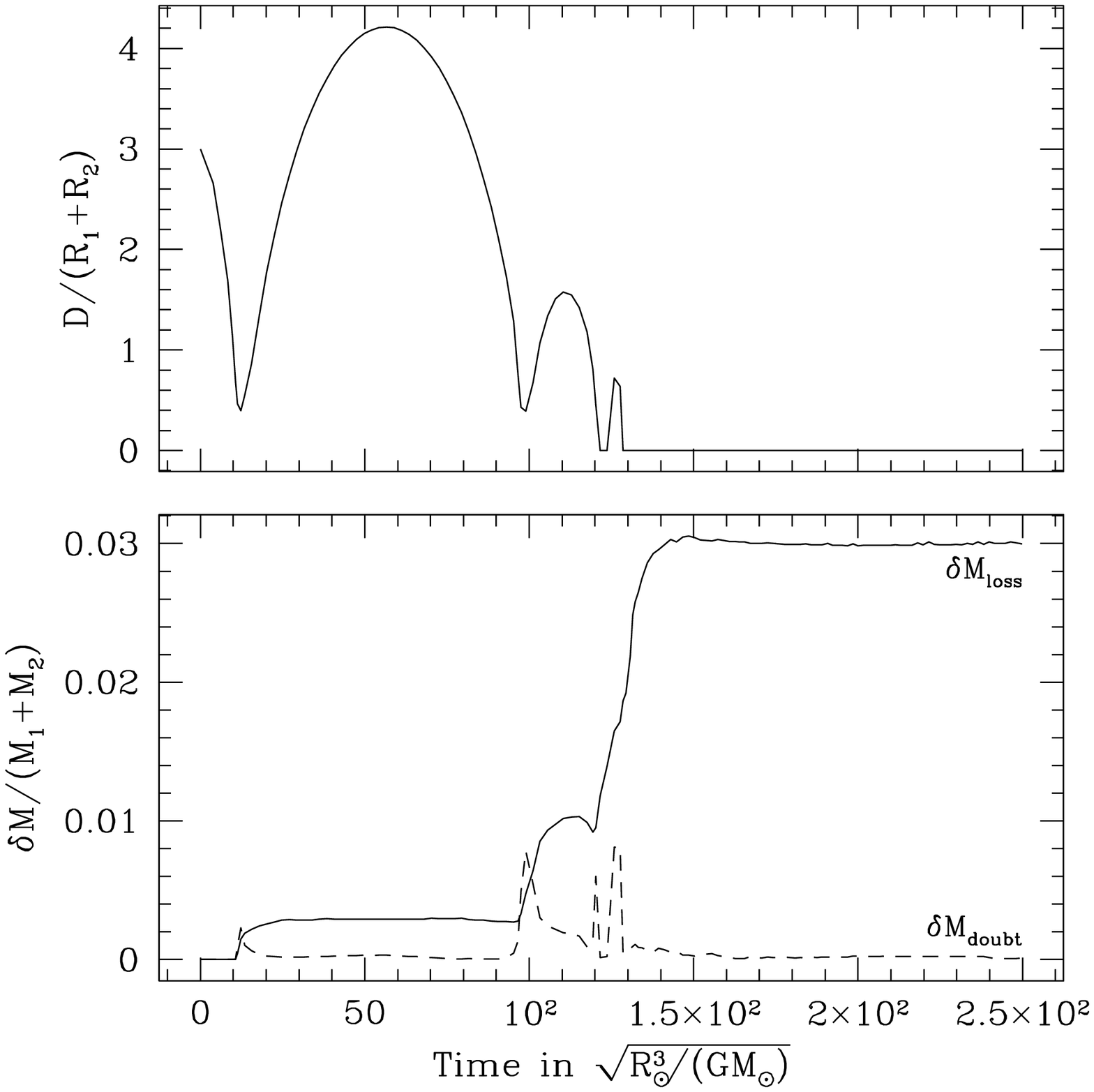}} \\
    \end{tabular}
  \caption{%
    Collision between stars with $M_1=1\,{\Msun}$, $M_2=3\,{\Msun}$,
    $V_\mathrm{rel}^{\infty}=0.07V_{\ast}=43.7\,\mathrm{km}\,\mathrm{s}^{-1}$,
    $d_\mathrm{min}/(R_1+R_2)=0.39$.  Panel (a): Trajectories of
    colliding stars, as identified by the algorithm used to analyse
    the outcome of collisions. This encounter leads to the formation
    of a binary which coalesce after two orbits. The analysis
    algorithm is unable to tell one star from the other during the
    final merging.  Panel (b), top: Evolution of the separation
    between both stars. Panel (b), bottom: Evolution of the amount of
    gas unbound to any star, with positive or negative energy in the
    centre-of-mass reference frame ($\delta M_\mathrm{loss}$ and
    $\delta M_\mathrm{doubt}$ respectively). This diagram illustrates
    how the mass loss increases abruptly at each periastron passage
    and reaches a steady value after complete merging. The amount of
    gas with doubtful fate also gets quickly to a vanishingly small
    value. This ensures that the interaction has been integrated for a
    sufficiently long time.  }
  \label{fig:evol_ResColl_1}
\end{figure*}

In the vast majority of simulations, we only run the analysis software
just described on the final SPH file. As mentioned above, if, for that
configuration, $M_\mathrm{doubt}$ exceeds some fraction of the total
mass (1\,\%) or wrong attribution is seen, we compute the interaction
for a longer physical time.  When the integration is deemed over and
the properties and kinematics of the surviving star(s) have been
determined, we assume that the stars' masses have reached constant
values and that the subsequent orbital evolution is purely Keplerian
again. This allows to compute $\theta_\mathrm{coll}$ as an asymptotic
value. The physical time $t_\mathrm{end}$ over which the SPH
simulation is computed has thus to be long enough for the strong
hydrodynamical regime to be over. On the other hand, choosing too
large a value for $t_\mathrm{end}$, is not only computationally
expensive but could result in inaccurate results due to the
accumulation of small numerical errors. Hence, it is of interest to
analyse a few typical collisions at a number of increasing times
during the SPH computation to test whether the outcome quantities have
reached steady values and whether these values show sign of numerical
drift at large $t$. Fig.~\ref{fig:evol_ResColl_1} is an example of
such computations. The plot of the trajectories (panel (a)) testifies
that, in most cases, the analysis procedure identifies the stars
correctly, even during close interaction. The curves for the evolution
of predicted mass and energy losses show abrupt increases at
periastron passages and stay nearly constant quickly after the last
close encounter (leading to a merger) is over. \Added{Although the
  analyse software gets confused when the stars penetrate each other,
  this is of no practical concern because it is only a transitory
  situation. For fly-bys (including the case the small star emerges as
  an unbound, expanding cloud), we integrate until the stars are again
  very well separated; when stars capture each other, the analysis is
  only done after a merged object has formed or when the stars,
  forming a binary, do not overlap.}  We conclude that the way we
terminate SPH collisions and analyse their results is sound.

\subsection{Building a comprehensive table of collisions}
\label{subsec:buildingtable}

\begin{figure*}
  \resizebox{\hsize}{0.9\vsize}{\includegraphics{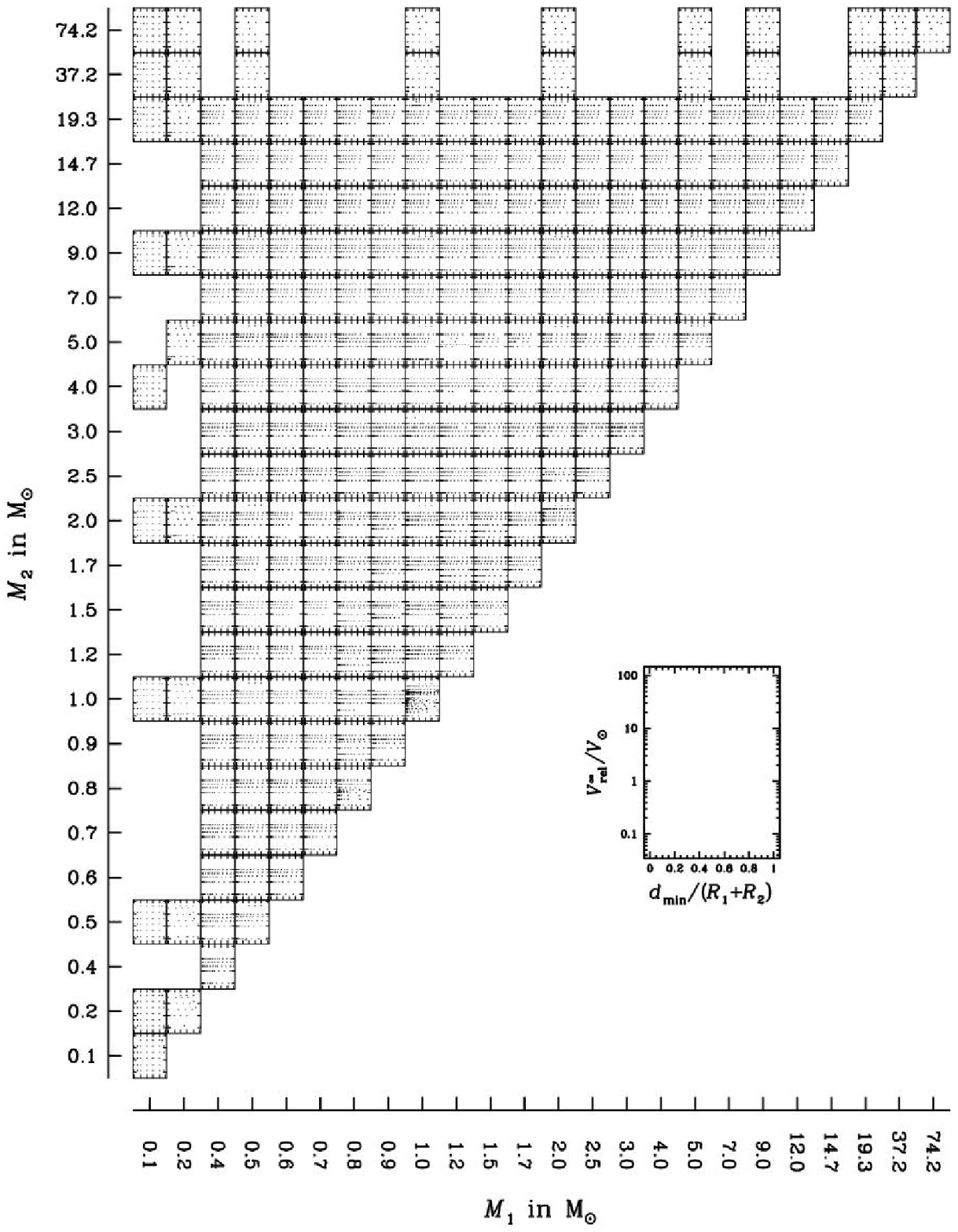}}
\caption{ Diagram depicting the initial conditions for all the 
  collision simulations we performed. Dots in each small box represent
  the pericentre distance $d_{\rm min}$ (in units of the stellar
  radii) and relative velocity at infinity $V_{\rm rel}^\infty$ (in
  units of $V_\odot=436.5\,\kms$) of all simulations for a given
  $(M_1,\,M_2)$ pair. The enlarged box displays the $(d_{\rm
    min},\,V_{\rm rel}^\infty)$ plotting area. Note that the masses
  axis are neither linear nor logarithmic but simply represents the
  differents masses in a sequence.
\label{fig:all_ini_cond}
}
\end{figure*}

This study was first embarked on as a sub-project. It is part of a
work aimed at simulating the stellar dynamical evolution of dense
galactic nuclei hosting black holes. To this end, a new Monte Carlo
code for cluster dynamics has been developed \citep{FB01a,FB02b}. In
order to incorporate the effects of stellar collisions with a high
level of realism into it, we decided to compute a large number of SPH
simulations spanning all the relevant values of the initial
conditions. Our hope was then to extract fitting formulae from this
database of results to get an efficient description of the outcome of
any arbitrary collision that could occur during a cluster simulation
run.\footnote{A collision requires a few hours to a few days of CPU
  time to be simulated by the SPH code on a standard workstation and
  some simulated high density nuclei experience many thousands of
  these events during a run spanning a physical duration of
  $10^{10}$\,years. It is consequently clearly impossible to switch to
  on-the-fly SPH integrations when collisions are detected in the
  cluster simulations!} Figuring out such mathematical descriptions
proved too difficult and we have resorted to an interpolation
procedure. This will be explained in Sec.~\ref{subsec:use_results}.

Contrary to globular clusters where all collisions are essentially
parabolic due to the velocity dispersion being much lower than the
escape velocity from a stellar surface, galactic nuclei may have deep
potential wells, or even harbour massive black holes and thus force
some of their inhabiting stars to collide on high-velocity hyperbolic
trajectories. For instance, at the centre of the Milky Way, ``SO''
stars are on orbits with pericentre velocities of up to $\sim
12\,000\,\kms$ \citep{GhezEtAl03,SchoedelEtAl03} and even higher values
will probably show up in future higher resolution observations
reaching closer to the $\sim 3-4\times 10^6\,{\Msun}$ black hole
{\SgrA}. Hence, we cannot restrict ourselves to collisions with
$V_\mathrm{rel}^\infty\simeq 0$ but have to go up to a few thousands
of {\kms}.

Moreover, the population in galactic nuclei does not consist of old
stars all born at the same time but may include MS stars with an
extended range of masses. High mass stars are particularly important
in the first phases of the system evolution: relaxation-induced mass
segregation may quickly concentrate them in the high density central
regions where, having large cross sections, they get relatively high
collision probabilities, despite their overall scarcity and their
short MS lifetimes. Consequently, we have also to span a large range
of initial masses, extending far beyond the $\sim 1\,{\Msun}$ tun-off
mass that would be sufficient for a study of collisions in present-day
Galactic globular clusters.

Finally, to further extend the domain in parameter space to be
explored, we note that stars of different masses have very different
internal structures (see density profiles in appendix) so we cannot hope to scale out the absolute mass from the
collision process. For instance, we would expect the (dimensionless)
results to depend only on the mass ratio $M_1/M_2$ only if stellar
structures were homologous to each other \emph{and} a power-law
$M$--$R$ relation was obeyed. As this is not true, we have to consider
the masses of the in-coming stars as two independent variables.

Summing it up, we have to deal with a fully 4-dimensional parameter
space in which we have sampled a domain which is more or less the
following:
\begin{itemize}
\item Stellar masses from 0.1 to 74.3\,${\Msun}$ (the latter value
  corresponds to a ZAMS mass of 85\,${\Msun}$).
\item Relative velocities in the range
  $V_\mathrm{rel}^\infty/V_\ast\simeq 0.03$--$30$.
\item Impact parameters corresponding to
  $d_\mathrm{min}/(R_1+R_2)=0$--$0.9$.
\end{itemize}

A mere 10~points resolution for each parameter already turns to a
total of 10\,000 collisions to be computed, a number clearly beyond
what can be managed ``by hand''. This high number grounded our
decision to neglect other, ``second order''parameters such as
metallicity, rotation or evolutionary stage along the MS track. A
complete software system, consisting of many UNIX shell scripts has
been developed to run these SPH simulation in a (nearly) completely
automatic way. The system looks through a table for collision
simulations that have not yet been computed to their end and makes
them run on idle workstations available through the local computer
network. The system interrupts a simulation job when the computer on
which it's running ceases to be ``available'' (basically during
daytime) and calls the analysis software when a run is over. If no
further integration is required, the results are added to an output
table. Supervising this automatic system is not as painless as it may
sound: due to the number of simulations that run concurrently (10 to
50, typically), ``exceptional'' problems mainly originating from
malfunctions in the local network occur nearly every day and have to
be fixed manually.  All in all, obtaining a system reasonably
crash-proof revealed itself to be unexpectedly difficult. This paper
reports on the results of the $\sim$14\,000 simulations we managed to
compute with this approach. \Added{On Fig.~\ref{fig:all_ini_cond}, we
  attempt to show the initial conditions for all simulations.}

\subsection{Formation of binaries through tidal interactions}
\label{subsec:tid_bin}

Even when the periastron distance is larger than the sum of the
stellar radii, close encounters at low relative velocity can rise
tides in the interacting stars and lead to the formation of a bound
binary. As already pointed out by \citet{FPR75}, in globular clusters,
the cross section for such tidal captures is a factor 1--2~times as
large as for collisions (assuming a typical relative velocity of
10\,{\kms}). \Added{Determining through SPH simulations the critical
  impact parameter for tidal captures in (quasi-)parabolic,
  non-touching encounters is a demanding task, requiring high
  resolution of the stellar envelopes where tides transfer energy from
  the orbital motion to stellar oscillations.} This phenomenon is not
treated in this paper because, in typical galactic nuclei, the
relative velocities are in excess of 50\,{\kms}, a regime where tidal
binaries can form only for very close encounters, requiring {\bf
  contact} interaction in most cases, with the possible exception of
less concentrated, low-mass stars \citep{LN88,KL99}. Hence, we
restricted ourselves to the range $d_\mathrm{min}<(R_1+R_2)$.

\section{Results}
\label{sec:results}

\subsection{Overall survey of the results}
\label{subsec:results_overall}

Trying to get a complete coverage of collision parameter space implies
a huge volume of simulation results. The difficulty of our approach is
to extract useful information in manageable form out of these data.
As the database was nearing completion, we looked for mathematical
relations between various input and output quantities. Due to the
deterministic nature of collisions, many strong correlations are
clearly visible but finding fitting formulae for them eluded us. The
basic difficulty stems from dimensionality of the initial parameter
space which seems to be genuinely 4D.

\Added{Here we do not show the results from specific collision
  simulations nor discuss the physical mechanism at play during them,
  as this has been done extensively in previous works
  \citep{BH87,BH92,LRS93,LRS96}. For the interested reader a few
  specific simulations are presented in the appendix. What concerns us
  here is a description of the simulation database as a whole.}

\begin{figure*}
  \resizebox{!}{23cm}{\includegraphics{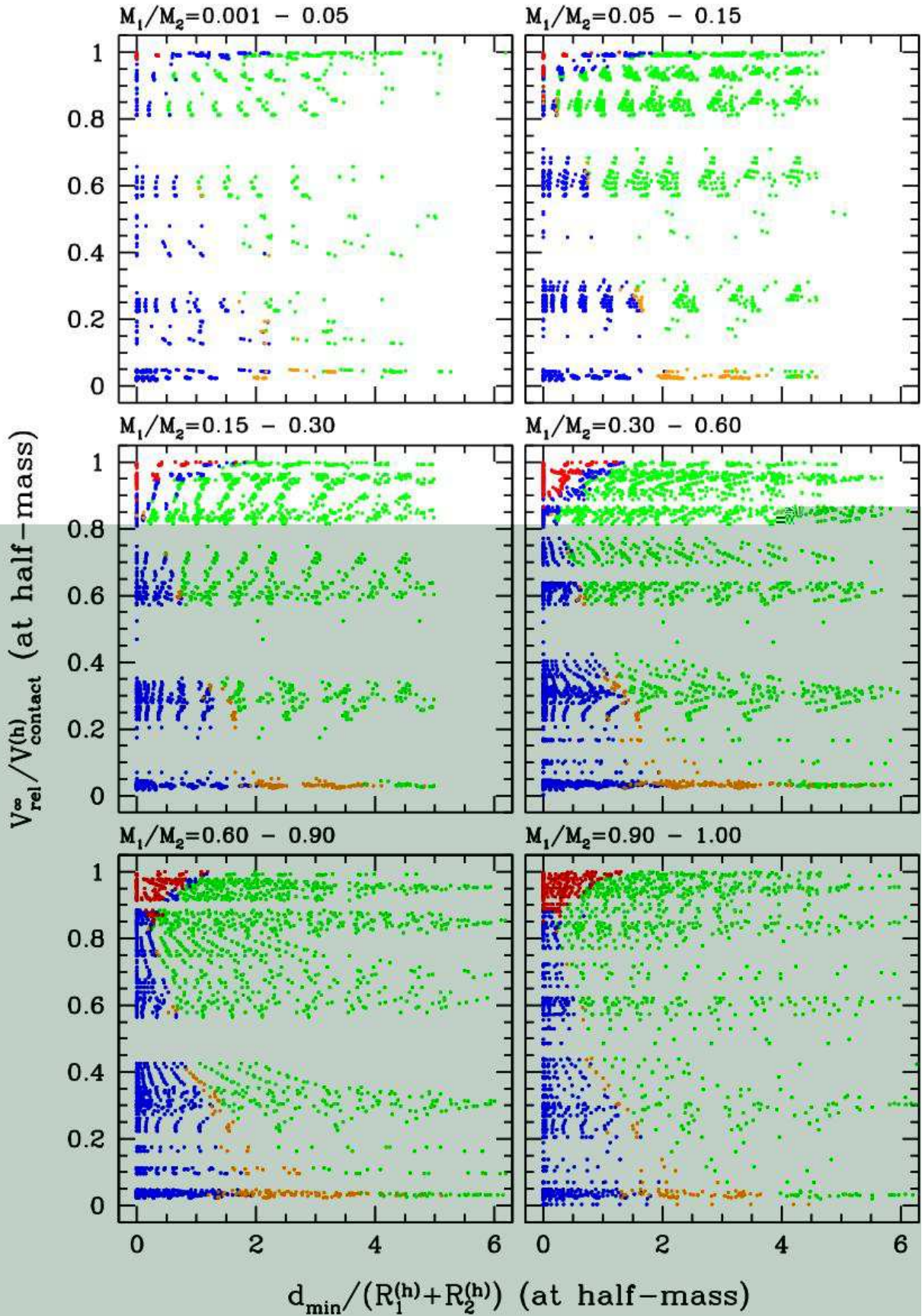}}
  \caption{%
    Number of stars surviving collisions. Red dots are for collisions
    leading to complete disruption, blue dots for cases with one
    surviving star, orange dots for tidal binaries (very likely to
    eventually merge) and green dots for encounters with two surviving
    stars. See text for further explanations and comments. }
  \label{fig:Nstars}
\end{figure*}

The simplest, most qualitative, description of the collisional outcome
is the number of outgoing star(s). For given initial masses, we can
plot a 2D diagram indicating this number for all collision simulations
performed, as a function of the impact parameter and the relative
velocity (Fig.~\ref{fig:Nstars}). 

Before we comment on that figure, some explanations are called for.
$V_\mathrm{contact}^{(\mathrm{h})}$ is an approximate value of the
relative velocity at ``half-mass contact''. It is defined through
\begin{equation}
  V_\mathrm{contact}^{(\mathrm{h})}= \sqrt{ \left(V_\mathrm{rel}^\infty\right)^2 +
    \left(V_{\ast}^{(\mathrm{h})}\right)^2 }.
\end{equation}
It should be noticed that such a ``deep'' contact does not occur
during encounter with large impact parameters; this value only serves
as a convenient parameterisation that allows to map the $[0,\infty[$
$V_\mathrm{rel}^\infty$ range onto $[0,1[$. In these plots, each dot
represent one SPH simulation. Green dots are collisions survived by
both stars (although significant amount of mass loss may have
occurred). Blue dots indicate that there is only one star left at the
end of the encounter. Orange dots stand for tidal binaries and red
dots for complete disruption of both stars. One sees that, for this
half-mass parametrisation of the initial conditions, the borders
between these various regimes are primarily set by the mass ration
$q=M_1/M_2$, quite independently of the actual masses. Unfortunately,
as will be stressed below, this appears generally not to hold for more
quantitative results.

These diagrams provide a division of the collisions into a few
different regimes.  Most of the $(d_{\mathrm{min}},V_{\mathrm{rel}})$
plane is occupied by ``fly-bys'', i.e. encounters from which two
unbound stars escape. In some cases, this domain extends to
$d_{\mathrm{min}}=0$ like a small wedge between the merger regime
(lower velocities) and disruptions (higher velocities). It is thus
possible for a small star to pass right through the centre of a larger
one and not being disrupted. We detected about 250 such cases in our
survey, all with $q$ between 0.04 and 0.25 and $M_1$ (small star)
between 0.7 and $2\,{\Msun}$.  Moreover, in about one third of these
simulations (with $\nu<1.7$), the small star \emph{gains} mass during
the interaction while the larger star always suffers from important
mass loss. It seems even possible that in some collisions, the small
star, acting like a bullet, shatters its target but remains nearly
intact. Similar outcomes were obtained by BH92 for $n=1.5$ polytropes
with $q=0.2$. LRS93 did not find any head-on collision with a
surviving small star. As pointed out by these author, such
discrepancies --as well as other differences between our results and
published data, see Sec.~\ref{subsec:comp_lit}-- probably originate in
the fact that different stellar models have been used. The ratio of
stellar central densities is likely to be a key parameter in allowing
such ``fly-through'' collisions. In all the cases identified by us,
this ratio exceeds 6. However, the astrophysical importance of this
phenomenon is low because, at large relative velocities, collisions
with small $d_{\mathrm{min}}$ are unlikely as gravitational focusing
is quenched.

Mergers or bound binaries are formed during encounters with low
relative velocities and impact parameter below some critical
$\lambda_{\rm merg}$. \Added{This value depends on the relative
velocity and the masses (mostly through the mass ratio). It is
apparent as the transition between green and orange or blue dots on
Fig.~\ref{fig:Nstars}. It is generally larger than $R_1^{({\rm h})} +
R_2^{({\rm h})}$ for $\nu < 0.6$ and smaller at larger velocities.  An
ad-hoc analytical parametrisation of $\lambda_{\rm merg}$ as a
function of $M_1$, $M_2$ and $\nu$ will be published soon
\citep{FRB05}.}  Remarkably, the maximum velocity for a head-on
collision that still leads to merger is $\nu
\simeq 1.7-2.1$, quite independently of the stellar models. The border
between this region and the ``fly-by'' regime at higher
$d_{\mathrm{min}}$ is also rather well defined if half-mass variables
are used.

\begin{figure}
  \resizebox{\hsize}{!}{%
    \includegraphics{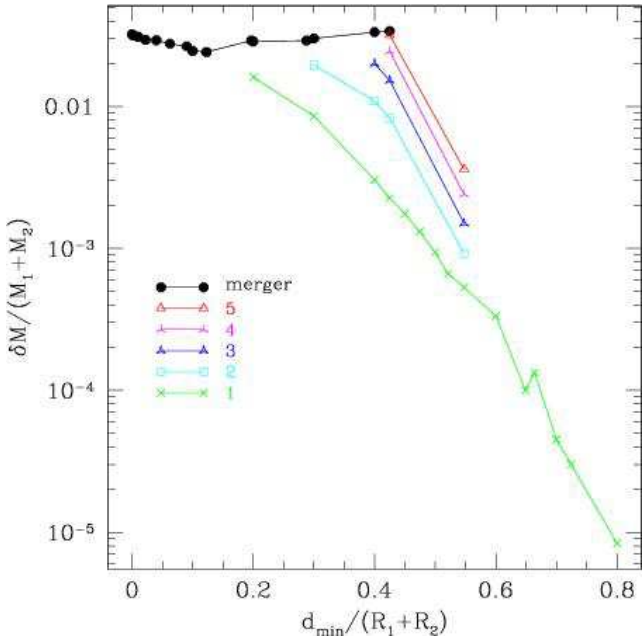}%
    }
  \caption{%
    Fractional mass loss for collisions between stars of masses $1$
    and $5\,{\Msun}$ with a relative velocity at infinity of
    $43.7$\,km\,s$^{-1}$. We indicate the mass loss for each
    successive periastron passage. \Added{The line with crosses (green
      in the colour on-line version) shows the mass loss after the
      first passage, the line with open squares (cyan) the mass loss
      after the second passage and so on until the stars have merged.
      The total mass loss is shown by the line with round dots
      (black).}  For the most central collisions
    ($d_{\mathrm{min}}<0.425$), the evolution was integrated up to
    final merging. This was not possible for more distant encounter
    due to the important increase in computing time this would require
    and the loss of numerical precision to be expected in such long
    SPH integrations.}
  \label{fig:delta_Npassages}
\end{figure}

All binaries formed in our simulations will presumably merge into
single stars after a few orbits. The reason for this is that, at each
periastron passage with $d_{\mathrm{min}}<(R_1+R_2)$, some orbital
energy is converted into heat by shocks and the stellar radii expand
so that at next periastron passage the hydrodynamical interaction is
stronger and more energy is dissipated \citep{BHT89}.  Hence, the fate
of these binaries is not as complex an issue as the long-term orbital
evolution of systems formed through more distant encounters
\citep{Mardling95a,Mardling95b}. Thus, the border between the regions
of merging and binary formation probably results from the criteria we
use to stop the SPH computations and has no strong physical meaning.
If it were possible to integrate the evolution for many orbital
periods, there is little doubt that any binary would eventually merge.
Fig.~\ref{fig:delta_Npassages} illustrates this point. To produce this
diagram we computed a set of collisions with increasing
$d_{\mathrm{min}}$ for given stellar models and a fixed value of
$V_{\mathrm{rel}}^{\infty}$ which is sufficiently low that every
collision leads either to direct merger or binary formation.  Unlike
the bulk of our simulations for which we analysed only the ``final''
state, here we report the mass loss after each successive periastron
passage. Obviously, as $d_{\mathrm{min}}$ is increased, the number of
orbits preceding the final coalescence get higher and higher, as does
the orbital period. Consequently, the required CPU time grows up to
unacceptable values. A noticeable feature of
Fig.~\ref{fig:delta_Npassages} is that all the collisions apparently
converge to nearly the same total mass loss at merging. The reason for
this behaviour is unknown to us.

Apart from the low velocity merging zone, another region with one
surviving star is present in the diagrams of Fig.~\ref{fig:Nstars}.
This second zone is more or less confined between cases where stars
are completely destroyed (for lower impact parameters) or both survive
(for higher impact parameter). This ``one-star band'', which does not
show up when the two stars are (nearly) identical, is populated by
collisions during which the small impactor is disrupted without being
accreted into the large star. In such high velocity collisions, the
small star accumulates so much thermal energy as it flies through the
massive one, that it turns into an unbound, expanding gas cloud.

The most spectacular collisions are those that lead to complete
disruption of both stars. However, to achieve this result, we note
that both a high relative velocity and a small impact parameter are
required, a combination made unlikely by the absence of gravitational
focusing at such high velocities so it is clear that neither mergers
nor complete disruptions are likely outcome in galactic nuclei, as
confirmed by Monte Carlo simulations \citep{FB02b,FGR04c}.

\subsection{Comparison with literature}
\label{subsec:comp_lit}

\begin{figure*}
      \resizebox{0.95\hsize}{!}{%
      \includegraphics{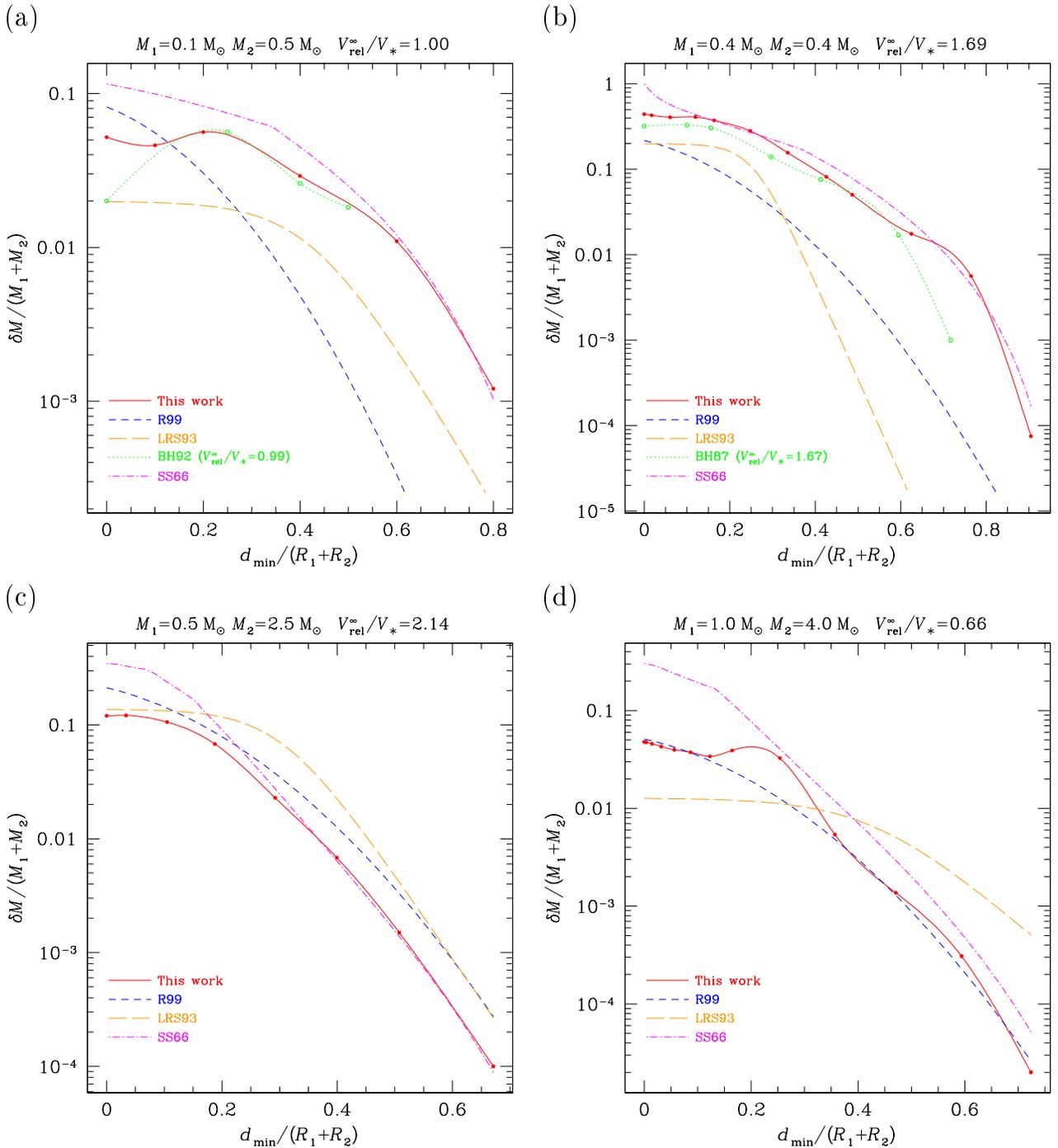}%
      }
  \caption{
    Collisional fractional mass loss. We compare some of our
    simulations (dots and solid line spline interpolation) with
    results from the literature (see text). To obtain the ``Spitzer \&
    Saslaw 66'' (SS66) curves we applied the method of these authors
    to our stellar models. Panels (a) and (b): for such small mass
    stars, the structure is quasi-identical to a $n=1.5$ polytrope.
    This is why we get a good agreement with BH87 and BH92 but big
    discrepancies with formulae from LRS93 and R99, as these authors
    use more concentrated $n=3$ structures. Note that the SS66
    prescription gives very satisfying prediction for off-axis
    encounters! Panel~(b) This is a case with relatively good
    agreement with published formulae. Still, the predictions from R99
    and LSR93 are 2--3 times larger than our mass losses. The
    agreement with SS66 is excellent as soon as
    $d_{\mathrm{min}}/(R_1+R_2)>0.15$. Panel~(d): Here, the best
    agreement is obtain with the R99 formula, despite the velocity
    being a bit lower than the range explored for this work. SS66
    gives satisfactory results, but not LRS93. The reason for this
    discrepancy is unknown. See caption of
    Fig.~\protect\ref{fig:comp_dm_polytr} for the probable explanation
    of the bump at $d_{\mathrm{min}}/(R_1+R_2)\simeq 0.2$ in our
    curve.  }
  \label{fig:comp_lit_a}
\end{figure*}

\begin{figure*}
      \resizebox{0.95\hsize}{!}{%
      \includegraphics{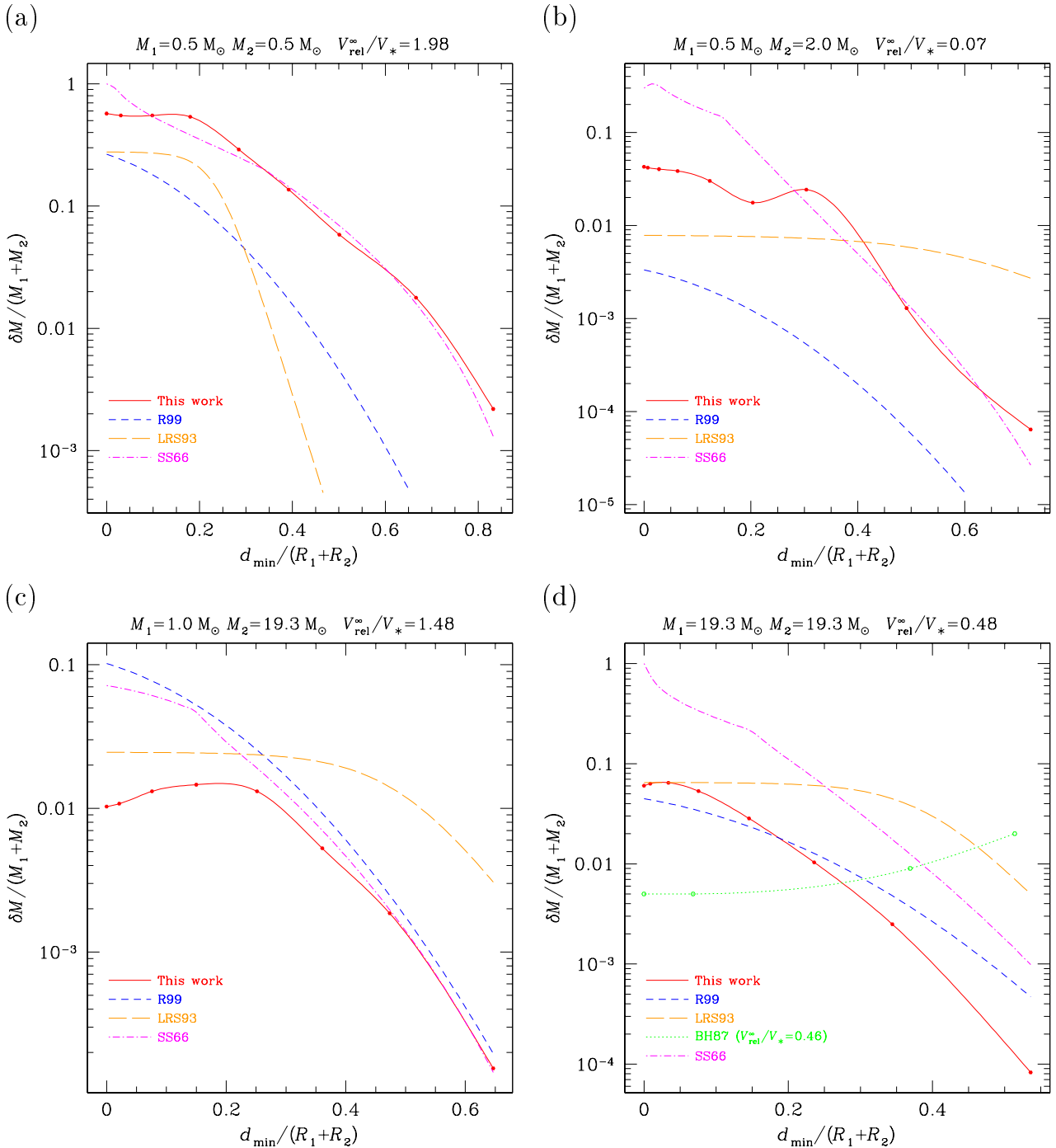}%
      }
  \caption{
    Similar to Fig.~\protect\ref{fig:comp_lit_a}. Here we push results
    from the literature somewhat beyond their natural range to test
    for their predictive power. Panel~(a): the poor agreement between
    us and R99 and LRS93 is due to $0.5\,{\Msun}$ MS stars
    being much less concentrated than $n=3$ polytropes. Panel~(b): the
    disagreement with R99 originates in the low velocity we use. The
    same may be true for LRS93. Note that our result for large impact
    parameter is probably an under-estimate. For such low initial
    velocities, we expect the formation of a binary and a subsequent
    merging to occur. However, it is likely that we did not integrate
    past the first pericentre passage (see
    Fig.\protect\ref{fig:delta_Npassages}). The flatness of the curve
    of LRS93 may reflect this phenomenon. Panel~(c): here, we have
    smaller mass ratio than any simulations from LRS93 and R99. The
    agreement with R99 at large $d_{\mathrm{min}}$ is probably
    fortuitous. Panel~(d) The discrepancy with BH87 stems from our use
    of completely different stellar models. R99 provides not so bad an
    agreement, given the low value of velocity. The mismatch with
    LSR93 is of more mysterious nature. This is one of the few cases
    where SS66 prescription fails at large impact parameters. }
  \label{fig:comp_lit_b}
\end{figure*}

\begin{figure*}
      \resizebox{0.95\hsize}{!}{%
      \includegraphics{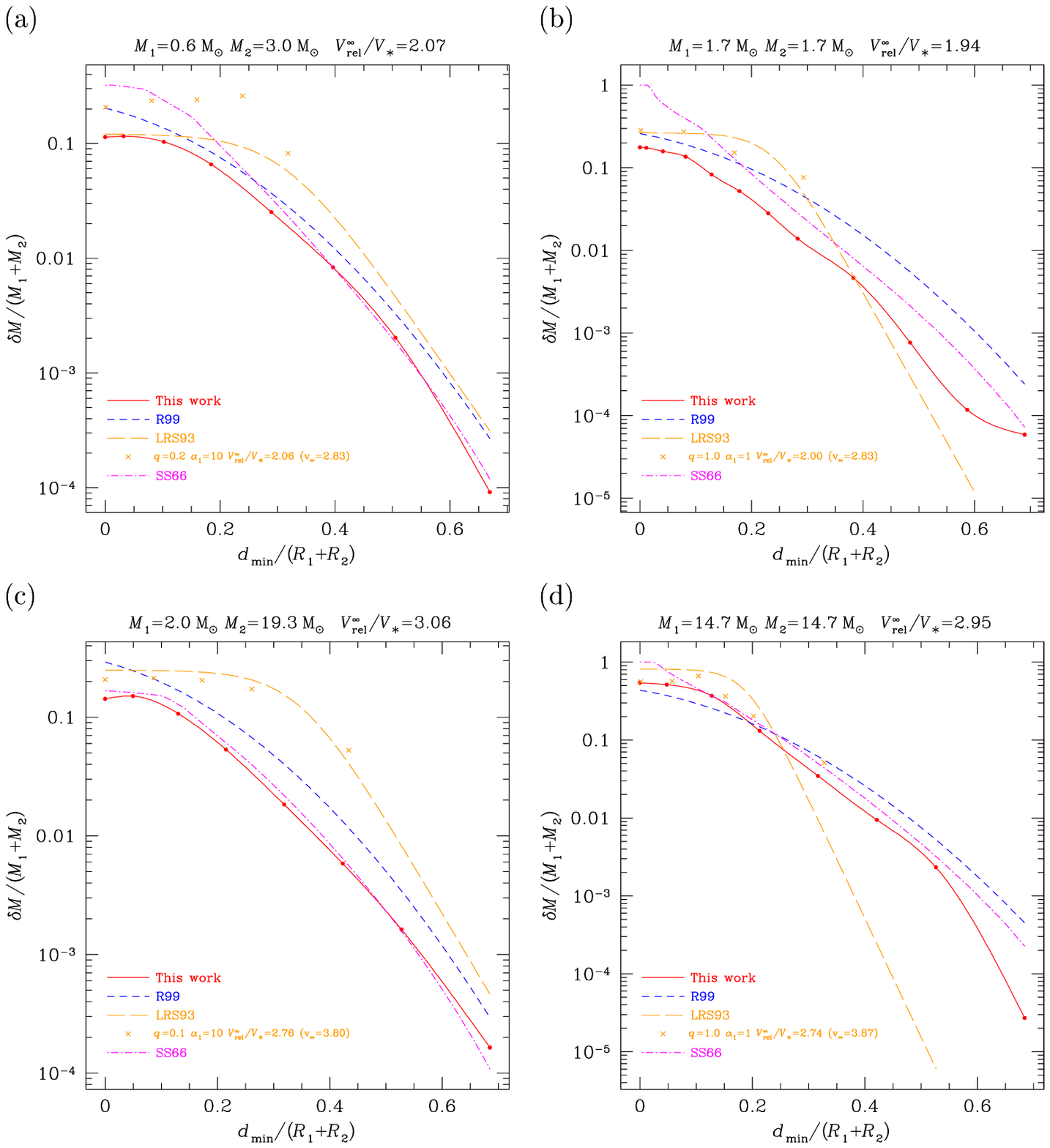}%
      }
  \caption{ Similar to Figures~\protect\ref{fig:comp_lit_a} and
  \ref{fig:comp_lit_b}.  In these diagrams, we make comparisons with
  individual simulation results from LRS93 (crosses, from their
  Fig.~13). In the legends for the LRS93 data, $q$ is the mass ratio,
  $\alpha_1$ ($=\alpha_>$ in our text) the $\alpha$ value (see text)
  of the {\em more massive star} (beware that, contrary to us, LRS93
  use ``1'' to designate the the massive star), $v_\infty=V_{\rm
  rel}^\infty(GM_>/R_>)^{-1/2}$ where $M_>$, $R_>$ are the mass and
  radius of the massive star. Note that, when applying LRS93's
  formulae for the mass loss (long dashes), we determine $\beta_>$
  (equivalent to $\alpha_>$, see text) for our stellar models, through
  the relation $\beta\simeq 2E/W$ where $E$ is the total energy of the
  star (thermal plus gravitational) and $W$ its gravitational
  energy. In general, this corresponds to a value of $\alpha$
  different from the one used in the LRS93 simulations. This explains
  the possible mismatch between the long-dashed line and the crosses.}
  \label{fig:comp_lit_c}
\end{figure*}

In this section, we perform critical comparisons between our results
and data and methods previously published (see
Sec.~\ref{subsec:prev_works}).

The first attempt at quantitatively predicting the outcome of off-axis
stellar collisions was presented by \citetalias{SS66}. As it is both
elegant and simple (but also very approximate), we implemented it for
comparison purposes. This allowed us to apply it to the same stellar
models that we used in SPH computations. With no particular
optimisation or numerical tricks, this algorithm computes the results
of 50 stellar collisions in less than 3~seconds on a standard
workstation! In comparison, a typical SPH run takes about one day of
CPU time.  In our version of this method, which is nearly identical to
that of \citet{MCD91}, we consider that the stars encounter on
straight line trajectories with an impact parameter (distance between
parallel trajectories) set to $d_{\mathrm{min}}$ (Eq.~\ref{eq:dmin})
and a relative velocity equal to $V_{\mathrm{rel}} = V_{\mathrm{max}}
= V_{\mathrm{rel}}^{\infty}(b/d_{\mathrm{min}})$ (see
Sec.~\ref{subsec:def} for the definitions of these quantities). We
then proceed by dividing both colliding stars into small sticks of
square cross-section that are parallel to the trajectory. The result
of the collision, in terms of mass and energy loss, is computed by
considering completely inelastic (i.e. ``sticking'') collisions
between one mass stick from each star in the overlapping
cross-section.  Stick $i$ of star 1 collides with stick $j$ of star 2
if they have coincident position in the plane perpendicular to the
rectilinear trajectories. No energy or momentum exchange is taken into
account between stick $i$ and other mass elements from its ``parent''
or the other star. We further assume that all kinetic energy to be
dissipated to merge $i$ and $j$ is converted into thermal energy to be
shared between these two elements and that there is no heat exchange
between them so that the thermal energy is given to $i$ and $j$ in
proportion to their pre-collision kinetic energies in the collision
centre-of-mass frame. Finally, the condition for mass element $i$ to
be liberated is that its acquired specific energy is larger than the
initial escape velocity of its parent star, $V_{\ast}^{(1)}$. As
demonstrated by \citet{MCD91}, this results in the following simple
escape condition for element $i$ of star 1:
\begin{equation}
  \frac{\Delta m_j}{\Delta m_i + \Delta m_j} >
  \frac{V_{\ast}^{(1)}}{V_{\mathrm{rel}}}
\end{equation} 
where $\Delta m_{i,j}$ are the masses of sticks i and j. For a given
collision, we iterate this procedure a few times with increasing
resolution (decreasing the cross-section of the sticks) until the
result converges to some prescribed precision level. As can be judged
from this description, the number and importance of simplifications in
this approach is impressive. It is thus difficult to figure out the
regime(s) in which we expect them to hold true. The assumptions on
rectilinear motion, the use of $V_{\mathrm{max}}$, and the escape
criterion leave little hope that sensible results can be obtained
either for low velocity encounters, or for nearly head-on collisions
or for cases where high fractional mass-loss is expected (high
$V_{\mathrm{rel}}$ but small $d_{\mathrm{min}}$). In an attempt to get
better prediction at low impact parameters, we implemented the
following trick, inspired by \citet{Sanders70b}. For each star, a
``core radius'' is defined; it it the radius enclosing $1/4$ of the
total mass. An ``effective'' transverse distance is used instead of
$d_{\rm min}$, $d_{\rm eff}=\min(d_{\rm min},R_{\rm core,1}+R_{\rm
  core,2})$. $d_{\rm eff}$ is used to determine the overlapping
sections of the stars and to set the effective relative velocity
during the collision, through $V_{\rm rel}^2 =
(V_{\mathrm{rel}}^{\infty})^2 + 2G(M_1+M_2)/d_{\rm eff}$. This recipe
is admittedly quite arbitrary and, if SS66-like treatment of collision
is to be used in stellar dynamical simulations, one should experiment
with other similar prescriptions to find the most satisfying one.

\Removed{One important limitation of the SS66 prescription is that it
  does not naturally tell us whether the stars merge or not because
  the matter elements that are not expelled from the stars are not
  supposed to lose any kinetic energy at all.}

All the other literature results included in our comparison were
obtained through SPH simulations. The pioneers in this field were
\citet{BH87,BH92}. They did not try to describe their results with
fitting formulae, so we can only compare their simulations to cases
with very similar initial conditions. \citet{LRS93} performed a more
extended numerical survey from which they devised a general empirical
mathematical description to represent the fractional mass loss as well
as the critical $d_{\mathrm{min}}$ for merger/binary formation.
Although it is already clear from the figures of their paper that this
all-encompassing fit does not provide a very precise adjustment of
their mass-loss results, we use it anyway for our comparison. This
permits an assessment of the utility of such formulae as an
interpolation tool. To the best of our knowledge these formulae have
never been adopted to incorporate the effect of collisions in stellar
dynamics simulations. For his study of the collisional evolution of a
star cluster bound to a supermassive black hole, \citet{Rauch99}
derived another set of fitting formulae from a set of collision
simulations performed by Melvyn Davies. Individual results from these
simulations are not published but it's worth mentioning that
Rauch-Davies' formulae give not only the mass loss but also the energy
loss and the (non-Keplerian) angle of deviation for the trajectories.

These comparisons are motivated by two complementary goals:
\begin{enumerate}
\item To test our results. Although the SPH code has be throughly
  tested in the past, we had to develop new tools for the present
  work. For instance, we developed the program to compute initial
  conditions and stellar structures and the one that carries out the
  analysis of the stellar outcome at the end of the simulation. To
  perform this check, we have to choose, in our runs and in the
  literature, cases that have initial conditions and stellar
  structures agreeing as closely as possible with each other.
\item To assess whether already published results, which covered only
  a limited region in the parameter space, still yield meaningful
  results when extended beyond this zone. We thus dare to compare some
  of our results with data obtained using quite different stellar
  models or with prediction of formulae that we apply outside the
  parameter domain for which they were established. Such
  confrontations should certainly not be seen as a way to cast doubt
  on those published results but as an {\em a posteriori} motivation
  for our own work.
\end{enumerate}
 
All our comparisons focus on the fractional mass loss. This quantity
is presumably the most important for inclusion of the effect of
star-star collisions in stellar dynamics models and it is given in all
previous works. In Sec.~\ref{subsec:use_results}, we explain that, in a
general case, the description of the outcome of a collision requires
at least 4 quantities.

In Fig.~\ref{fig:comp_lit_a}, we show some selected cases for which we
expect a good agreement with the literature results. There are
however some exceptions that we explain in the caption of this figure.
In Fig~\ref{fig:comp_lit_b}, more extreme comparisons are made. With
Fig.~\ref{fig:comp_lit_c}, we concentrate on comparisons with
simulation results of LRS93.

After inspection of these plots and many others not shown here, the
following comments can be made:
\begin{enumerate}
\item When comparing some of our simulations to other individual
  computations with very similar initial parameters, a comforting, if
  not surprising, agreement emerges. This is particularly true of
  results from BH87 and BH92\Added{\footnote{BH87 made use of an
      earlier, much simpler version of our SPH code. The smoothing
      length had a unique, non-evolving value, an exponential kernel
      was used and the gravity was computed by direct summation. The
      code used by BH92 included essentially the same features as ours
      but all particles had the same mass (as in BH87).}}. The
  situation with LRS93 (Fig.~\ref{fig:comp_lit_c}) is more complicated
  and we discuss it in detail below.
\item The initial stellar structure plays a central role in
  determining the results. But this strong dependency may probably be
  compensated to a large amount by some ``clever'' parameterisation of
  the initial conditions (see below).
  \label{item:init_struct}
\item Fitting formulae can not be used as extrapolation tools. This
  means not only that we cannot trust them when applied to larger or
  smaller velocity, masses or impact parameter values than the ones
  the have been forged for, but also that they will fail at predicting
  outcomes for other stellar models.
  \label{item:fit_for}
\item Predictions from LRS93 and R99 formulae are generally quite
  different, even when applied to the parameter domain in which they
  should both be relevant. This may be due to variations in the
  stellar structure (the $M$--$R$ relation) and/or amplified from
  small differences at the SPH level by the fitting procedures
  themselves. This is another indications that such formulae should be
  use with extreme caution.
\item An unexpected result from these confrontations is that the best
  match at $d_{\mathrm{min}}/(R_1+R_2)>0.15$ and
  $V_{\mathrm{rel}}^{\infty}\ge 1$ is obtained with the SS66 method,
  which incorporates nearly no real physics! Furthermore, some of the
  crudest assumptions it relies on, which are certainly to be blamed
  for its breakdown at low impact parameter may probably be improved
  on. An exploration of the real potentialities of this simple
  approach would be an interesting follow-up of the present study,
  mainly because it reduces stellar collisions to very simple
  considerations about momentum and energy conservation and could thus
  be a useful guide toward a deeper insight into these processes.
  Once again, this unexpectedly good agreement strongly hints toward
  the central importance of the stellar structure in collision
  simulations.  We should add that the SS66 approach also apparently
  breaks down for very high velocities $V_{\mathrm{rel}}^{\infty} >
  10$ where it yields too small a mass loss as compared to our
  simulations. It is interesting to note that the parameter domain for
  which SS66 gives very good results is well suited for collisions
  occurring in dense galactic nuclei. It may thus be that this recipe,
  when complemented with some prescription describing the domain of
  complete disruption, can be made into a useful ingredient for the
  study of such systems.
\end{enumerate}

Let's now focus on to comparison with results of LRS93, illustrated by
Fig.~\ref{fig:comp_lit_c}. \Added{This work is of special importance
  as it constitutes the only survey of some breadth, also including
  high-velocity encounters, published so far.} These authors used
Eddington models with $n=3$ polytropic density profiles and assumed
$R_\ast \propto M_\ast^{0.8}$.  Eddington models have a constant
$\beta=P_{\rm gas}/P_{\rm tot}$; they can be parametrised by $\alpha =
7.89(1-\beta)^{1/2}\beta^{-2}$. LRS93 further assume
$q=M_</M_>=\alpha_</\alpha_>\le 1$ where subscripts $<$ and $>$ indicate
the more and less massive star in the encounter, respectively. LRS93
have parametrised their results through a set of formulae that give
the fractional mass loss as a function of $q$, $\alpha_>$,
$v_\infty:=V_{\rm rel}^\infty(GM_>/R_>)^{-1/2}$ and $d_{\rm
  min}/(R_<+R_>)$. When comparing our results to this parametrisation,
we set $\beta_>:=2E_>/W_>$ where $E_>$ is the total energy of the
massive star (thermal plus gravitational) and $W_>$ the gravitational
contribution. This relation is exact for Eddington models and is used
here to define some ``effective'' $\beta$ parameter. $\beta$ is very
close to 1 for $M_\ast<10\,\Msun$, leading to small $\alpha$ values.
Hence, most LRS93 results (with $\alpha_>=10$, panels a and c of
Fig.~\ref{fig:comp_lit_c}) are adapted to $M_>\gg 10\,\Msun$. This
probably explains why LRS93 get considerably more mass loss than us;
their stellar model have little binding energy compared to our
realistic MS stars. Indeed, the best agreement is reached with the few
$\alpha_>=1$ models, see Fig.~\ref{fig:comp_lit_c}b,d. Also, we stress
again that $n=3$ polytropes do not represent in a satisfactory way the
mass distribution of {\em any} (evolved) MS star except, maybe, around
$M_\ast=0.9$ (see appendix).

\begin{figure*}
      \resizebox{\hsize}{!}{
	\includegraphics{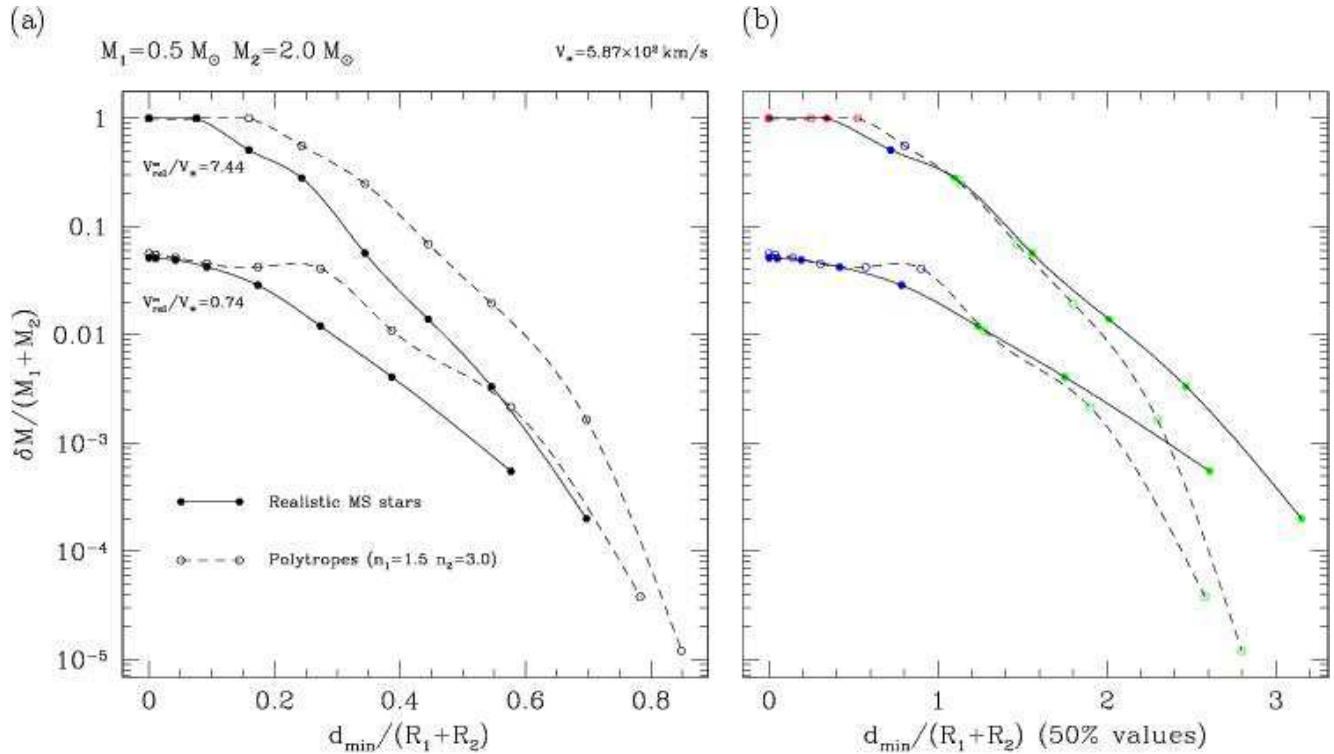} }
      \caption{ Fractional mass loss as obtained in simulations with
      polytropes (dashed lines) and realistic stellar models (solid
      lines). For the simulations with polytropes, we used models of
      indices $n=1.5$ and 3 for the small and the large star,
      respectively. The same M-R relation was used for both sets of
      simulations. In panel (a), we normalised the Keplerian closest
      approach distance by the sum of the total stellar radii, We note
      that, except for head-on collisions which result in the same
      mass-loss, polytropes lead to a systematic over-estimate of
      $\delta M$. This is probably due of less concentrated density
      structure of the $n=3$ polytrope as compared to a ``real''
      $2\,{\Msun}$ star. In panel (b), we use the half-mass radii as a
      normalisation. In this representation, the agreement is much
      better up to $\sim
      2(R_1^{\mathrm{(h)}}+R_2^{\mathrm{(h)}})$. The small bump on the
      low velocity curve for polytropes at $\sim
      0.9(R_1^{\mathrm{(h)}}+R_2^{\mathrm{(h)}})$ is probably the sign
      that this collision is a two-stage merger, i.e., that a
      short-lived binary is formed at first periastron passage that
      merge into a single object at second passage. The symbol type
      indicates the outcome of the collision: triangle for a complete
      disruption, open square for a binary, filled square for a fly-by
      and round dot when only one star remains (merger or disruption
      of the smaller star).}  \label{fig:comp_dm_polytr}
\end{figure*}

We now turn to an examination of the impact of the stellar models on
collision results. In Fig.~\ref{fig:comp_dm_polytr}, we compare two
sets of simulations.  In both series, we computed collisions between
stars of masses $0.5$ and $2.0\,{\Msun}$ for two different relative
velocities and a sequence of impact parameters. In the first set, we
used realistic stellar models, while in the second series, the small
star is represented as a $n=1.5$ polytrope (which is a very good
approximation) and the large one as a $n=3$ polytrope (a poorer
model). Panel~(a) in this diagram strongly confirms
point~\ref{item:init_struct} of the previous enumeration.  Except for
head-on encounters, the polytropic models systematically overestimate
the mass-loss by factors as large as 5! This seems to strongly justify
our use of realistic stellar structure instead of the traditional
polytropes but panel~(b) slightly weakens this statement. There we use
the half-mass radii to normalise $d_{\mathrm{min}}$. This simple
change of parameter, scales out the discrepancy to a large amount.
Only for large $d_{\mathrm{min}}$ is the mismatch still strong
(actually stronger!)\footnote{We use the same $M$--$R$ relation in
  both sets of simulations.
  $(R_1+R_2)/(R_1^{\mathrm{(h)}}+R_1^{\mathrm{(h)}})$ is equal to 4.52
  for the realistic stars and to 3.30 for the polytropes. Normalising
  by the half-mass radii amounts to a relative contraction of the
  polytropic models by a factor $3.30/4.52=0.73$.}. This fact suggests
that it could be possible to scale out much of the dependency on the
stellar structure by use of some subtle parameterisation of the
``closeness'' of the collision that is a better representation than
$d_{\mathrm{min}}/(R_1+R_2)$ of the amount of stellar matter which is
highly affected by the collision. In cases with stars of very
different sizes, a good variable could be the mass fraction of the
larger star inside $d_{\mathrm{min}}$ or some more realistic closest
approach distance that includes corrections for non-Keplerian effects
at small distances. In the same spirit, rather than using
$V_{\mathrm{rel}}^{\infty}/V_{\ast}$ (or the half-mass version of this
quantity), we could look for a parameterisation of the encounter's
severity that reflects the energy input compared to the total binding
energy of the stars, for instance. In other words, our only hope to
find a general description of our results that is both relatively
simple and robust enough to allow some amount of extrapolation, is to
trade apparent complexity in the results for physically motivated
complexity in the parameters! At any rate clever parameterisations can
possibly bring together the results of collisions for different
stellar structures only as long as general quantities such as the mass
and energy losses are concerned. Because the entropy and chemical
profile of an evolved MS star is very different from a homogeneous
polytrope, the structure and evolution of the collision products
strongly depend on the use of realistic initial models, as
demonstrated by \citet{SL97}.


Such remarks, as well as our comments on the strong limitations to the
use of published fitting formulae (point~\ref{item:fit_for}, above)
convinced us that any successful mathematical description of the
collisions' outcome should stem from physical considerations if it has
to be used not only as a handy summary of the computed collisions but
also to extrapolate to somewhat different initial conditions.
Unfortunately, due to the complexity of the physical processes at play
during collisions, such a ``unifying'' description seems very
difficult to find and has escaped us so far. This pushed us to cover
as completely as possible the relevant domain of initial conditions
and motivated the use of an interpolation algorithm to determine the
outcome of any given collision with parameters inside this domain.

\subsection{Using the collision results in stellar dynamics
  simulations}

\label{subsec:use_results}

\subsubsection{The struggle for fitting formulae}

\begin{figure*}
      \resizebox{\hsize}{!}{%
      \includegraphics{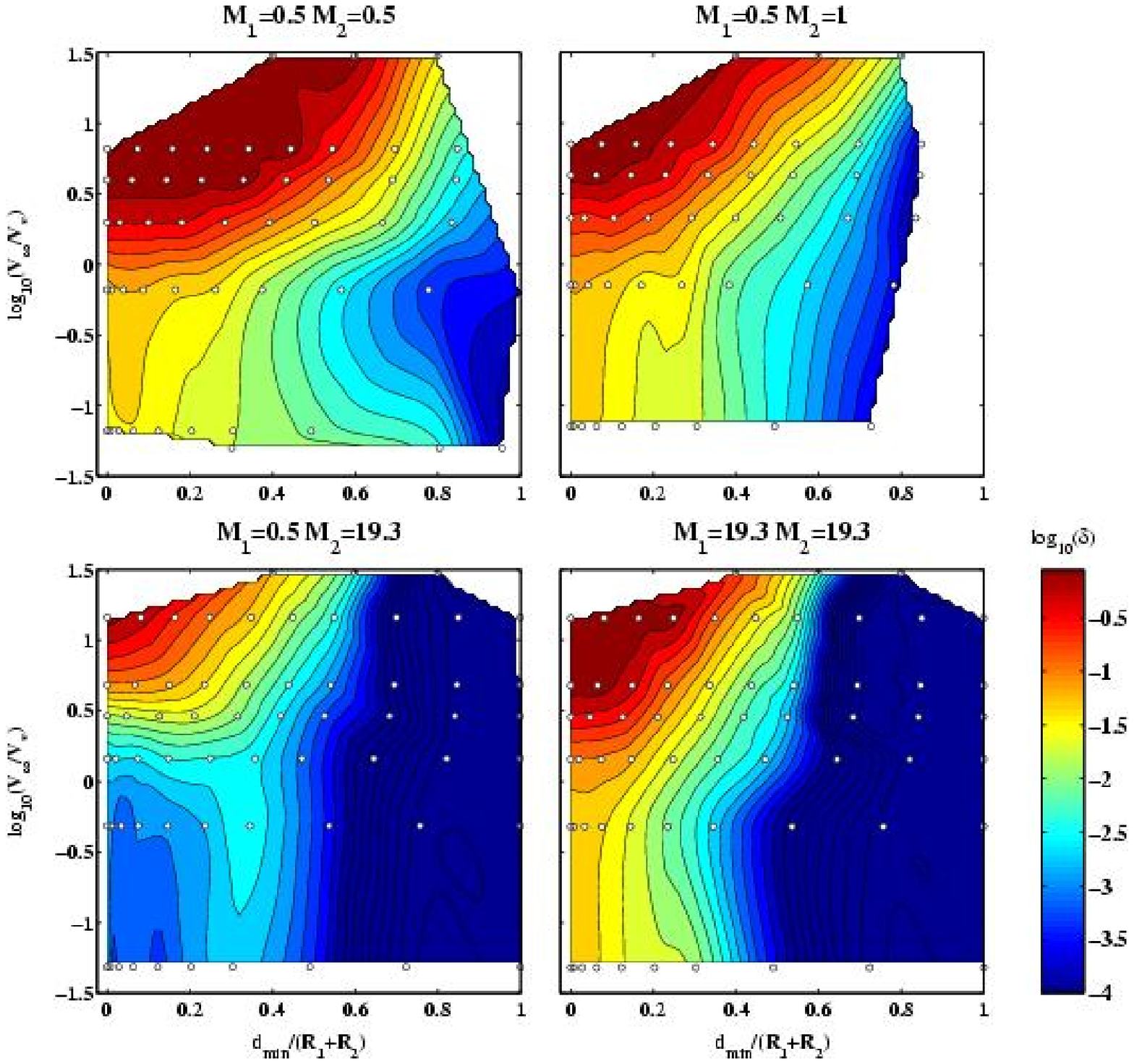}%
      }
  \caption{
    Collisional fractional mass losses for four different $(M_1,M_2)$
    pairs (values in ${\Msun}$). Simulation Data. White dots show
    the SPH simulations. The contours and colour maps are a bi-cubic
    interpolation of the SPH results. $\delta$ is the fractional mass
    loss $\delta=\delta M/(M_1+M2)$. Masses are in units of ${\Msun}$.
    In each frame, the upper left contour indicates fractional mass
    loss larger than 85\,\%.
    }
  \label{fig:delta_data}
\end{figure*}

\begin{figure}
      \resizebox{\hsize}{!}{%
      \includegraphics{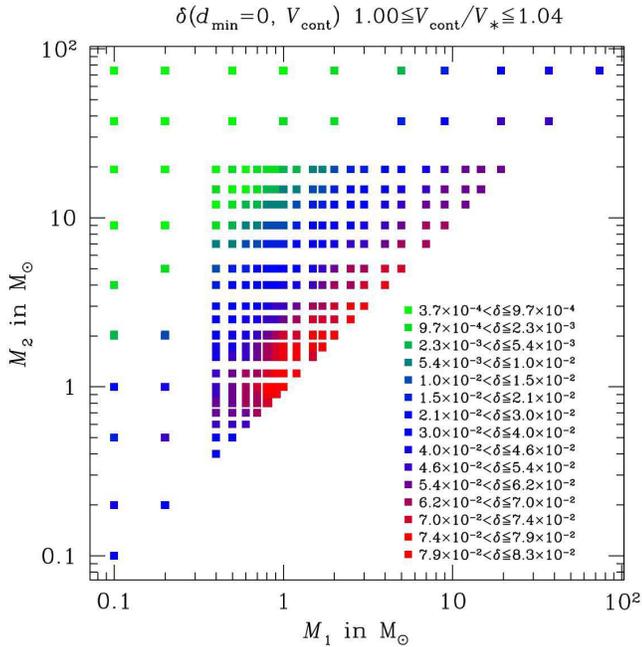}%
      }
    \caption{
      Fractional mass loss ($\delta$) for all head-on,
      ``zero-velocity'' collisions. Here, $V_{\mathrm{cont}}$ is the
      contact velocity (at separation $R_1+R_2$). $\delta$ is not
      constant on lines of constant $M_1/M_2$. }
    \label{fig:delta00}
\end{figure}

\begin{figure}
      \resizebox{\hsize}{!}{%
      \includegraphics{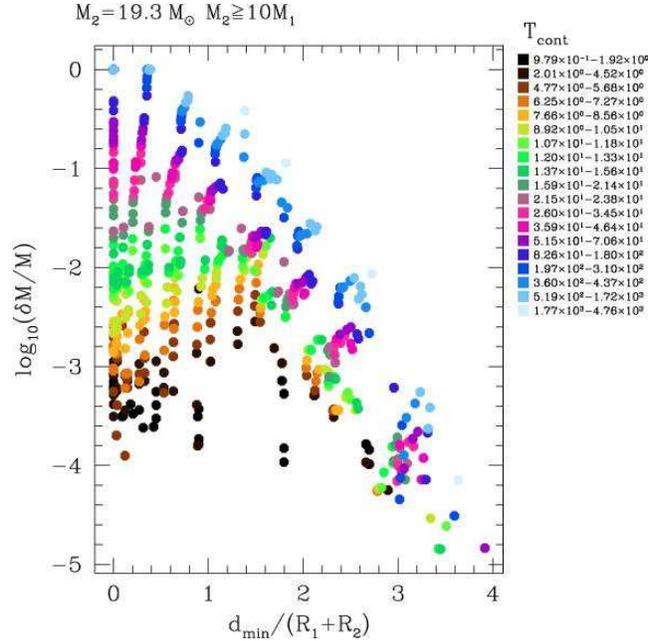}%
      }
    \caption{
      Fractional mass loss for all collisions between a star of mass
      $19.3\,{\Msun}$ and stars that are at least 10 times less
      massive. Here we normalise the distance by $R_2$, the total
      radius of the large star. Points are colour-coded according to
      the kinetic energy of the impactor at contact (in solar units, 
      $G\Msun^2/\Rsun$).
      }
    \label{fig:delta_Tcont}
\end{figure}

\begin{figure*}
  \resizebox{\hsize}{!}{\includegraphics{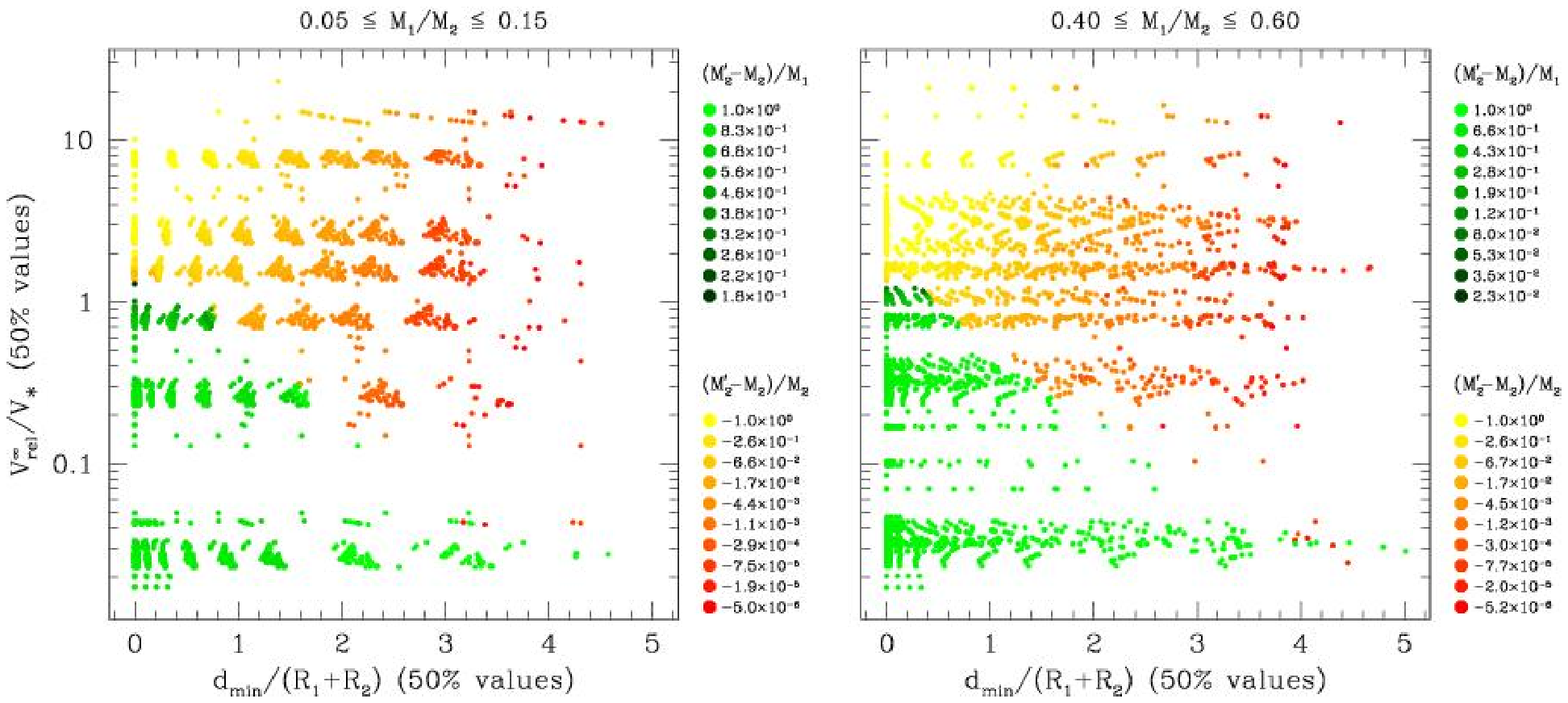}}
  \caption{%
    Plots of the relative modification of the mass $M_2$ of the larger
    colliding star as a function of $d_{\mathrm{min}}$ and
    $V_{\mathrm{rel}}^{\infty}$. Mass decreases, colour-coded in red
    to yellow, are normalised as fractions of $M_2$. Mass increases,
    colour-coded in green tones are normalised as fractions of $M_1$.
    \Added{We choose this two different normalisations so these relative mass
    changes are always comprised between 0 and 1 in absolute value.
    $(M^\prime_2-M_2)/M_2$ can me interpreted as the
    ``fractional damage'' caused by the ``bullet'' (small star) to the
    target, while $(M^\prime_2-M_2)/M_1$ is the ``efficiency'' by
    which mass of the small star is added to the more massive one.}}
  \label{fig:delta2}
\end{figure*}

The result of a collision is described through a small set of
quantities: the fractional mass loss
$(M_1+M_2-{M_1}^{\prime}-{M_2}^{\prime})/(M_1+M_2)$, the new mass
ratio, the fractional loss of orbital energy and the angle of
deviation of the relative velocity. Note that these values completely
describe the kinematic outcome of a collision only if the
centre-of-mass reference frame for the resulting star(s) (not
including ejected gas) is the same as before the collision.
Asymmetrical mass ejection violates this simplifying assumption by
giving the resulting star(s) a global kick \citep{BH87}. However, we
checked that the kick velocity is generally much lower than the
relative velocity. Thus, this simplification, which greatly reduces the
complexity of the situation, should not lead to an important bias in
the global influence of collisions in the energy balance of a star
cluster.

We have kept the final SPH particle configuration for (nearly) all our
simulations. This would allow us to re-analyse these files and extract
other quantities of interest, like the amount of rotation imparted by
the collision, a quantity worth investigating because it can deeply
influence the subsequent evolution of the star(s) \citep{MM00,SFLRW00}
and lead to observational signatures that would reveal the importance
of collisions and close encounters in given environments \citep{AK00}.
Another interesting issue is the resulting internal stellar structure.
This is key to a prediction of the subsequent evolution and
observational detectability of collision products \cite[for
instance]{SLBDRS97,SFLRW00}.  Unfortunately, according to
\citet{LSRS99}, low resolution and use of particles of unequal masses
can lead to important spurious particle diffusion in SPH simulations
so that our models are probably not well suited for a study of the
amount of collisional mixing, for instance.

Fig.~\ref{fig:delta_data} shows the (interpolated) fractional mass
loss in the $(d_{\mathrm{min}},\,V_{\mathrm{rel}}^{\infty})$ plane for various
$(M_1,M_2)$ pairs. Note how the ``landscape'' changes from one choice
of $(M_1,M_2)$ to another one. \Added{Such relatively complex
  structure obviously is a challenge to attempts of describing the
  results by fitting formulae.}

Let's report some unsuccessful attempts at finding easy-to-express
regularities in the simulation data. We first convinced ourselves that
the outcome of a collision does depend on both stellar masses and not
only on the mass ratio $q=M_1/M_2$. This is demonstrated in
Fig~\ref{fig:delta00} in which we plot the total fractional mass loss
for head-on mergers with $V_{\mathrm{rel}}^{\infty}\simeq 0$. If this
quantity depended on $M_1$ and $M_2$ only through $q$, we would obtain
constant $\delta$ values along diagonals, which is not the case.
Fig.~\ref{fig:delta_Tcont} depicts another wrong guess, namely that
for stars of very different sizes, the mass loss would only depends on
the kinetic energy of the impactor (and on $d_{\mathrm{min}}$) and not
on its mass.  There is not much interest in explaining in detail all
the strategies we have tried to reduce our huge dataset to a more
manageable mathematical formulation. As a last illustration of the
difficulty of such a programme, let's mention our attempt at a global
parameterisation of the mass-loss curves. We found a 3-parameter formula
that allowed good fits of individual $d_{min}/(R_1+R_2)
\longrightarrow \delta M/(M_1+M_2)$ relations (for fixed $M_1$, $M_2$
and $V_{\mathrm{rel}}^{\infty}$)\footnote{To achieve these fits, we
  removed all points corresponding to the formation of binaries,
  because our parameterised function was monotonically decreasing with
  increasing $d_{\mathrm{min}}$ and could not reproduce extra mass
  loss due to subsequent periastron passages.}. This looked very
promising but we were left with 1180 sets of parameters to be adjusted
in turn by some ``meta-formula'', with the added difficulty that they
displayed a lesser level of regularity than the raw collisional data
itself. This proved unmanageable. Furthermore, this parameterisation 
had no sound physical justification.

To end this subsection on a more positive note, let's turn to
Fig.~\ref{fig:delta2}. In this diagram, we plot the relative mass gain
or loss for the larger star, $\delta_2$, as a function of the usual
half-mass normalised $\lambda$ and $\nu$. The figures are remarkably
smooth in the sense that collisions with comparable mass ratios, but
otherwise different $M_1$ and $M_2$, and same $(\lambda,\nu)$, produce
very similar $\delta_2$. There is thus some hope that, in further
investigations, we could discover some ``universal''
$\delta_2=\delta_2(q,\nu,\lambda)$ relation to describe this
regularity. Such a description would be particularly useful to
explore, with analytical or semi-analytical models, the possibility of
run-away merging sequences in the evolutions of dense clusters. Using
realistic SPH results to re-examine these scenarios is one important
application of the present work (see Sec.~\ref{sec:StellCollInNat}).

\subsubsection{Interpolation of the collision results}

\begin{figure}
      \resizebox{\hsize}{!}{%
      \includegraphics{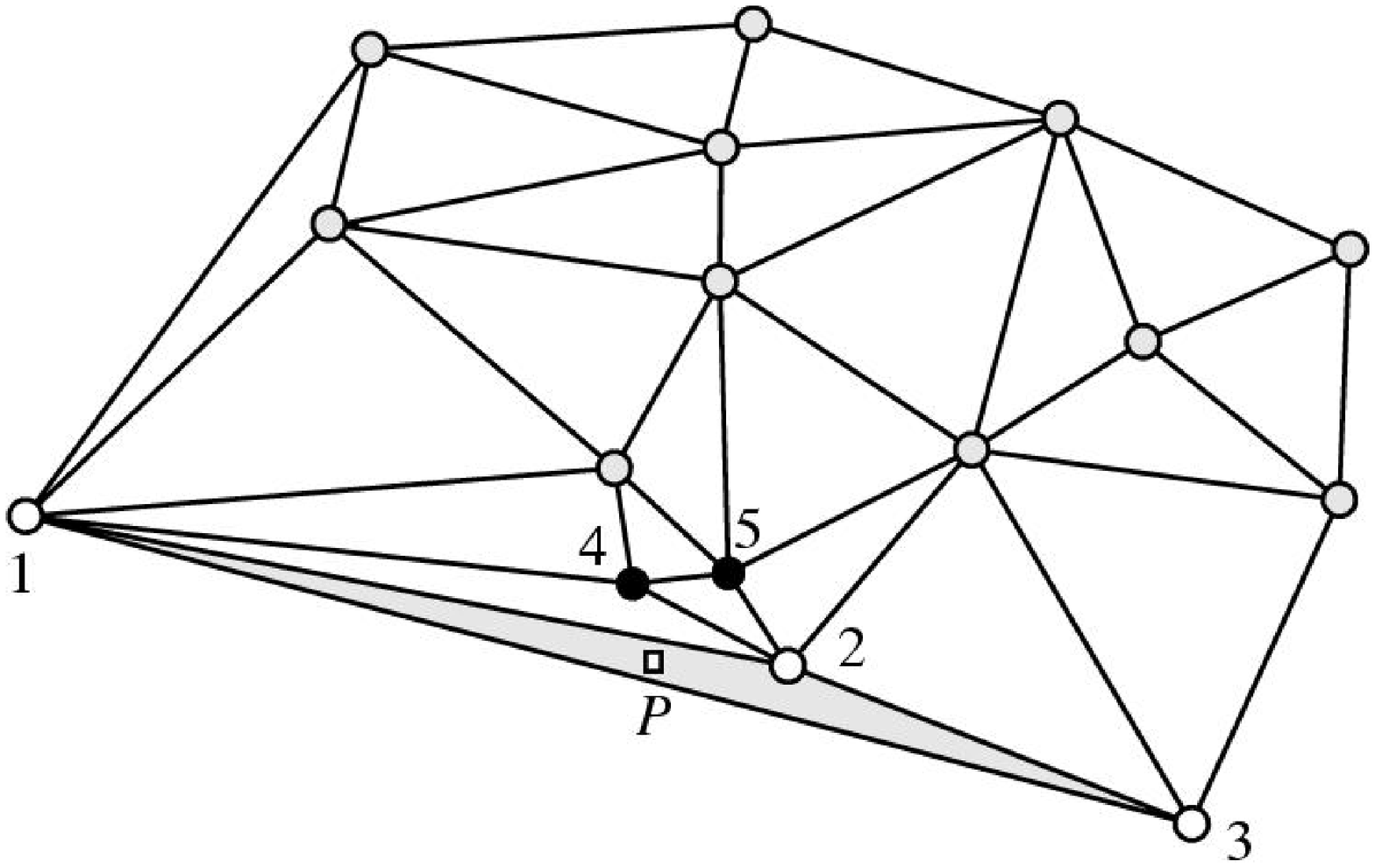}%
      }
    \caption{
      Delaunay triangulation in two dimensions. In our interpolation
      method we would get the value for the point $P$ (small square)
      from data points 1, 2 and 3 that are the vertices of the
      triangle $P$ lays in. Although they are much closer to $P$ than
      1 or 3, points 4 and 5 would not contribute at all!  }
  \label{fig:delaunay}
\end{figure}

\begin{figure*}
      \resizebox{\hsize}{!}{%
      \includegraphics{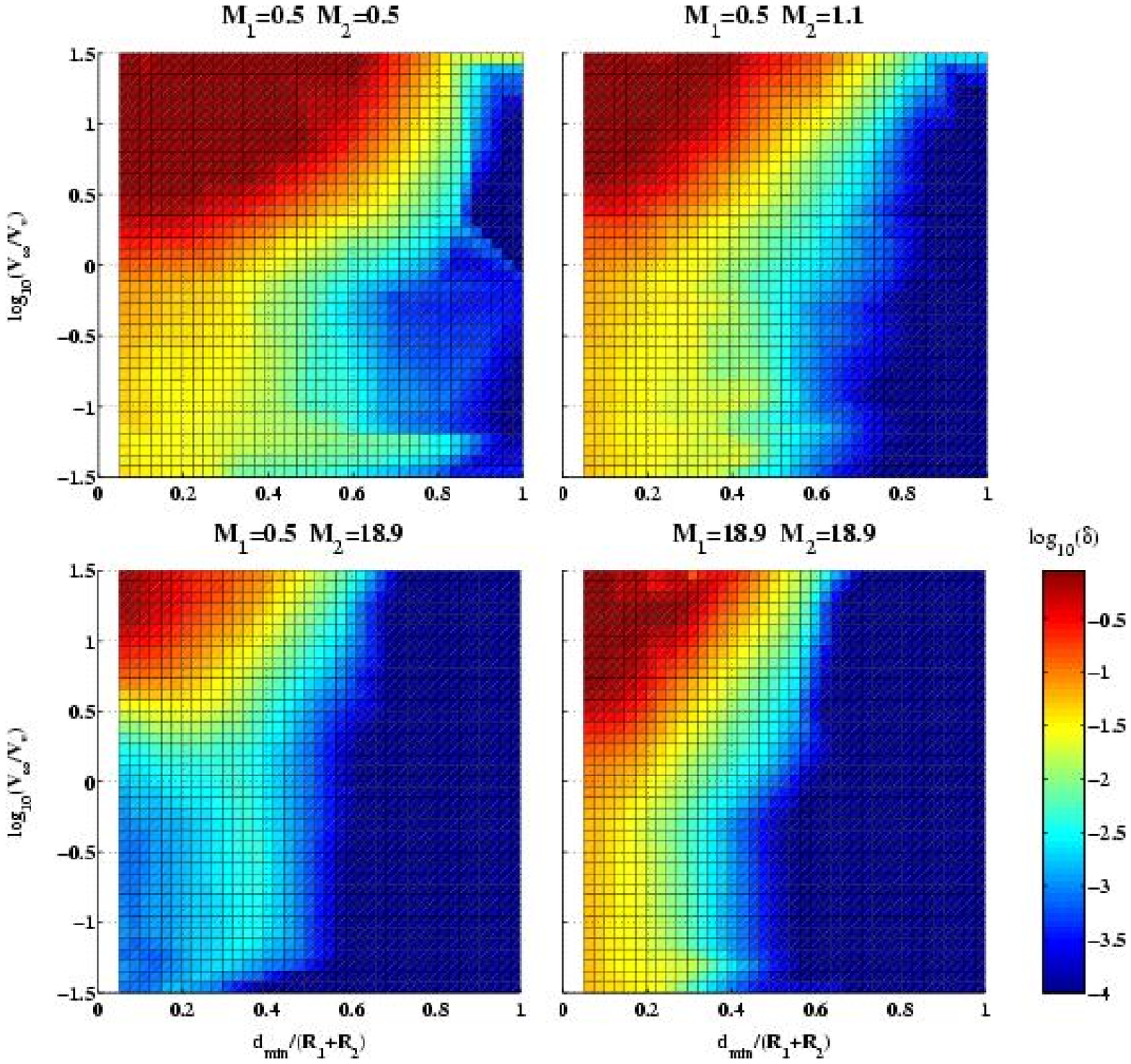}%
      }
    \caption{
      Four slices through our interpolation grid for the collisional
      fractional mass loss.  We performed cuts that correspond to
      $(M_1,M_2)$ values close to those of
      Fig.~\protect\ref{fig:delta_data}.  $\delta$ is the fractional
      mass loss $\delta=\delta M/(M_1+M2)$.  Masses are in units of
      ${\Msun}$. The Delaunay interpolation produces some artifacts
      at low and high relative velocities, in particular for the
      top-left and bottom-right panels (compare with
      Fig.~\protect\ref{fig:delta_data}). This is due to the simplices
      being very elongated near the border of the convex hull of our
      data (in four dimensional space).  }
  \label{fig:delta_grid}
\end{figure*}

Being unable to distillate the results of our SPH simulations into any
compact mathematical formulation without losing most of the
information, we resorted to the following interpolation strategy.  In
the 4D initial parameter space, the simulations form a irregular grid
of points. We compute a Delaunay triangulation of this set
\cite[Chap.~28]{Sedgewick88} using the program
\verb|QHULL|\footnote{Available at \url{http://www.qhull.org/}}
\citep{BDH96} which allows us to interpolate the results onto a
regular 4D grid. To evaluate the value of any of the four quantities
that summarise the outcome of a collision, $\mathfrak{Q}$, we first
find the simplex $\mathcal{S}$ of the triangulation, if any, that
contains the 4D point $P$ of the initial conditions of the collision.
This simplex has 5 vertices: $Q_i, i=1\ldots 5$. By removing one of
these vertices, say $Q_k$ and replacing it by $P$, one forms another,
smaller simplex $\mathcal{S}_k$ that is contained in $\mathcal{S}$. We
compute the interpolated value of $\mathfrak{Q}$ at point $P$,
$\tilde{\mathfrak{Q}}(P)$ from its values at the vertices $Q_i$,
$\mathfrak{Q}(Q_i)$ by linear combination with weights
$V_{\mathcal{S}_k}/V_{\mathcal{S}}$ where $V_{\mathcal{S}}$ stands for
the (hyper-)volume of $\mathcal{S}$. Of course, this procedure does
not allow \emph{extrapolation} outside the convex hull of our
simulation initial conditions. Another, more tricky problem is that,
near the borders of this convex hull, simplices can be very elongated
which means that the interpolations can be done with data points
corresponding to very remote initial conditions instead of using more
local information. This is illustrated, for two dimensions, in
Fig.~\ref{fig:delaunay}.

However, we did not find a better procedure. We tried to use a
kernel-based method, {\it \`a la} SPH, but it produced very poor
results. The main problem with this class of algorithms is to adapt
locally the 4 independent axis of the kernel ellipsoid in such a way
that only neighbouring data points contribute to the evaluation at a
given point.  Defining ``neighbours'' is unfortunately not obvious in
a parameter space with no natural metrics\footnote{This problem about
  the metrics being ill-defined actually also plagues the Delaunay
  method, but in a less visible and apparently less harmful way! To
  get more acquainted with Delaunay triangulation, see the interactive
  demonstration at
  \url{http://wwwpi6.fernuni-hagen.de/GeomLab/VoroGlide/}.}.

\Added{The quality of the data obtained through our interpolation
  mechanism is illustrated on Fig.~\ref{fig:delta_grid}. It shows four
  2D slices of the fractional mass loss interpolated onto the 4D grid.
  Each slice corresponds to a $(M_1,\,M_2)$ pair chosen as to be as
  close as possible to the values used for Fig.~\ref{fig:delta_data},
  allowing a direct comparison. The general dependency of the mass
  loss on impact parameter and relative velocity is well reproduced
  but some details are smoothed out while small artifacts have
  appeared near the borders of the domains for the reason explained
  above. We interpret the horizontal ``peninsula'' of high mass loss
  for $M_1=M_2=0.5\,\Msun$ as the result of the interpolation between
  a simulation which was integrated long enough for complete merger
  (visible on Fig.~\ref{fig:delta_data} for the lowest relative
  velocity value at $d_{\rm min}/(R_1+R_2)=0.8$) and led to relatively
  high mass loss and others which were stopped at an earlier phase
  (unmerged binary).}

The table thus computed is the backbone of the routine that implements
stellar collisions in our Monte Carlo simulations of stellar clusters.
Collision outcome quantities are indeed easily obtained through a
second, much quicker, interpolation stage using this regular grid. Of
course, extrapolation prescriptions have to be specified for events
whose initial conditions fall outside the convex hull of the SPH
simulation points. Most commonly, this happens when a collisionally
produced star with mass outside the $0.1$--$74\,{\Msun}$ range
experience a further collision. In such cases, we try to re-scale both
masses while preserving $M_1/M_2$ to get a ``surrogate collision''
lying in the domain covered by the SPH simulations. If
$V_{\mathrm{rel}}$ is too low or too high, we increase or decrease it
to enter the simulation domain\footnote{All this fiddling does not
violate mass or energy conservation as collision results are coded in
a dimensionless fashion in the interpolation grid and are scaled back
to the real physical masses and velocities before they are applied to
the particles in a stellar dynamics simulation.}. In its present
state, this treatment of ``extreme'' velocities is not completely
satisfying. At very high $V_{\mathrm{rel}}$, our data do not show
convergence toward a unique mass loss curve. Instead, the domain of
complete disruption keeps extending to higher and higher impact
parameters with a progressive steepening of the mass loss curves for
``fly-bys''. At very small velocities, the values of the table can be
trusted only when no binary has formed or if the binary evolution has
been followed up to merging. In case of binary formation, some
constant fractional mass loss could be used in order to reflect our
finding that the process of binary merging, that requires more and
more pericentre passages for larger and larger $d_{\mathrm{min}}$,
eventually leads to an amount of mass loss relatively independent of
this impact parameter. A more precise determination of the maximal
$d_{\mathrm{min}}$ that still leads to binary formation for small
initial velocities would allow us to know where to switch between this
prescription and interpolation in the table.  Finally, cases with too
high $d_{\mathrm{min}}$ are treated as purely Keplerian hyperbolic
encounters with no mass loss, which is a very good
approximation. Recently, in the frame of our work on collisional
run-away formation of very massive stars, we have implemented a few
more small tricks to complement our ``blind'' interpolation routine
and reduce its artifacts \citep{FRB05,FGR05}.

An important aspect of the work reported here is that we make the data
describing the initial conditions and outcome of all our simulations
available on the web, on the site of the ``MODEST'' working group on
stellar collisions at
\url{http://obswww.unige.ch/~freitag/MODEST_WG4/FB_Collision_Data/}.
We provide a description of the outcome of a given collision in terms
of the number and masses of star(s) at the end of the simulation and
their orbital properties. Colleagues are invited to develop their own
methods to use this data and share their experience with others,
including the authors of the present paper. Files containing detailed
information for all SPH particles at the end of a simulation are
available upon request to MF.

\section{Conclusions and future work}
\label{sec:concl}

In this article, we presented a large set of simulations of collisions
between two main sequence stars. More than 14\,000 SPH simulations
have been computed over about four years to complete that database.
Initial conditions span $M_\ast=0.1-75\,\Msun$ for the stellar masses,
impact parameters corresponding approximately to
$d_{\mathrm{min}}/(R_1+R_2) =0-0.9$ and relative velocities at
infinity ranging, more or less, from 0.03 to 30 times the stellar
escape velocity. This represents a effort of unprecedented breadth in
this field.

Our motivation in this work was to incorporate the effects of stellar
collisions into models of dense stellar systems like galactic nuclei
with as much realism as possible. To reach this goal we developed a
module that interpolates between our results to predict the outcome of
any collision with initial conditions inside the (large) domain of
parameter space we explored. Results of our dynamical simulations of
dense clusters including collisions are presented elsewhere
(\citealt{FB01d,FB02b,RFG03,FGR04b,FGR04c,FRB05,FGR05}).

The quest for a handy mathematical description of these results has
been unsuccessful so far. This was a source of disappointment but we
hope that further study of our simulation data will eventually reveal
some way of casting our results in a compact and physically
enlightening formulation. Exploring when and why the excellent match
with the SS66 prescription is found is a possible way to this goal.

Beyond the scientific exploitation of this important database,
--either to develop a better understanding of the physics of
collisions or as an ingredient for collisional stellar dynamics
studies-- we are aware that other kinds of stellar collisions, not
treated here, are also of great astrophysical interest. First, in
galactic nuclei, collisions with red giants are certainly more
frequent than MS-MS encounters. So we should assess the importance of
this process (for stellar evolution and stellar dynamics) and, if
needed, compute a set of simulations similar to the one presented
here. This would complement the work of \citet{BD99}. Collisions with
compact remnants should also be taken into account as they may be of
great importance in producing peculiar objects.

Our work is not well adapted to the physical conditions that are
typical in globular clusters. Indeed, we did not study with particular
care the low velocity, quasi parabolic encounters, by far the most
common in those environments. However, they have already been quite
thoroughly studied by other researchers. The conditions for tidal
binary formation are probably better determined using other methods
and their long term evolution, whose nature and result is still
debated \citep{Mardling95a,Mardling95b}, extends over much too many
hydrodynamical time scales to be tackled by SPH. Still in the domain
of globular clusters, the study of hydrodynamical effects (including
direct collisions) in interactions with or between binary stars is
still in its infancy \citep{GH91,DBH93,DBH94}.

Finally, going back to galactic nuclei in which we are mostly
interested, let's mention that the connected problem of the
destructive close encounter between a central massive black hole and a
star, even though it has been the focus of many papers \cite[and many
others]{CL83,BG83,EK89,LMZD93,Fulbright96,MLB96,ALP00,BogdanovicEtAl04},
has still not been explored systematically. For instance, all studies
so far have assumed simplified stellar models. In this problem,
however the main uncertainties probably lurk in the post-disruption
evolution of the stellar gas, rather than in the ``collision'' process
itself.

\appendix

\section{Smoothed Particle Hydrodynamics stellar models.}

\subsection{Stellar structure used in the simulations}

\begin{figure*}
  \resizebox{0.45\hsize}{!}{%
    \includegraphics{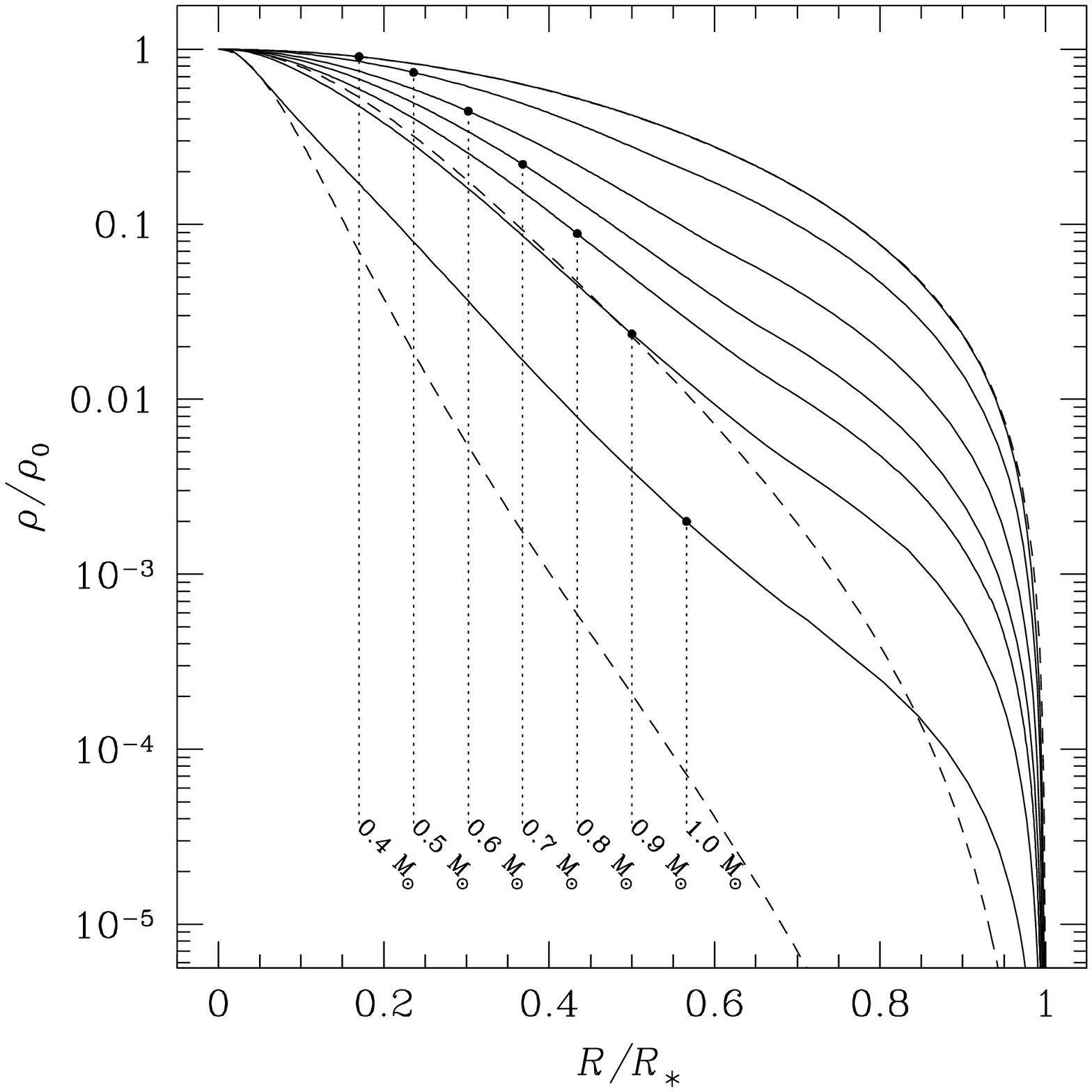}%
    }
  \resizebox{0.45\hsize}{!}{%
    \includegraphics{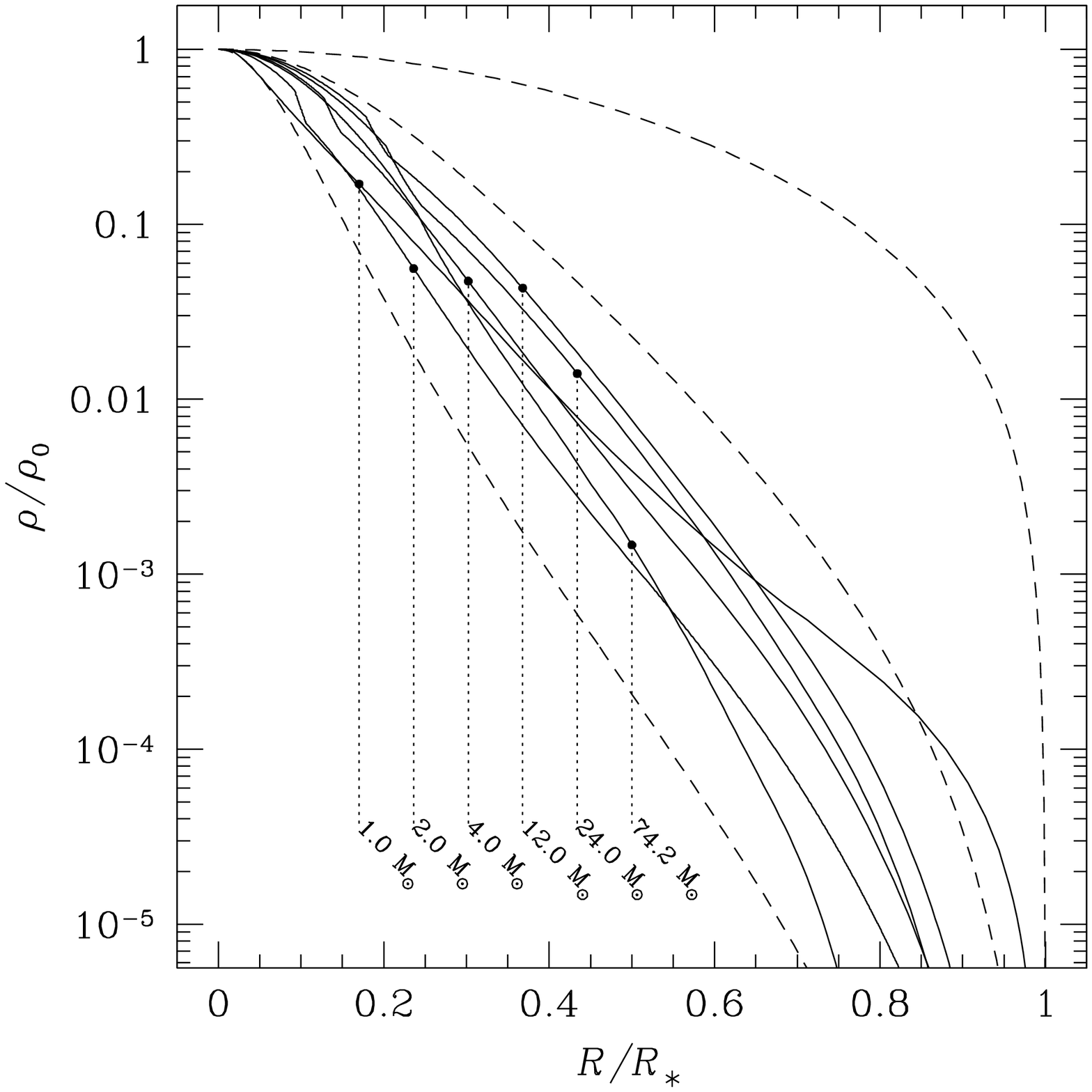}%
    }
  \caption{%
    Density profiles for realistic star models
    \citep{SSMM92,CDSBMMM99} for low (top) and high (bottom)
    mass stars. The dashed lines are polytropic models for $n=1.5$, 3
    and 4, in order of increasing concentration. Below
    $0.4\,M_{\sun}$, the density structure is well represented by a
    polytrope with $n=1.5$ but no good polytropic fit is possible for
    higher masses.
  \label{fig:rho_vs_r}
}
\end{figure*}

\begin{figure*}
  \resizebox{12cm}{!}{%
    \includegraphics{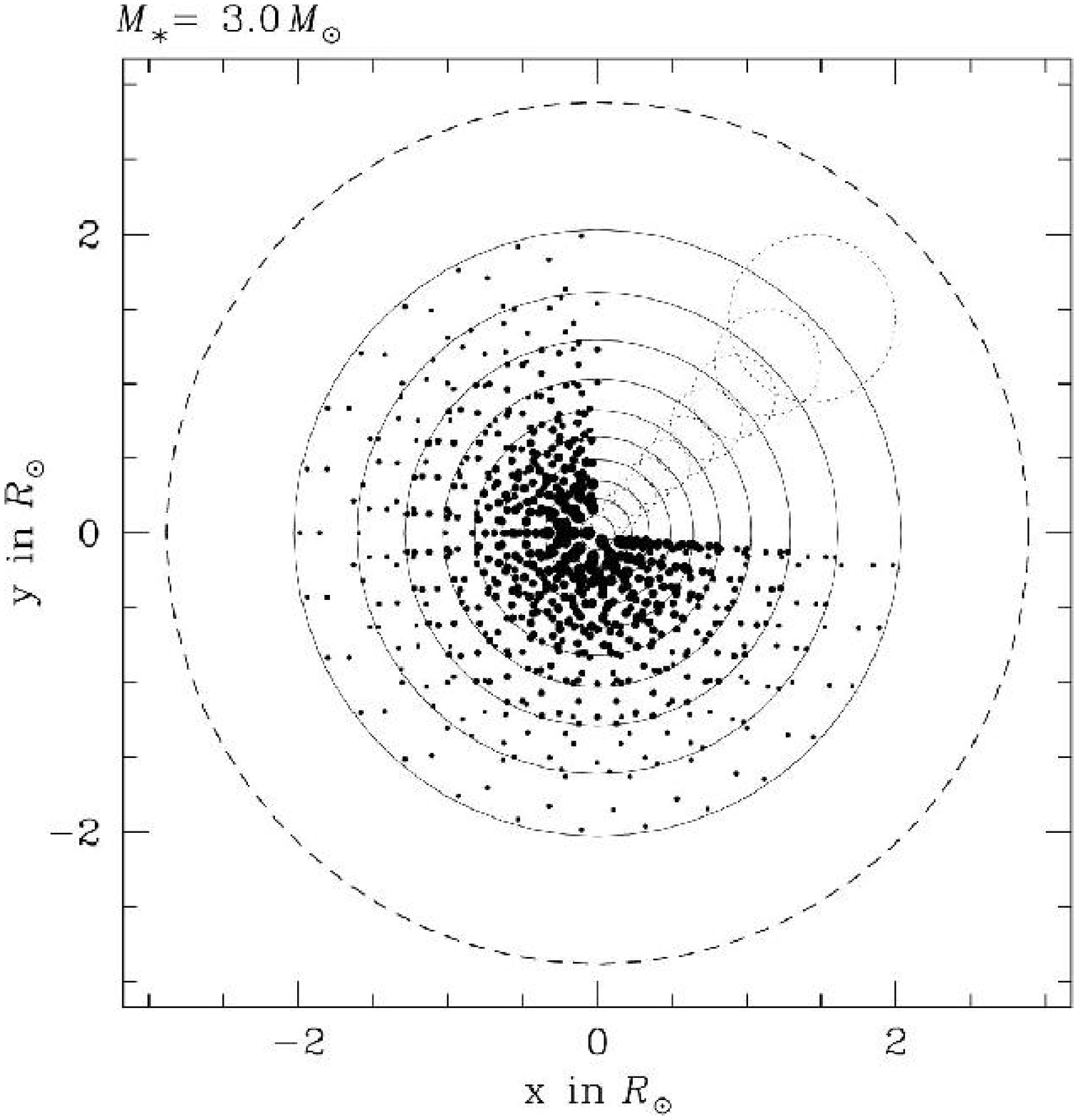}%
    }
  \hfill
  \parbox[b]{55mm}{
  \caption{%
    SPH realisation of a $3\,M_{\sun}$ stellar model with $\sim$2000
    particles. Round dots show the positions of SPH particles with a
    symbol surface proportional to the particle's mass. The big dashed
    circle shows the size of the star according to the structure
    model. Plain line circles depict the concentric spheres on the
    surface of which the particles' centres are placed. Particles on
    the $x,y>0$ corner of the diagram have been removed to show the
    actual half-size of particles on each sphere (dotted circles of radius
    $h$).
  \label{fig:SPH_star}
}
}
\end{figure*}

\begin{figure*}
  \resizebox{\hsize}{!}{%
    \includegraphics{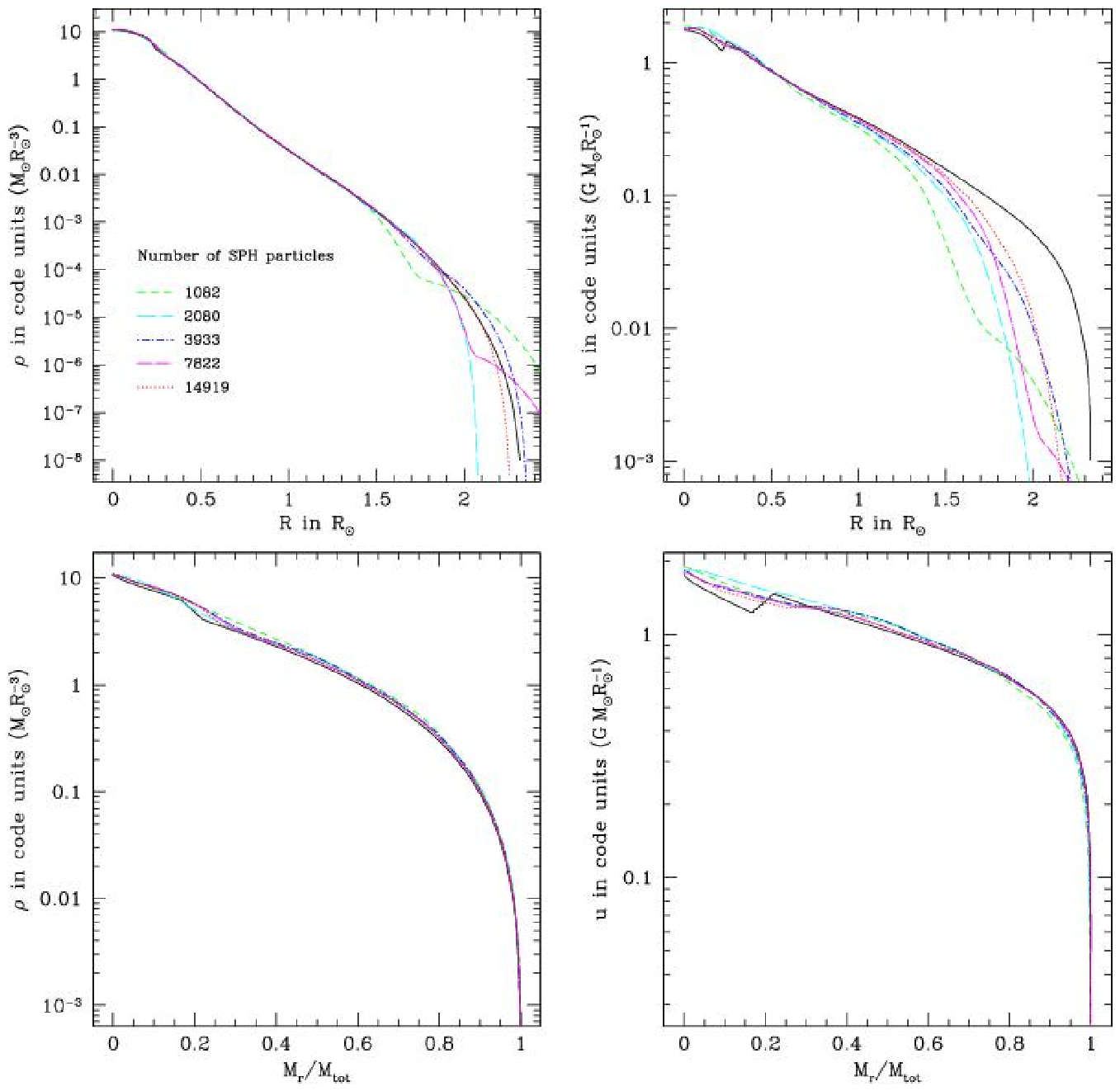}%
    }
    \caption{ Comparison between the theoretical internal structure of a
      $2\,M_{\sun}$ stellar model (solid lines) and SPH realisations
      of it with increasing number of particles (dashed and dotted
      lines). We show the mass density (left column) and the internal
      energy (right column). In the top row, we use the radius as the
      abscissa, while we use the enclosed mass for the diagrams of the
      bottom row. The top plots show that the outermost part of the
      star are poorly represented. However, it is clear from the
      bottom plots that only a tiny mass fraction suffers from
      this mismatch. The SPH profiles are the kernel-interpolated
      values along the line $z=0$.
    \label{fig:comp_si_sph}
  }
\end{figure*}

In Fig.~\ref{fig:rho_vs_r}, we show the density profiles for some
realistic models for MS stars and compare them with polytropic stars.
It is readily seen that for $M>0.4\,M_{\sun}$, stellar structures are
highly non-homologous and that polytropes do not match, even if
allowance is made for a $M$-variable $n$ index. If such a fit is
required anyway, a value of $n\simeq 3.5$ seems more appropriate for
$M\geq 1\,M_{\sun}$ than the commonly used $n=3$.

An SPH particle configuration for a star is illustrated on
Fig.~\ref{fig:SPH_star}. One sees that the outermost layers of the
star are very poorly modelled, with a clear failure at precisely
reproducing the real stellar radius. Fig.~\ref{fig:comp_si_sph} is a
comparison between the density and internal energy profiles of two
stellar models and their SPH approximation for increasing number of
particles.

\subsection{Choice of the particle number}

\begin{figure*}
  \resizebox{\hsize}{!}{%
    \includegraphics{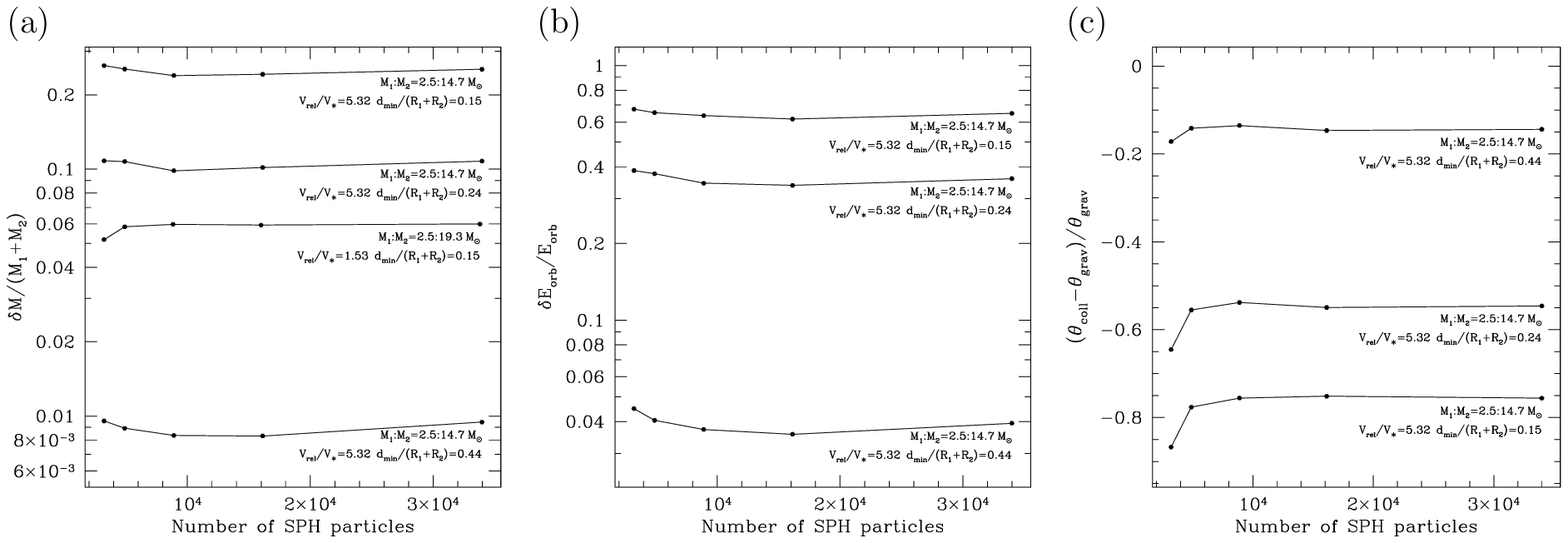}%
    }
  \caption{%
    Study of the dependency of collision results on the particle number
    for 4 sets of collision simulations. (a) Fractional mass loss.
    (b) Fractional loss in orbital energy. The simulation set with
    $M_2=19.3\,M_{\sun}$ is not reported here because it results in a
    merger ($\delta E_\mathrm{orb}/E_\mathrm{orb}=1$). (c) Deviation
    of the collisional deflection angle $\theta_{\rm coll}$ from the value for pure Keplerian
    point-mass trajectory $\theta_{\rm grav}$.}
  \label{fig:Npart_study}
\end{figure*}

In Fig.~\ref{fig:Npart_study}, we show how the overall results of SPH
collision simulations (mass and energy loss, deflection angle) depend
on the resolution, i.e. the number of particles used to represent the
stars. We considered resolution ranging from 1000$+$2000 to
2000$+$32\,000 particles.

\section{Results of SPH simulations}

\subsection{A few specific collision simulations}

\begin{figure*}
  \begin{center}
    \resizebox{!}{20.5cm}{%
      \includegraphics{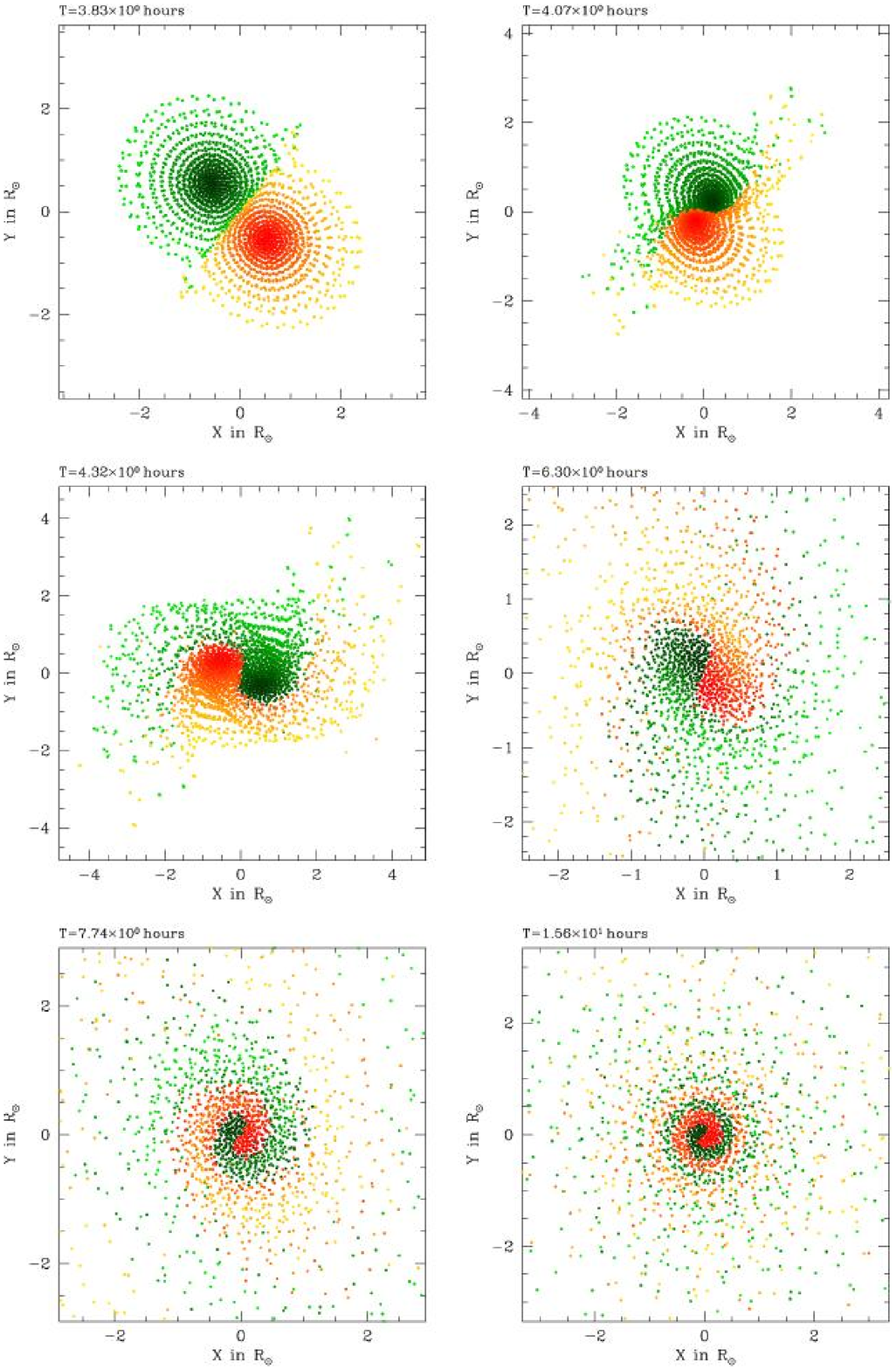}%
      }
  \end{center}
  \caption{%
    Collision between two $2.0\,M_{\sun}$ stars  at
    $V_{\mathrm{rel}}^{\infty}/V_{\ast}=0.77$ ($465$\,km\,s$^{-1}$)
    and $d_{\mathrm{min}}/(R_1+R_2)=0.1$. Each plot show the position,
    projected onto the orbital plane, of SPH particles that lie close
    to this plane. Beware that the length scale 
    may change from frame to frame. This collision creates a
    merged star with mass $3.81\,M_{\sun}$. The particles are
    colour-coded according to their rank in the initial stellar models.
    }
  \label{fig:seq_prod15274}
\end{figure*}

\begin{figure*}
  \begin{center}
    \resizebox{!}{20.5cm}{%
      \includegraphics{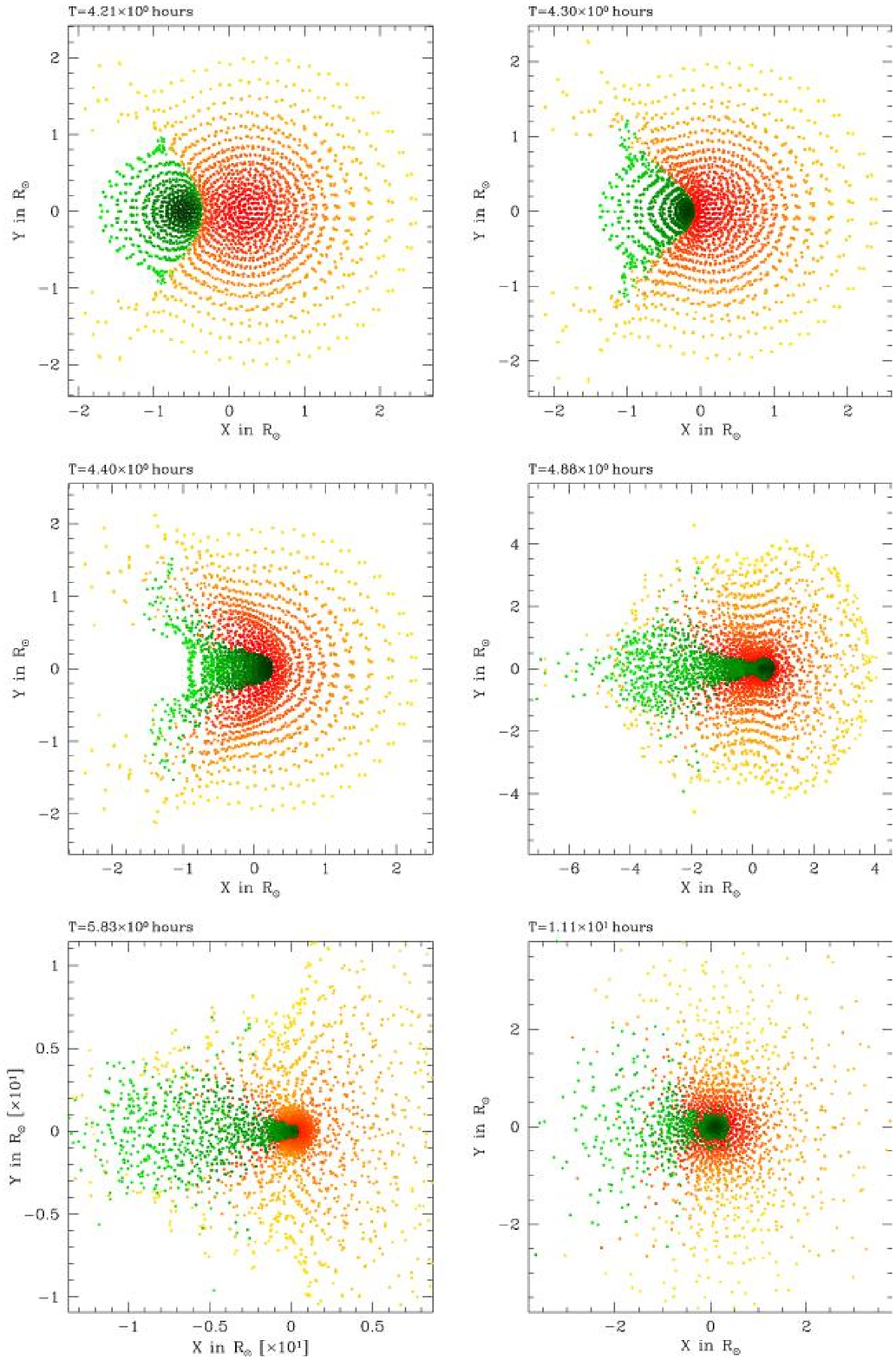}%
      }
  \end{center}
  \caption{%
    Collision between stars of masses 1.0 and $3.0\,M_{\sun}$ at
    $V_{\mathrm{rel}}^{\infty}/V_{\ast}=0.07$ ($44$\,km\,s$^{-1}$) and
    $d_{\mathrm{min}}/(R_1+R_2)=0.0$. This collision creates a merged
    star with mass $3.80\,M_{\sun}$. The particles are colour-coded
    according to their rank in the initial stellar models.  }
  \label{fig:seq_prod15273}
\end{figure*}

\begin{figure*}
  \begin{center}
    \resizebox{\hsize}{20.5cm}{%
      \includegraphics{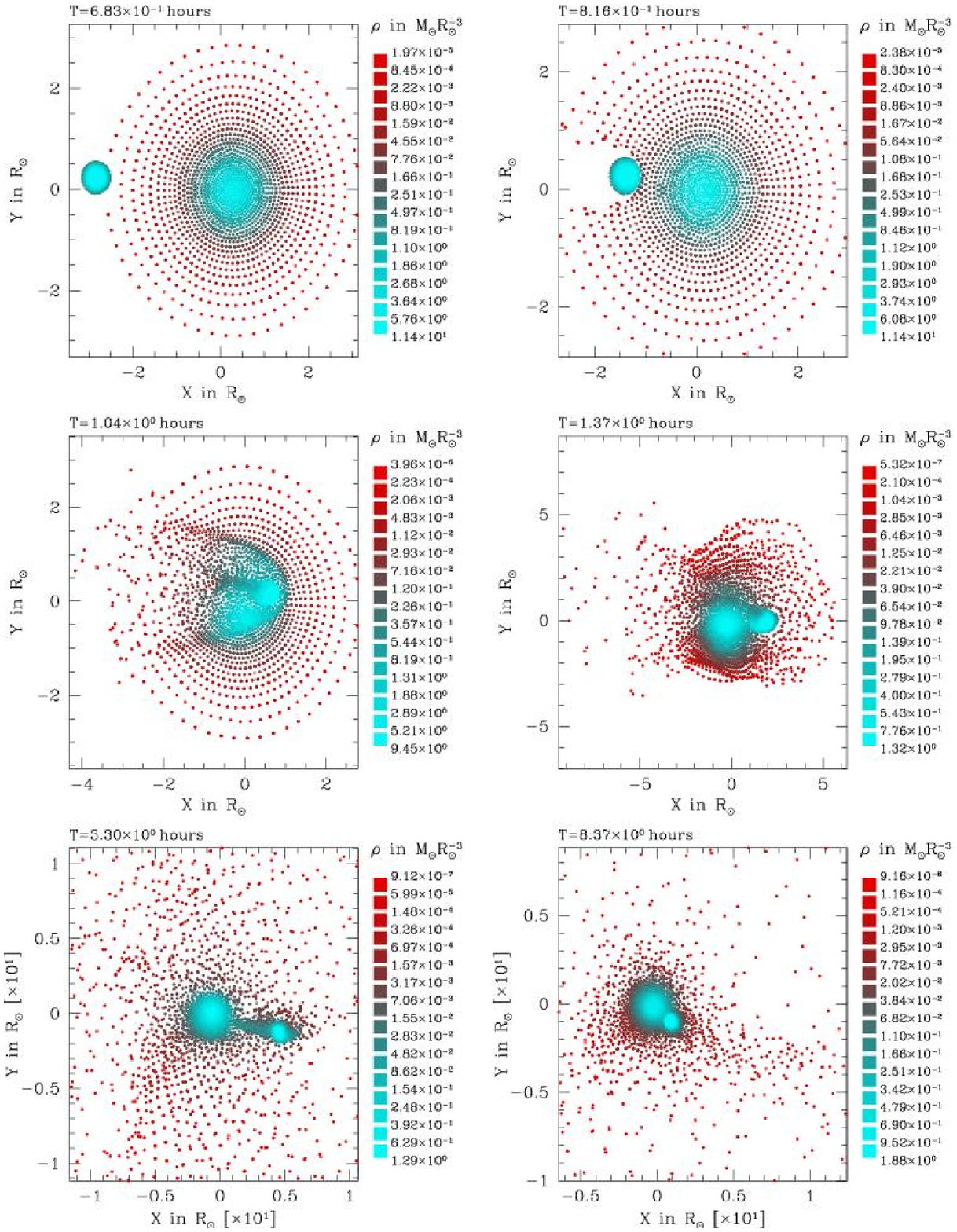}%
      }
  \end{center}
  \caption{%
    Collision between stars of masses $0.4\,M_{\sun}$ and
    $4.0\,M_{\sun}$ at $V_{\mathrm{rel}}^{\infty}/V_{\ast}=3.24$
    ($2180$\,km\,s$^{-1}$)
    and $d_{\mathrm{min}}/(R_1+R_2)=0.04$. 
    The colours code the gas density. This collision results in a
    merged star with mass $3.75\,M_{\odot}$.}
  \label{fig:seq_prod15885}
\end{figure*}

\begin{figure*}
  \resizebox{\hsize}{!}{%
    \includegraphics{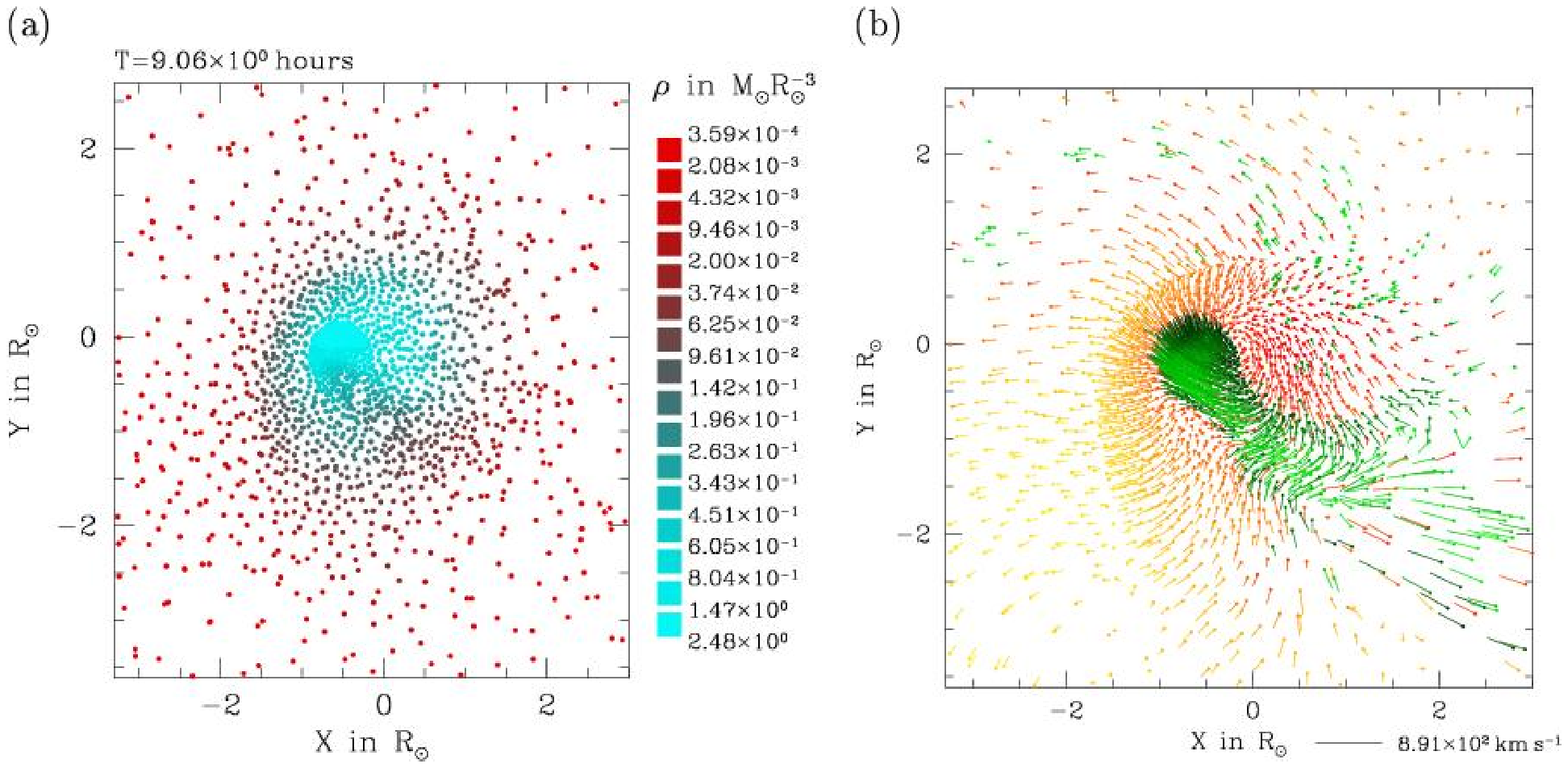}%
    }
  \caption{%
    Snapshot of the collision simulation of
    Fig.~\protect\ref{fig:seq_prod15885} at later stage. One sees the
    small star spiralling into the centre of the larger
    star. Panel~(a): density plot.  Panel~(b): Velocity plot with
    colours indicating the radial position of each particle in the
    initial stellar models. Particles from the large star are coded
    in red to yellow, from centre to surface. Particles from the small 
    star are coded in dark to light green. The small dots are the position 
    of the particles. The velocity scale is given by the horizontal line 
    segment at the bottom right of the diagram.
    }
  \label{fig:end_prod15885}
\end{figure*}

\begin{figure*}
  \resizebox{\hsize}{20.5cm}{%
    \includegraphics{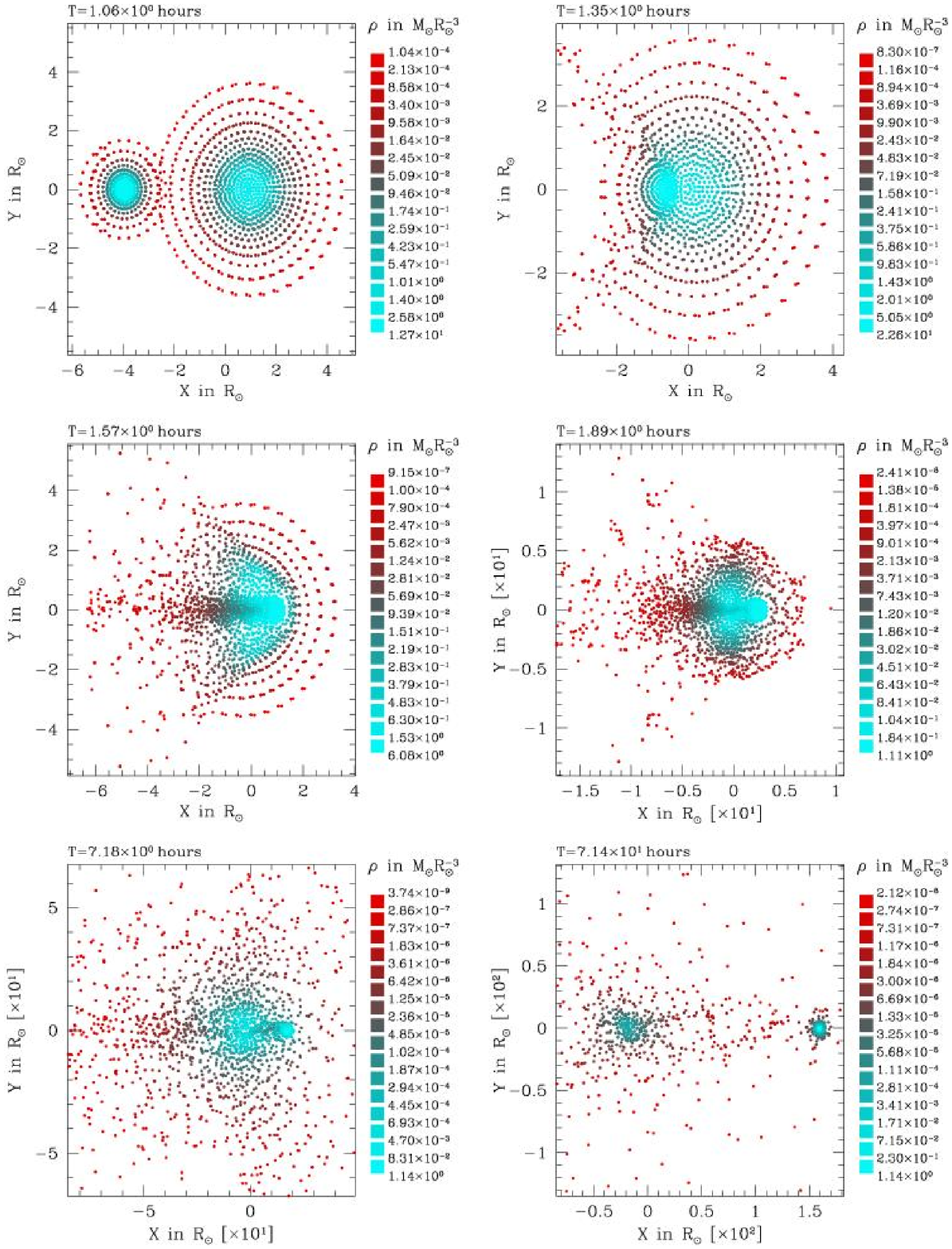}%
    }
  \caption{%
    Collision between stars of masses $1.7\,M_{\sun}$ and
    $7.0\,M_{\sun}$ at $V_{\mathrm{rel}}^{\infty}/V_{\ast}=3.68$
    (2620\,km\,s$^{-1}$) and $d_{\mathrm{min}}/(R_1+R_2)=0.0$. Not
    only is the small star still bound as it emerges from the
    collision, but it has also accreted some gas from the larger star
    so that its final mass is $1.74\,M_{\sun}$! Much damage has been
    caused to the larger star, though, which has lost all but
    $1.94\,M_{\sun}$.  }
  \label{fig:seq_prod15288}
\end{figure*}

\begin{figure*}
  \resizebox{\hsize}{20.5cm}{%
    \includegraphics{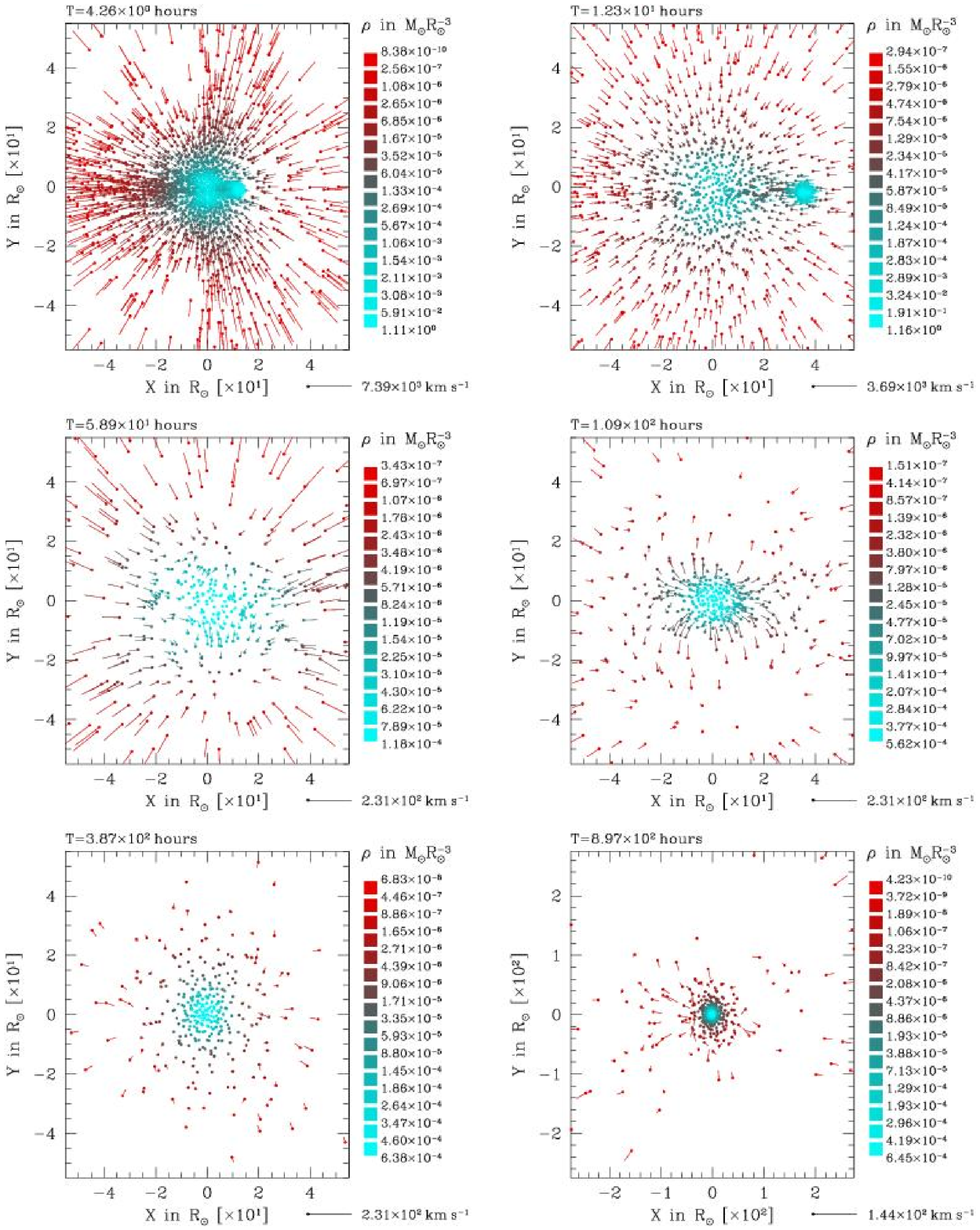}%
    }
  \caption{%
    Same collision as in Fig.~\protect\ref{fig:seq_prod15288}. In this 
    series of plots, Positions and densities are relative to the
    particle that lies at the centre of the larger star at the end of
    the simulations. A constant length scale is applied to all
    diagrams but the last one which shows a larger view. The
    velocity scale is adapted from frame to frame.
    }
  \label{fig:seq_prod15288_vit}
\end{figure*}

\begin{figure*}
  \resizebox{\hsize}{20.5cm}{%
    \includegraphics{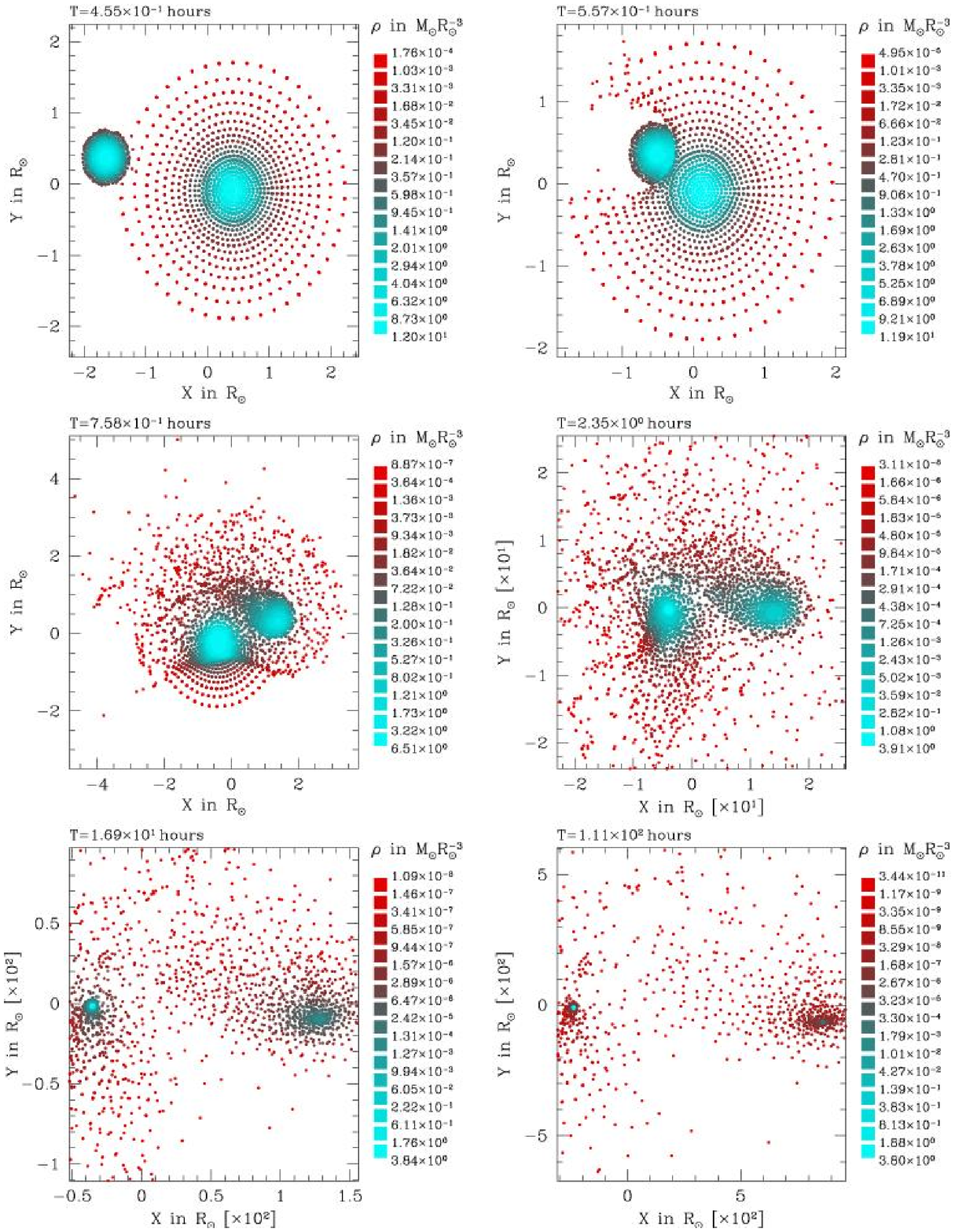}%
    }
  \caption{%
    Collision between stars of masses $0.5\,M_{\sun}$ and
    $2.0\,M_{\sun}$ at
    $V_{\mathrm{rel}}^{\infty}/V_{\ast}=4.48$ (2620\,km\,s$^{-1}$) and
    $d_{\mathrm{min}}/(R_1+R_2)=0.15$. }
  \label{fig:seq_prod15886}
\end{figure*}

\begin{figure*}
  \resizebox{\hsize}{7.5cm}{%
    \includegraphics{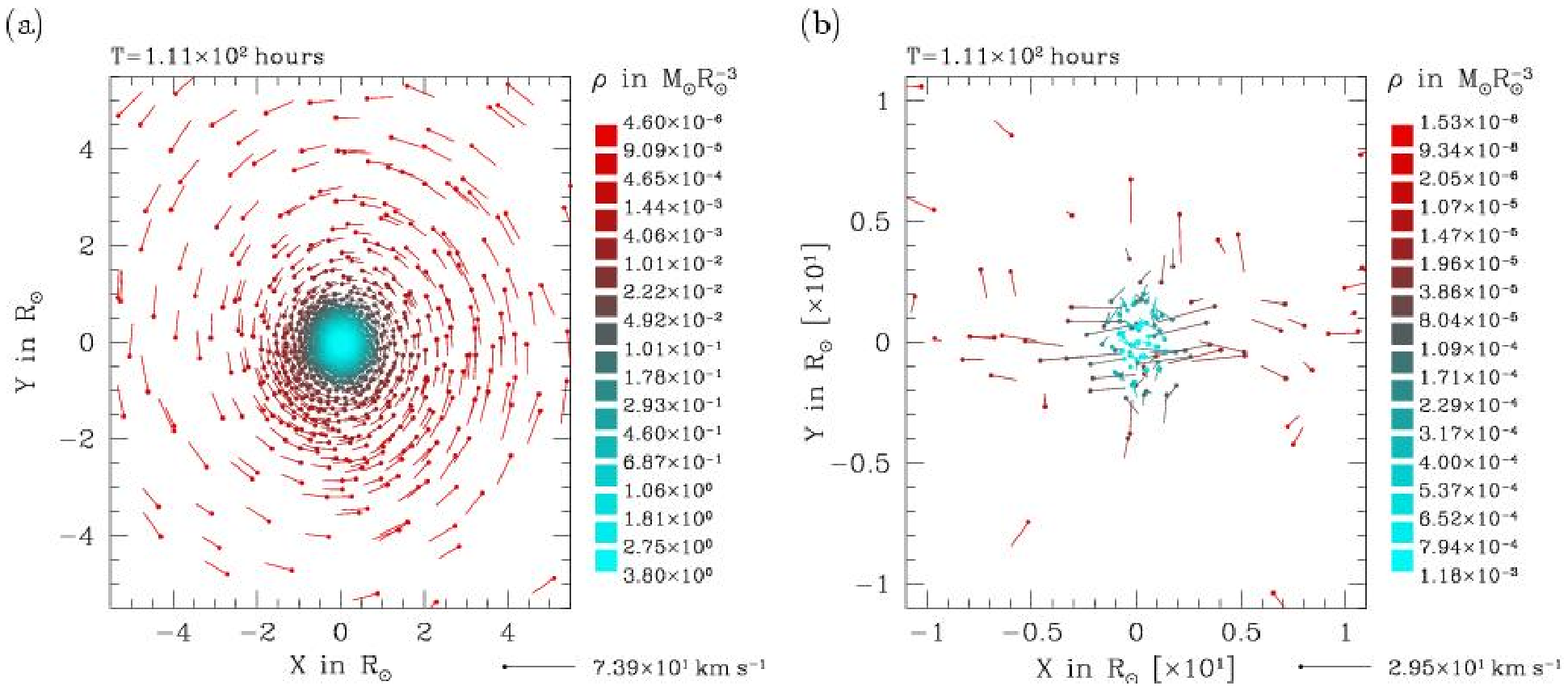}%
    }
  \caption{%
    Enlargements of the last frame of
    Fig.~\protect\ref{fig:seq_prod15886}. Position and velocities
    are relative to the particle with the highest density in each
    panel. Panel~(a): remaining of the
    large star ($1.53\,M_{\sun}$). Panel~(b): remaining of the
    small star ($0.05\,M_{\sun}$).  
    }
  \label{fig:end_prod15886}
\end{figure*}

\begin{figure*}
  \begin{center}
    \resizebox{\hsize}{20.5cm}{%
      \includegraphics{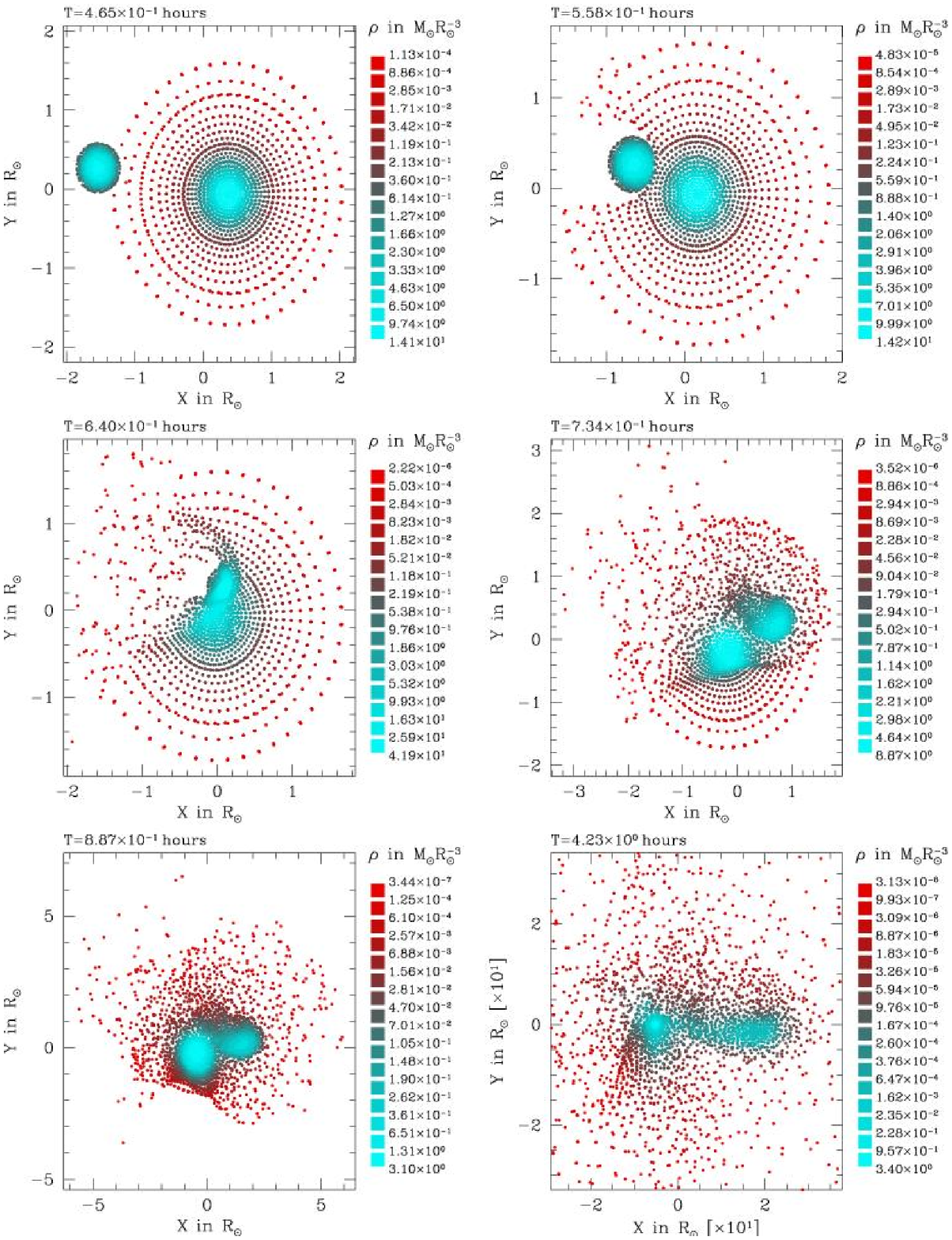}%
      }
  \end{center}
  \caption{%
    Collision between stars of masses $0.4\,M_{\sun}$ and
    $1.7\,M_{\sun}$ at
    $V_{\mathrm{rel}}^{\infty}/V_{\ast}=3.80$ (2180\,km\,s$^{-1}$) and
    $d_{\mathrm{min}}/(R_1+R_2)=0.11$.}
  \label{fig:seq_prod15884a}
\end{figure*}

\begin{figure*}
  \begin{center}
    \resizebox{\hsize}{7.5cm}{%
    \includegraphics{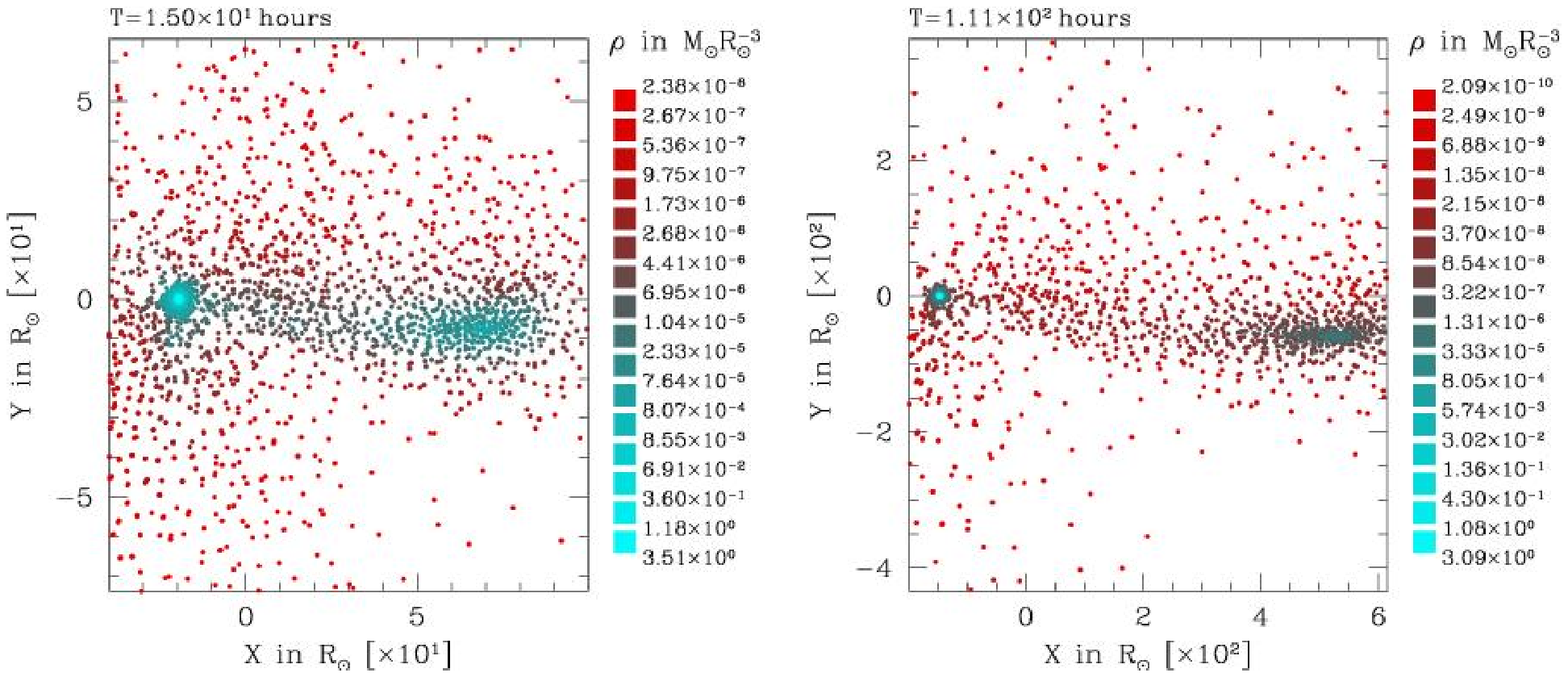}%
    }
  \end{center}
  \caption{%
    Continuation of the sequence of Fig.~\protect\ref{fig:seq_prod15884a}. }
  \label{fig:seq_prod15884b}
\end{figure*}

\begin{figure*}
      \resizebox{\hsize}{7.5cm}{%
      \includegraphics{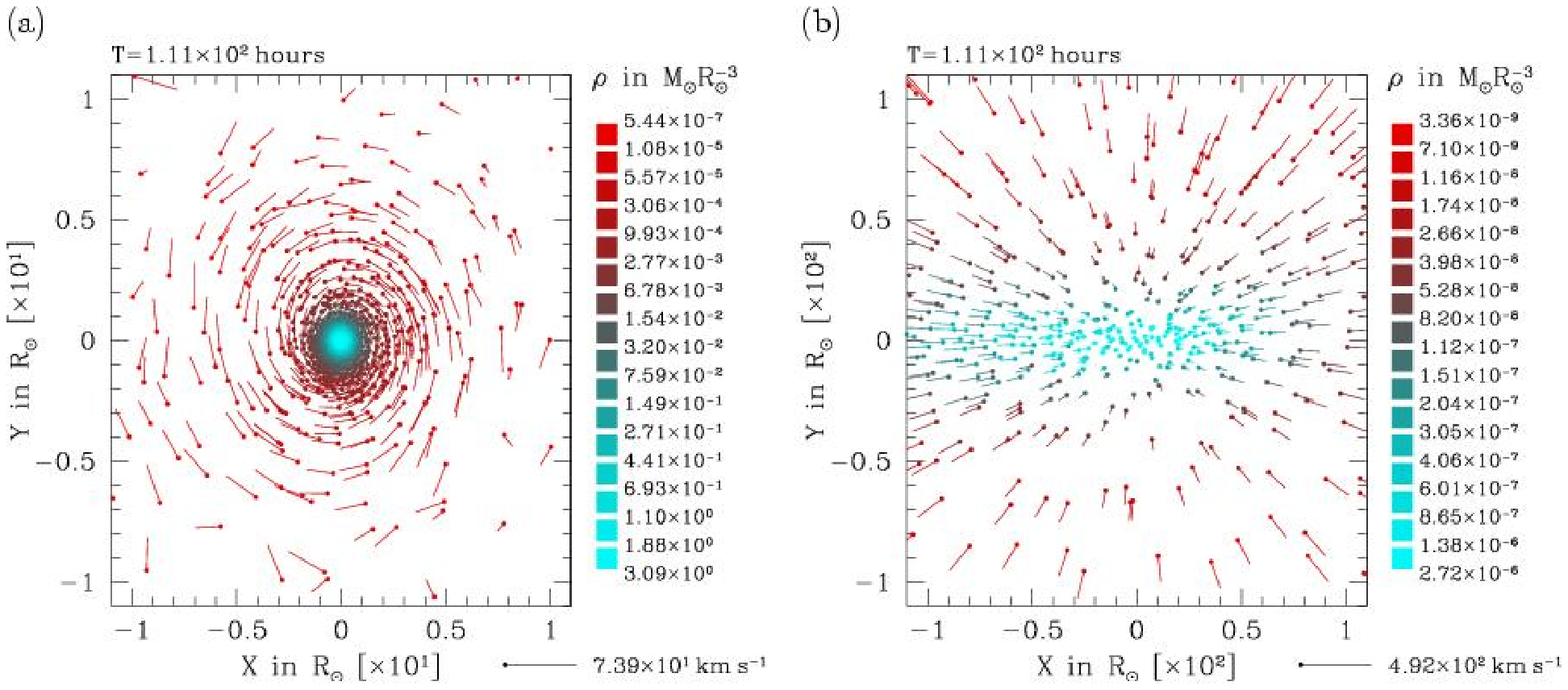}%
      }
  \caption{%
    Enlargements of the last frame of
    Fig.~\protect\ref{fig:seq_prod15884b}. Position and velocities
    are relative to the particle with the highest density in each
    panel.  Panel~(a): surviving core of the large star, a
    $1.26\,M_{\sun}$ rotating star. Panel~(b): remaining of the core
    of the small star, an unbound expanding gas cloud.  The velocity
    of this cloud in the centre-of-mass reference frame of the
    collision is nearly 1000\,km\,s$^{-1}$ while the velocity of the
    small star was initially 1770\,km\,s$^{-1}$.  }
  \label{fig:end_prod15884}
\end{figure*}

Precise descriptions of the physical mechanisms at play during stellar
collisions have already been published \citep{BH87,BH92,LRS93,LRS96}.
In this subsection, we just highlight a few particular collisions from
our survey for illustration purposes.  We do not particularly
concentrate on ``classical'' typical cases because they have been well
covered in these previous works. Instead, we concentrate on
simulations with parameters lying on the border-lines of the various
regimes. Many of them have been re-computed in order to test
surprising results. Indeed, for lack of sufficient disk space for data
storage, only the final ``state'' of each SPH simulation was conserved
for most runs. So, when any doubtful result appeared, we had to
re-compute the complete simulation and write the data to disk
frequently in order to understand the evolution of the
system.

In Fig.~\ref{fig:seq_prod15274}, we show an off-axis low velocity
encounter between identical stars. As the impact parameter is small,
the stars merge together after the first periastron passage. The colour
mapping used in these diagrams allow to trace each particle back to
its initial radial position in the colliding stars. Despite the rather
low resolution (about 8000 particles in total) a tighter and tighter
spiral pattern is clearly visible. As explained by Lombardi and
collaborators, in low velocity collisions, specific entropy $s$ is
nearly conserved as shock heating is weak, and, as stability of the
resulting star imposes $\mathrm{d}s/\mathrm{d}r>0$, low entropy
material that was initially at the centre of the stars, settles in the
centre of the merged object. In fact, for such gentle encounters, one
can even predicts the final material stratification by sorting mass
elements from the two stars in order of increasing entropy
\citep{LRS96,LWRSW02,LombardiEtAl03}. A consequence of this mechanism is that, in
mergers between stars with unequal masses, the core of the smaller
one, having the lowest entropy, sinks to the centre of the merger.
This is also what happens in the two collisions depicted in
Figures~\ref{fig:seq_prod15273}, \ref{fig:seq_prod15885} and
\ref{fig:end_prod15885}. The second collision is an example of a high
velocity merger which produce an object with a total mass slightly
\emph{lower} than those of the initial larger star. Such a case lies
in the tip of merger region in a diagram like those in
Fig.~9 of the main paper.

In Figures~\ref{fig:seq_prod15288} and \ref{fig:seq_prod15288_vit}, we
display snapshots from one of the few head-on collisions in which the
small star pass through the large one and remains essentially intact.
A further peculiarity of this collision is its relatively large mass
ratio: $q=0.24$. No collision with a larger $q$ resulted in a similar
outcome.  Figures~\ref{fig:seq_prod15886} and \ref{fig:end_prod15886}
depict a more typical ``fly-by'' in the sense that it has
non-vanishing impact parameter. However, this high velocity encounter
lies very close to the strip of complete disruption of the small star.
For this particular simulation, the small star loses more than 89\,\%
of its mass! The remaining cloud has a very low central density,
around $10^{-4}$\,g\,cm$^{-3}$. It is made of only 187 particles so
simulations with higher resolution are clearly needed to confirm that
the production of such tiny survivors is not a numerical artifact. It
is however unlikely that such small, rarely formed objects, may have
important astrophysical relevance, either as detectable ``exotic''
stars or dynamically.

We finally present a collision from the high velocity, 1-star branch,
i.e. a case of collisionally induced evaporation of the small star. We
particularly checked that this kind of outcome was real and not some
artifact cause by our analysis software. Indeed, during these
verifications, we noted that many cases of nearly complete destruction
of the small star like the one described in
Figures~\ref{fig:seq_prod15288} and \ref{fig:seq_prod15288_vit} were
mis-interpreted because our code missed the second, much lower,
density peak. Consequently, we had to re-analyse all high $\nu$
collisions that were reported to result in the disruption of the small
star. We conclude that although the precise location of its right
(large $d_{\mathrm{min}}$) edge may depend on numerical issues
(resolution, analysis procedure), the 1-star branch is real.
Inspecting the last frames of Figures~\ref{fig:seq_prod15884a} and
\ref{fig:seq_prod15884b} and Fig.~\ref{fig:end_prod15884} makes this
fact obvious. Furthermore, such collision results have been reported
by \citet{LRS93}.

\subsection{Examples of mass and energy loss results}

\begin{figure*}
      \resizebox{\hsize}{!}{%
      \includegraphics{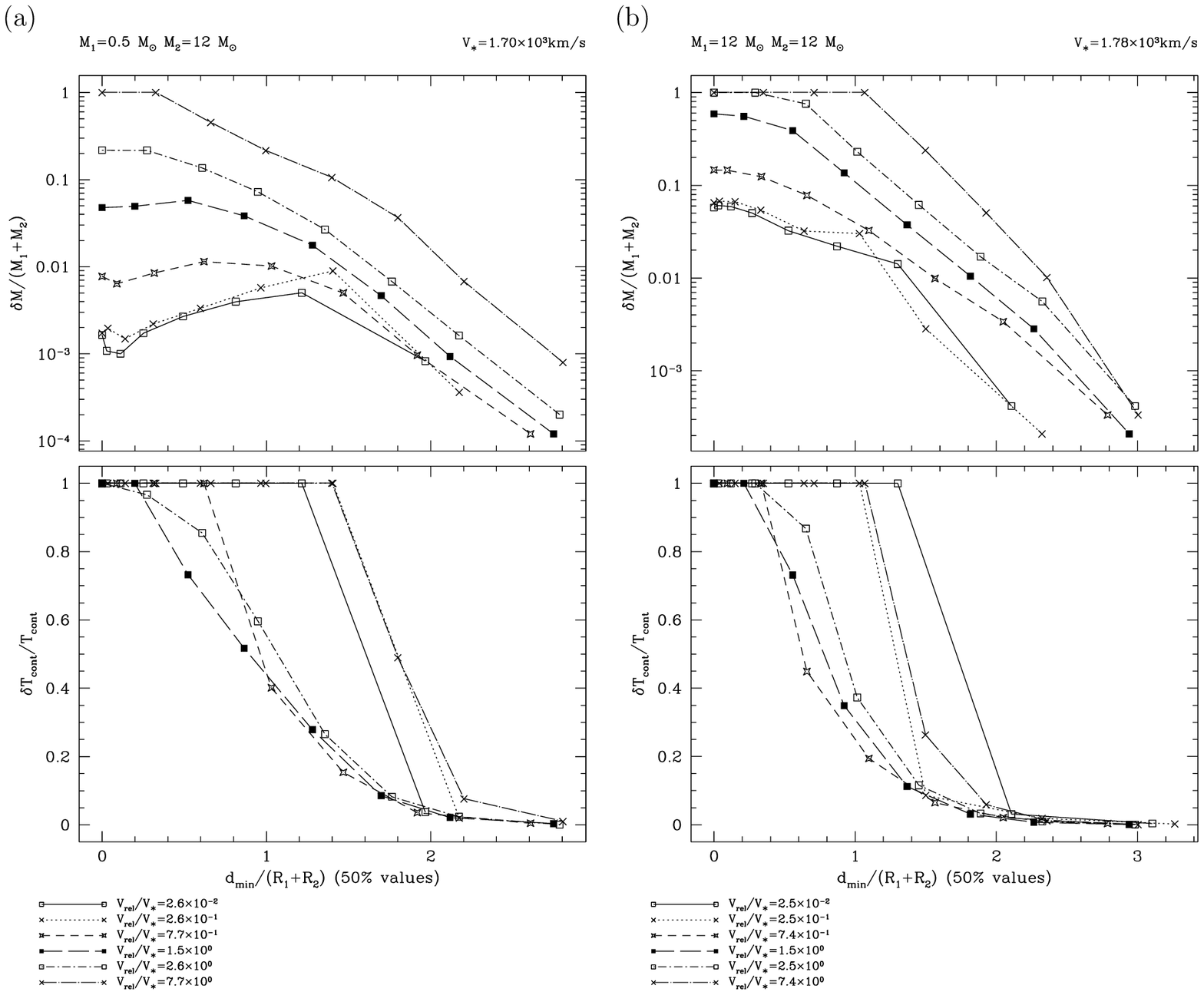}%
      }
  \caption{%
    Relative mass and kinetic energy losses for all collision
    simulations between stars of masses $0.5$ and $12\,\Msun$ (column
    (a)) and $12$ and $12\,\Msun$ (column (b)).  Half-mass radii are
    used to normalise parameters. $T_{\mathrm{cont}}$ is the orbital
    kinetic energy at ``half-mass contact'' (separation equal to
    $R_1^{\mathrm{(h)}}+R_2^{\mathrm{(h)}}$), assuming purely
    Keplerian acceleration. For very small
    $V_{\mathrm{rel}}^{\infty}$, all encounters result either in
    mergers or in bound binaries that should eventually merge
    together, with $\delta T_{\mathrm{cont}}/T_{\mathrm{cont}}=1$ and
    a higher $\delta M/(M_1+M_2)$ as consequences. At high velocities,
    the domain of 100\,\% complete energy loss extends to
    $d_{\mathrm{min}}$ values where mass loss is only partial.  This
    is due to the complete disruption of the smaller star after it
    emerges from the large one. The shoulders on the low velocity mass
    loss curves are due to the formation and subsequent merging of a
    binary.}
  \label{fig:delta_M_E}
\end{figure*}

Fig.~\ref{fig:delta_M_E} shows the energy and mass loss curves for the
simulations of collisions between stars with
$(M_1,\,M_2)=(0.5,\,12)\,\Msun$, and $(12,\,12)\,\Msun$. Similar
curves for other $(M_1,\,M_2)$ couples are available upon request to
MF. The can also easily be drawn using the complete tables of
collision results available on-line.

\section*{Acknowledgements}

We thank Isabelle Baraffe, Corinne Charbonnel and Georges Meynet for
providing stellar structure models and detailed explanations about
them. We are also grateful to the anonymous referee who produced a
very thorough and constructive report; this greatly helped improve the
quality of the article. The work of MF reported here was started at
Geneva Observatory thanks to a grant of the Swiss National Science
Foundation and was supported by Sonder\-forschungs\-bereich (SFB)
439/A5 of German Science Foundation (DFG) at the University of
Heidelberg at the time of the submission of this article.

\bibliographystyle{mn}
\bibliography{aamnem99,biblio}

\begin{thebibliography}{128}
\expandafter\ifx\csname natexlab\endcsname\relax\def\natexlab#1{#1}\fi

\bibitem[{{Alexander}(1999)}]{Alexander99}
{Alexander} T., 1999, ApJ, 527, 835

\bibitem[{{Alexander} \& {Kumar}(2001)}]{AK00}
{Alexander} T., {Kumar} P., 2001, ApJ, 549, 948

\bibitem[{{Arabadjis}(1997)}]{Arab97}
{Arabadjis} J.~S., 1997, PhD thesis, University of Michigan

\bibitem[{{Ayal} {et~al.}(2000){Ayal}, {Livio}, \& {Piran}}]{ALP00}
{Ayal} S., {Livio} M., {Piran} T., 2000, ApJ, 545, 772

\bibitem[{{Bahcall} \& {Wolf}(1976)}]{BW76}
{Bahcall} J.~N., {Wolf} R.~A., 1976, ApJ, 209, 214

\bibitem[{{Bahcall} \& {Wolf}(1977)}]{BW77}
---, 1977, ApJ, 216, 883

\bibitem[{{Bailey} \& {Davies}(1999)}]{BD99}
{Bailey} V.~C., {Davies} M.~B., 1999, MNRAS, 308, 257

\bibitem[{{Barber} {et~al.}(1996){Barber}, {Dobkin}, \& {Huhdanpaa}}]{BDH96}
{Barber} C.~B., {Dobkin} D.~P., {Huhdanpaa} H.~T., 1996, ACM Transactions on
  Mathematical Software, 22, 469

\bibitem[{{Barth}(2004)}]{Barth04}
{Barth} A.~J., 2004, in Coevolution of Black Holes and Galaxies, from the
  Carnegie Observatories Centennial Symposia., {Ho} L., ed., Cambridge
  University Press, p.~21

\bibitem[{{Barth} {et~al.}(2004){Barth}, {Ho}, {Rutledge}, \&
  {Sargent}}]{BHS04}
{Barth} A.~J., {Ho} L.~C., {Rutledge} R.~E., {Sargent} W.~L.~W., 2004, ApJ,
  607, 90

\bibitem[{{Begelman} \& {Rees}(1978)}]{BR78}
{Begelman} M.~C., {Rees} M.~J., 1978, MNRAS, 185, 847

\bibitem[{{Benz}(1990)}]{Benz90}
{Benz} W., 1990, in Numerical Modelling of Nonlinear Stellar Pulsations
  Problems and Prospects, {Buchler} J.~R., ed., pp. 269--288

\bibitem[{{Benz} {et~al.}(1990){Benz}, {Cameron}, {Press}, \&
  {Bowers}}]{BCPB90}
{Benz} W., {Cameron} A. G.~W., {Press} W.~H., {Bowers} R.~L., 1990, ApJ, 348,
  647

\bibitem[{{Benz} \& {Hills}(1987)}]{BH87}
{Benz} W., {Hills} J.~G., 1987, ApJ, 323, 614

\bibitem[{{Benz} \& {Hills}(1992)}]{BH92}
---, 1992, ApJ, 389, 546

\bibitem[{{Benz} {et~al.}(1989){Benz}, {Hills}, \& {Thielemann}}]{BHT89}
{Benz} W., {Hills} J.~G., {Thielemann}, 1989, ApJ, 342, 986

\bibitem[{{Bicknell} \& {Gingold}(1983)}]{BG83}
{Bicknell} G.~V., {Gingold} R.~A., 1983, ApJ, 273, 749

\bibitem[{{Binney} \& {Tremaine}(1987)}]{BT87}
{Binney} J., {Tremaine} S., 1987, Galactic Dynamics. Princeton University Press

\bibitem[{{Bogdanovi{\' c}} {et~al.}(2004){Bogdanovi{\' c}}, {Eracleous},
  {Mahadevan}, {Sigurdsson}, \& {Laguna}}]{BogdanovicEtAl04}
{Bogdanovi{\' c}} T., {Eracleous} M., {Mahadevan} S., {Sigurdsson} S., {Laguna}
  P., 2004, ApJ, 610, 707

\bibitem[{{Bressan} {et~al.}(1993){Bressan}, {Fagotto}, {Bertelli}, \&
  {Chiosi}}]{BFBC93}
{Bressan} A., {Fagotto} F., {Bertelli} G., {Chiosi} C., 1993, A\&AS, 100, 647

\bibitem[{{Carter} \& {Luminet}(1983)}]{CL83}
{Carter} B., {Luminet} J.-P., 1983, A\&A, 121, 97

\bibitem[{{Chabrier} \& {Baraffe}(1997)}]{CB97}
{Chabrier} G., {Baraffe} I., 1997, A\&A, 327, 1039

\bibitem[{{Chabrier} \& {Baraffe}(2000)}]{CB00}
---, 2000, ARA\&A, 38, 337

\bibitem[{{Charbonnel} {et~al.}(1999){Charbonnel}, {D\"{a}ppen}, {Schaerer},
  {Bernasconi}, {Maeder}, {Meynet}, \& {Mowlavi}}]{CDSBMMM99}
{Charbonnel} C., {D\"{a}ppen} W., {Schaerer} D., {Bernasconi} P.~A., {Maeder}
  A., {Meynet} G., {Mowlavi} N., 1999, A\&AS, 135, 405

\bibitem[{{Colgate}(1967)}]{Colgate67}
{Colgate} S.~A., 1967, ApJ, 150, 163

\bibitem[{{Combes}(2001)}]{Combes01}
{Combes} F., 2001, in Advanced Lectures on the Starburst-AGN Connection, INAOE,
  June 2000, {Aretxaga} I., {M{\'{u}}jica} R., {Kunth} D., eds., World
  Scientific, p. 223

\bibitem[{{Courvoisier} {et~al.}(1996){Courvoisier}, {Paltani}, \&
  {Walter}}]{CPW96}
{Courvoisier} T. J.-L., {Paltani} S., {Walter} R., 1996, A\&A, 308, L17

\bibitem[{{David} {et~al.}(1987{\natexlab{a}}){David}, {Durisen}, \&
  {Cohn}}]{DDC87a}
{David} L.~P., {Durisen} R.~H., {Cohn} H.~N., 1987{\natexlab{a}}, ApJ, 313, 556

\bibitem[{{David} {et~al.}(1987{\natexlab{b}}){David}, {Durisen}, \&
  {Cohn}}]{DDC87b}
---, 1987{\natexlab{b}}, ApJ, 316, 505

\bibitem[{{Davies}(1996)}]{Davies96}
{Davies} M.~B., 1996, in IAU Symp. 174: Dynamical Evolution of Star Clusters:
  Confrontation of Theory and Observations, {Hut} P., {Makino} J., eds., p. 243

\bibitem[{{Davies} {et~al.}(1993){Davies}, {Benz}, \& {Hills}}]{DBH93}
{Davies} M.~B., {Benz} W., {Hills} J.~G., 1993, ApJ, 411, 285

\bibitem[{{Davies} {et~al.}(1994){Davies}, {Benz}, \& {Hills}}]{DBH94}
---, 1994, ApJ, 424, 870

\bibitem[{{Davies} {et~al.}(1998){Davies}, {Blackwell}, {Bailey}, \&
  {Sigurdsson}}]{DBBS98}
{Davies} M.~B., {Blackwell} R., {Bailey} V.~C., {Sigurdsson} S., 1998, MNRAS,
  301, 745

\bibitem[{{De{Y}oung}(1968)}]{DeYoung68}
{De{Y}oung} D.~S., 1968, ApJ, 153, 633

\bibitem[{{Dokuchaev} {et~al.}(1993){Dokuchaev}, {Karakula}, \&
  {Tkaczyk}}]{DKT93}
{Dokuchaev} V.~I., {Karakula} S., {Tkaczyk} W., 1993, A\&AS, 97, 109

\bibitem[{{Dokuchaev} \& {Ozernoi}(1977{\natexlab{a}})}]{DO77a}
{Dokuchaev} V.~I., {Ozernoi} L.~M., 1977{\natexlab{a}}, Sov. Astron. Lett., 3,
  112

\bibitem[{{Dokuchaev} \& {Ozernoi}(1977{\natexlab{b}})}]{DO77b}
---, 1977{\natexlab{b}}, Sov. Astron. Lett., 3, 157

\bibitem[{{Duncan} \& {Shapiro}(1983)}]{DS83}
{Duncan} M.~J., {Shapiro} S.~L., 1983, ApJ, 268, 565

\bibitem[{{Eisenstein} \& {Hut}(1998)}]{EH98}
{Eisenstein} D.~J., {Hut} P., 1998, ApJ, 498, 137

\bibitem[{{Evans} \& {Kochanek}(1989)}]{EK89}
{Evans} C.~R., {Kochanek} C.~S., 1989, ApJ Lett., 346, L13

\bibitem[{{Faber} {et~al.}(1997){Faber}, {Tremaine}, {Ajhar}, {Byun},
  {Dressler}, {Gebhardt}, {Grillmair}, {Kormendy}, {Lauer}, \&
  {Richstone}}]{Faber97}
{Faber} S.~M., {Tremaine} S., {Ajhar} E.~A., {Byun} Y., {Dressler} A.,
  {Gebhardt} K., {Grillmair} C., {Kormendy} J., {Lauer} T.~R., {Richstone} D.,
  1997, AJ, 114, 1771

\bibitem[{{Fabian} {et~al.}(1975){Fabian}, {Pringle}, \& {Rees}}]{FPR75}
{Fabian} A.~C., {Pringle} J.~E., {Rees} M.~J., 1975, MNRAS, 172, 15

\bibitem[{{Ferrarese} \& {Ford}(2004)}]{FF04}
{Ferrarese} L., {Ford} H., 2004, {Supermassive Black Holes in Galactic Nuclei:
  Past, Present and Future Research}. preprint, astro-ph/0411247

\bibitem[{{Frank}(1978)}]{Frank78}
{Frank} J., 1978, MNRAS, 184, 87

\bibitem[{{Freitag} \& {Benz}(2001{\natexlab{a}})}]{FB01d}
{Freitag} M., {Benz} W., 2001{\natexlab{a}}, in ASP Conf. Ser. 228: Dynamics of
  Star Clusters and the Milky Way, {Deiters} S., {Fuchs} B., {Just} R.,
  {Spurzem} R., eds., pp. 428--430

\bibitem[{{Freitag} \& {Benz}(2001{\natexlab{b}})}]{FB01a}
---, 2001{\natexlab{b}}, A\&A, 375, 711

\bibitem[{{Freitag} \& {Benz}(2002)}]{FB02b}
---, 2002, A\&A, 394, 345

\bibitem[{{Freitag} {et~al.}(2004{\natexlab{a}}){Freitag}, {G\"urkan}, \&
  {Rasio}}]{FGR04c}
{Freitag} M., {G\"urkan} M.~A., {Rasio} F.~A., 2004{\natexlab{a}}, {Collisions
  Between Single Stars in Dense Clusters: Runaway Formation of a Massive
  Object}. To appear in St-Louis, N. and Moffat, A., eds., Massive Stars in
  Interacting Binaries, astro-ph/0410327

\bibitem[{{Freitag} {et~al.}(2004{\natexlab{b}}){Freitag}, {G\"urkan}, \&
  {Rasio}}]{FGR04b}
---, 2004{\natexlab{b}}, {Run-away formation of intermediate-mass black holes
  in dense star clusters}. To appear in Storchi Bergmann T., Ho L. C. and
  Schmitt H. R., eds., The interplay among Black Holes, Stars and ISM in
  Galactic Nuclei, IAU Symp. 222, astro-ph/0403703

\bibitem[{{Freitag} {et~al.}(2005{\natexlab{a}}){Freitag}, {G\"urkan}, \&
  {Rasio}}]{FGR05}
---, 2005{\natexlab{a}}, Runaway collisions in young star clusters.
  {II}.~numerical results. submitted to MNRAS, astro-ph/0503130

\bibitem[{{Freitag} {et~al.}(2005{\natexlab{b}}){Freitag}, {Rasio}, \&
  {Baumgardt}}]{FRB05}
{Freitag} M., {Rasio} F.~A., {Baumgardt} H., 2005{\natexlab{b}}, Runaway
  collisions in young star clusters. {I}.~methods and tests. submitted to
  MNRAS, astro-ph/0503129

\bibitem[{{Fulbright}(1996)}]{Fulbright96}
{Fulbright} M.~S., 1996, PhD thesis, University of Arizona

\bibitem[{{G{\" u}rkan} {et~al.}(2004){G{\" u}rkan}, {Freitag}, \&
  {Rasio}}]{GFR04}
{G{\" u}rkan} M.~A., {Freitag} M., {Rasio} F.~A., 2004, ApJ, 604

\bibitem[{{Gebhardt} {et~al.}(2003){Gebhardt}, {Richstone}, {Tremaine},
  {Lauer}, {Bender}, {Bower}, {Dressler}, {Faber}, {Filippenko}, {Green},
  {Grillmair}, {Ho}, {Kormendy}, {Magorrian}, \& {Pinkney}}]{GebhardtEtAl03}
{Gebhardt} K., {Richstone} D., {Tremaine} S., {Lauer} T.~R., {Bender} R.,
  {Bower} G., {Dressler} A., {Faber} S.~M., {Filippenko} A.~V., {Green} R.,
  {Grillmair} C., {Ho} L.~C., {Kormendy} J., {Magorrian} J., {Pinkney} J.,
  2003, ApJ, 583, 92

\bibitem[{{Genzel} {et~al.}(2000){Genzel}, {Pichon}, {Eckart}, {Gerhard}, \&
  {Ott}}]{GPEGO00}
{Genzel} R., {Pichon} C., {Eckart} A., {Gerhard} O.~E., {Ott} T., 2000, MNRAS,
  317, 348

\bibitem[{{Genzel} {et~al.}(2003){Genzel}, {Sch{\" o}del}, {Ott}, {Eisenhauer},
  {Hofmann}, {Lehnert}, {Eckart}, {Alexander}, {Sternberg}, {Lenzen}, {Cl{\'
  e}net}, {Lacombe}, {Rouan}, {Renzini}, \& {Tacconi-Garman}}]{GenzelEtAl03}
{Genzel} R., {Sch{\" o}del} R., {Ott} T., {Eisenhauer} F., {Hofmann} R.,
  {Lehnert} M., {Eckart} A., {Alexander} T., {Sternberg} A., {Lenzen} R.,
  {Cl{\' e}net} Y., {Lacombe} F., {Rouan} D., {Renzini} A., {Tacconi-Garman}
  L.~E., 2003, ApJ, 594, 812

\bibitem[{{Genzel} {et~al.}(1996){Genzel}, {Thatte}, {Krabbe}, {Kroker}, \&
  {Tacconi-Garman}}]{GTKKTG96}
{Genzel} R., {Thatte} N., {Krabbe} A., {Kroker} H., {Tacconi-Garman} L.~E.,
  1996, ApJ, 472, 153

\bibitem[{{Gerhard}(1994)}]{Gerhard94}
{Gerhard} O.~E., 1994, in NATO ASIC Proc. 445: The Nuclei of Normal Galaxies:
  Lessons from the Galactic Center, {Genzel} R., {Harris} A.~I., eds., pp.
  267--282

\bibitem[{{Ghez} {et~al.}(2003){Ghez}, {Duch{\^ e}ne}, {Matthews}, {Hornstein},
  {Tanner}, {Larkin}, {Morris}, {Becklin}, {Salim}, {Kremenek}, {Thompson},
  {Soifer}, {Neugebauer}, \& {McLean}}]{GhezEtAl03b}
{Ghez} A.~M., {Duch{\^ e}ne} G., {Matthews} K., {Hornstein} S.~D., {Tanner} A.,
  {Larkin} J., {Morris} M., {Becklin} E.~E., {Salim} S., {Kremenek} T.,
  {Thompson} D., {Soifer} B.~T., {Neugebauer} G., {McLean} I., 2003, ApJ Lett.,
  586, L127

\bibitem[{{Ghez} {et~al.}(2005){Ghez}, {Salim}, {Hornstein}, {Tanner}, {Lu},
  {Morris}, {Becklin}, \& {Duch{\^ e}ne}}]{GhezEtAl03}
{Ghez} A.~M., {Salim} S., {Hornstein} S.~D., {Tanner} A., {Lu} J.~R., {Morris}
  M., {Becklin} E.~E., {Duch{\^ e}ne} G., 2005, ApJ, 620, 744

\bibitem[{{Goodman} \& {Hernquist}(1991)}]{GH91}
{Goodman} J., {Hernquist} L., 1991, ApJ, 378, 637

\bibitem[{{Hansen} \& {Kawaler}(1994)}]{HK94}
{Hansen} C.~J., {Kawaler} S.~D., 1994, Stellar interiors: physical principles,
  structure, and evolution. New York: Springer-Verlag, 1994. 1st ed.

\bibitem[{{Keenan}(1978)}]{Keenan78}
{Keenan} D.~W., 1978, MNRAS, 185, 389

\bibitem[{{Kim} {et~al.}(2004){Kim}, {Figer}, \& {Morris}}]{KFM04}
{Kim} S.~S., {Figer} D.~F., {Morris} M., 2004, ApJ Lett., 607, L123

\bibitem[{{Kim} \& {Lee}(1999)}]{KL99}
{Kim} S.~S., {Lee} H.~M., 1999, A\&A, 347, 123

\bibitem[{{Kippenhahn} \& {Weigert}(1994)}]{KW94}
{Kippenhahn} R., {Weigert} A., 1994, Stellar Structure and Evolution.
  Springer-Verlag Berlin Heidelberg

\bibitem[{{Kormendy}(2004)}]{Kormendy04}
{Kormendy} J., 2004, in Coevolution of Black Holes and Galaxies, from the
  Carnegie Observatories Centennial Symposia., {Ho} L., ed., Cambridge
  University Press, p.~1

\bibitem[{{Kormendy} \& {Richstone}(1995)}]{KR95}
{Kormendy} J., {Richstone} D., 1995, ARA\&A, 33, 581

\bibitem[{{Laguna} {et~al.}(1993){Laguna}, {Miller}, {Zurek}, \&
  {Davies}}]{LMZD93}
{Laguna} P., {Miller} W.~A., {Zurek} W.~H., {Davies} M.~B., 1993, ApJ Lett.,
  410, L83

\bibitem[{{Lai} {et~al.}(1993){Lai}, {Rasio}, \& {Shapiro}}]{LRS93}
{Lai} D., {Rasio} F.~A., {Shapiro} S.~L., 1993, ApJ, 412, 593

\bibitem[{{Langbein} {et~al.}(1990){Langbein}, {Fricke}, {Spurzem}, \&
  {Yorke}}]{LFSY90}
{Langbein} T., {Fricke} K.~J., {Spurzem} R., {Yorke} H.~W., 1990, A\&A, 227,
  333

\bibitem[{{Lauer} {et~al.}(1998){Lauer}, {Faber}, {Ajhar}, {Grillmair}, \&
  {Scowen}}]{Lauer98}
{Lauer} T.~R., {Faber} S.~M., {Ajhar} E.~A., {Grillmair} C.~J., {Scowen} P.~A.,
  1998, AJ, 116, 2263

\bibitem[{{Lee}(1994)}]{Lee94}
{Lee} H.~M., 1994, in NATO ASIC Proc. 445: The Nuclei of Normal Galaxies:
  Lessons from the Galactic Center, {Genzel} R., {Harris} A.~I., eds., p. 335

\bibitem[{{Lee}(1996)}]{Lee96}
---, 1996, in IAU Symp. 169: Unsolved Problems of the Milky Way, {Blitz} L.,
  {Teuben} P., eds., Vol. 169, p. 215

\bibitem[{{Lee} \& {Nelson}(1988)}]{LN88}
{Lee} H.~M., {Nelson} L.~A., 1988, ApJ, 334, 688

\bibitem[{{Lejeune} \& {Schaerer}(2001)}]{LS01}
{Lejeune} T., {Schaerer} D., 2001, A\&A, 366, 538

\bibitem[{{Lombardi} {et~al.}(1996){Lombardi}, {Rasio}, \& {Shapiro}}]{LRS96}
{Lombardi} J.~C. J., {Rasio} F.~A., {Shapiro} S.~L., 1996, ApJ, 468, 797

\bibitem[{{Lombardi} {et~al.}(2003){Lombardi}, {Thrall}, {Deneva}, {Fleming},
  \& {Grabowski}}]{LombardiEtAl03}
{Lombardi} J.~C., {Thrall} A.~P., {Deneva} J.~S., {Fleming} S.~W., {Grabowski}
  P.~E., 2003, MNRAS, 345, 762

\bibitem[{{Lombardi} {et~al.}(2002){Lombardi}, {Warren}, {Rasio}, {Sills}, \&
  {Warren}}]{LWRSW02}
{Lombardi} J.~C., {Warren} J.~S., {Rasio} F.~A., {Sills} A., {Warren} A.~R.,
  2002, ApJ, 568, 939

\bibitem[{{Lombardi} {et~al.}(1999){Lombardi}, {Sills}, {Rasio}, \&
  {Shapiro}}]{LSRS99}
{Lombardi} J. C.~J., {Sills} A., {Rasio} F.~A., {Shapiro} S.~L., 1999, Journal
  of Computational Physics, 152, 687

\bibitem[{{Maeder}(1992)}]{Maeder92}
{Maeder} A., 1992, A\&A, 264, 105

\bibitem[{{Maeder} \& {Meynet}(2000)}]{MM00}
{Maeder} A.~., {Meynet} G., 2000, ARA\&A, 38, 143

\bibitem[{{Magorrian} {et~al.}(1998){Magorrian}, {Tremaine}, {Richstone},
  {Bender}, {Bower}, {Dressler}, {Faber}, {Gebhardt}, {Green}, {Grillmair},
  {Kormendy}, \& {Lauer}}]{Mag98}
{Magorrian} J., {Tremaine} S., {Richstone} D., {Bender} R., {Bower} G.,
  {Dressler} A., {Faber} S.~M., {Gebhardt} K., {Green} R., {Grillmair} C.,
  {Kormendy} J., {Lauer} T., 1998, AJ, 115, 2285

\bibitem[{{Marck} {et~al.}(1996){Marck}, {Lioure}, \& {Bonazzola}}]{MLB96}
{Marck} J.~A., {Lioure} A., {Bonazzola} S., 1996, A\&A, 306, 666

\bibitem[{{Mardling}(1995{\natexlab{a}})}]{Mardling95a}
{Mardling} R.~A., 1995{\natexlab{a}}, ApJ, 450, 722

\bibitem[{{Mardling}(1995{\natexlab{b}})}]{Mardling95b}
---, 1995{\natexlab{b}}, ApJ, 450, 732

\bibitem[{{Mathis}(1967)}]{Mathis67}
{Mathis} J.~S., 1967, ApJ, 147, 1050

\bibitem[{{McMillan} {et~al.}(1981){McMillan}, {Lightman}, \& {Cohn}}]{McMLC81}
{McMillan} S. L.~W., {Lightman} A.~P., {Cohn} H., 1981, ApJ, 251, 436

\bibitem[{{Meynet} {et~al.}(1994){Meynet}, {Maeder}, {Schaller}, {Schaerer}, \&
  {Charbonnel}}]{MMSSC94}
{Meynet} G., {Maeder} A., {Schaller} G., {Schaerer} D., {Charbonnel} C., 1994,
  A\&AS, 103, 97

\bibitem[{{Miralda-Escud{\'e}} \& {Gould}(2000)}]{MEG00}
{Miralda-Escud{\'e}} J., {Gould} A., 2000, ApJ, 545, 847

\bibitem[{{Monaghan}(1992)}]{Monaghan92}
{Monaghan} J.~J., 1992, ARA\&A, 30, 543

\bibitem[{{Monaghan}(1999)}]{Monaghan99}
---, 1999, in Numerical Astrophysics : Proceedings of the International
  Conference on Numerical Astrophysics 1998, {Miyama} S.~M., {Tomisaka} K.,
  {Hanawa} T., eds., pp. 357--366

\bibitem[{{Monaghan} \& {Lattanzio}(1985)}]{ML85}
{Monaghan} J.~J., {Lattanzio} J.~C., 1985, A\&A, 149, 135

\bibitem[{{Murphy} {et~al.}(1991){Murphy}, {Cohn}, \& {Durisen}}]{MCD91}
{Murphy} B.~W., {Cohn} H.~N., {Durisen} R.~H., 1991, ApJ, 370, 60

\bibitem[{{Peebles}(1972)}]{Peebles72}
{Peebles} P. J.~E., 1972, ApJ, 178, 371

\bibitem[{{Pinkney} {et~al.}(2003){Pinkney}, {Gebhardt}, {Bender}, {Bower},
  {Dressler}, {Faber}, {Filippenko}, {Green}, {Ho}, {Kormendy}, {Lauer},
  {Magorrian}, {Richstone}, \& {Tremaine}}]{Pinkney03}
{Pinkney} J., {Gebhardt} K., {Bender} R., {Bower} G., {Dressler} A., {Faber}
  S.~M., {Filippenko} A.~V., {Green} R., {Ho} L.~C., {Kormendy} J., {Lauer}
  T.~R., {Magorrian} J., {Richstone} D., {Tremaine} S., 2003, ApJ, 596, 903

\bibitem[{{Portegies Zwart} \& {McMillan}(2002)}]{PZMcM02}
{Portegies Zwart} S.~F., {McMillan} S. L.~W., 2002, ApJ, 576, 899

\bibitem[{{Press}(1986)}]{Press86}
{Press} W.~H., 1986, in The Use of Supercomputers in Stellar Dynamics, {Hut}
  P., {McMillan} S. L.~W., eds., Springer-Verlag, p. 184

\bibitem[{{Quinlan} \& {Shapiro}(1990)}]{QS90}
{Quinlan} G.~D., {Shapiro} S.~L., 1990, ApJ, 356, 483

\bibitem[{{Rasio} {et~al.}(2004){Rasio}, {Freitag}, \& {G{\" u}rkan}}]{RFG03}
{Rasio} F.~A., {Freitag} M., {G{\" u}rkan} M.~A., 2004, in Coevolution of Black
  Holes and Galaxies, from the Carnegie Observatories Centennial Symposia.,
  {Ho} L., ed., Cambridge University Press, p. 138

\bibitem[{{Rauch}(1999)}]{Rauch99}
{Rauch} K.~P., 1999, ApJ, 514, 725

\bibitem[{{R{\'{o}}{\.{z}}yczka} {et~al.}(1989){R{\'{o}}{\.{z}}yczka}, {Yorke},
  {Bodenheimer}, {M{\"{u}}ller}, \& {Hashimoto}}]{RYBMH89}
{R{\'{o}}{\.{z}}yczka} M., {Yorke} H.~W., {Bodenheimer} P., {M{\"{u}}ller} E.,
  {Hashimoto} M., 1989, A\&A, 208, 69

\bibitem[{{Sanders}(1970{\natexlab{a}})}]{Sanders70b}
{Sanders} R.~H., 1970{\natexlab{a}}, ApJ, 162, 791

\bibitem[{{Sanders}(1970{\natexlab{b}})}]{Sanders70a}
---, 1970{\natexlab{b}}, ApJ, 159, 1115

\bibitem[{{Sch{\" o}del} {et~al.}(2003){Sch{\" o}del}, {Ott}, {Genzel},
  {Eckart}, {Mouawad}, \& {Alexander}}]{SchoedelEtAl03}
{Sch{\" o}del} R., {Ott} T., {Genzel} R., {Eckart} A., {Mouawad} N.,
  {Alexander} T., 2003, ApJ, 596, 1015

\bibitem[{{Schaller} {et~al.}(1992){Schaller}, {Schaerer}, {Meynet}, \&
  {Maeder}}]{SSMM92}
{Schaller} G., {Schaerer} D., {Meynet} G., {Maeder} A., 1992, A\&AS, 96, 269

\bibitem[{{Sedgewick}(1988)}]{Sedgewick88}
{Sedgewick} R., 1988, Algorithms, Second Edition. Addison-Wesley

\bibitem[{{Seidl} \& {Cameron}(1972)}]{SC72}
{Seidl} F. G.~P., {Cameron} A. G.~W., 1972, Ap\&SS, 15, 44

\bibitem[{{Shapiro} \& {Lightman}(1976)}]{SL76}
{Shapiro} S.~L., {Lightman} A.~P., 1976, Nat, 262, 743

\bibitem[{{Shara}(2002)}]{Shara02}
{Shara} M., ed., 2002, ASP Conf. Ser. 263: Stellar collisions \& mergers and
  their consequences.

\bibitem[{{Shara}(1999)}]{Shara99}
{Shara} M.~M., 1999, Physics Reports, 311, 363

\bibitem[{{Shields} \& {Wheeler}(1978)}]{SW78}
{Shields} G.~A., {Wheeler} J.~C., 1978, ApJ, 222, 667

\bibitem[{{Shlosman} {et~al.}(1990){Shlosman}, {Begelman}, \& {Frank}}]{SBF90}
{Shlosman} I., {Begelman} M.~C., {Frank} J., 1990, Nat, 345, 679

\bibitem[{{Sills} {et~al.}(2002){Sills}, {Adams}, {Davies}, \& {Bate}}]{SADB02}
{Sills} A., {Adams} T., {Davies} M.~B., {Bate} M.~R., 2002, MNRAS, 332, 49

\bibitem[{{Sills} {et~al.}(2001){Sills}, {Faber}, {Lombardi}, {Rasio}, \&
  {Warren}}]{SFLRW00}
{Sills} A., {Faber} J.~A., {Lombardi} J.~C., {Rasio} F.~A., {Warren} A.~R.,
  2001, ApJ, 548, 323

\bibitem[{{Sills} \& {Lombardi}(1997)}]{SL97}
{Sills} A., {Lombardi} J.~C., 1997, ApJ Lett., 484, L51

\bibitem[{{Sills} {et~al.}(1997){Sills}, {Lombardi}, {Bailyn}, {Demarque},
  {Rasio}, \& {Shapiro}}]{SLBDRS97}
{Sills} A., {Lombardi} J.~C., {Bailyn} C.~D., {Demarque} P., {Rasio} F.~A.,
  {Shapiro} S.~L., 1997, ApJ, 487, 290

\bibitem[{{Spitzer} \& {Saslaw}(1966)}]{SS66}
{Spitzer} L. J., {Saslaw} W.~C., 1966, ApJ, 143, 400

\bibitem[{{Steinmetz} \& {M{\"{u}}ller}(1993)}]{SM93}
{Steinmetz} M., {M{\"{u}}ller} E., 1993, A\&A, 268, 391

\bibitem[{{Torricelli-Ciamponi} {et~al.}(2000){Torricelli-Ciamponi}, {Foellmi},
  {Courvoisier}, \& {Paltani}}]{TCFCP00}
{Torricelli-Ciamponi} G., {Foellmi} C., {Courvoisier} T. J.-L., {Paltani} S.,
  2000, A\&A, 358, 57

\bibitem[{{Tremaine} {et~al.}(2002){Tremaine}, {Gebhardt}, {Bender}, {Bower},
  {Dressler}, {Faber}, {Filippenko}, {Green}, {Grillmair}, {Ho}, {Kormendy},
  {Lauer}, {Magorrian}, {Pinkney}, \& {Richstone}}]{TremaineEtAl02}
{Tremaine} S., {Gebhardt} K., {Bender} R., {Bower} G., {Dressler} A., {Faber}
  S.~M., {Filippenko} A.~V., {Green} R., {Grillmair} C., {Ho} L.~C., {Kormendy}
  J., {Lauer} T.~R., {Magorrian} J., {Pinkney} J., {Richstone} D., 2002, ApJ,
  574, 740

\bibitem[{{van den Bergh}(1965)}]{vdBergh65}
{van den Bergh} S., 1965, AJ, 70, 124

\bibitem[{{van der Marel}(1999)}]{vdM99c}
{van der Marel} R.~P., 1999, AJ, 117, 744

\bibitem[{{Woltjer}(1964)}]{Woltjer64}
{Woltjer} L., 1964, Nat, 201, 803

\bibitem[{{Young}(1977{\natexlab{a}})}]{Young77b}
{Young} P.~J., 1977{\natexlab{a}}, ApJ, 215, 36

\bibitem[{{Young}(1977{\natexlab{b}})}]{Young77a}
---, 1977{\natexlab{b}}, ApJ, 217, 287

\bibitem[{{Young} {et~al.}(1977){Young}, {Shields}, \& {Wheeler}}]{YSW77}
{Young} P.~J., {Shields} G.~A., {Wheeler} J.~C., 1977, ApJ, 212, 367

\bibitem[{{Yu}(2003)}]{Yu03}
{Yu} Q., 2003, MNRAS, 339, 189

\end{thebibliography}

\label{lastpage}

\end{document}